\documentclass[12pt]{article}

\ifx\pdfoutput\undefined
\usepackage[dvips,bookmarks]{hyperref}
\else
\usepackage{hyperref}
\fi
\hypersetup{colorlinks=false,bookmarksopen,bookmarksnumbered,citecolor=blue,
   pdfstartview=FitH}

\usepackage{latexsym}
\usepackage{amssymb,amsfonts,amsmath}
\usepackage{graphicx}
\usepackage{indentfirst}
\usepackage{bbm}
\usepackage{mathrsfs}

\parskip=\medskipamount

\newcommand {\cA}{{\cal A}}
\newcommand {\cB}{{\cal B}}

\newcommand {\cE}{{\cal E}}
\newcommand {\cF}{{\cal F}}

\newcommand {\cI}{{\cal I}}
\newcommand {\cJ}{{\cal J}}

\newcommand {\cN}{{\cal N}}
\newcommand {\cO}{{\cal O}}
\newcommand {\cP}{{\cal P}}

\newcommand {\cS}{{\cal S}}

\def\a{\alpha}

\def\b{\beta}

\def\d{\delta}
\def\e{\epsilon}

\def\g{\gamma}

\def\j{\psi}
\def\k{\kappa}
\def\l{\lambda}
\def\m{\mu}
\def\n{\nu}
\def\o{\omega}
\def\p{\pi}
\def\q{\theta}
\def\r{\rho}
\def\s{\sigma}
\def\t{\tau}

\def\x{\xi}
\def\z{\zeta}

\def\F{\Phi}
\def\J{\Psi}
\def\L{\Lambda}
\def\O{\Omega}
\def\P{\Pi}

\def\S{\Sigma}
\def\U{\Upsilon}
\def\X{\Xi}
\def\tr{{\rm tr}}
\def\rd{{\rm d}}
\def\ri{{\rm i}}
\def\re{{\rm e}}

\newcommand{\ad}{{\dot{\alpha}}}                           
\newcommand{\bd}{{\dot{\beta}}}                            
\newcommand{\ve}{\varepsilon}                            
\newcommand{\DB}{\bar{D}}

\newcommand{\pa}{\partial}                           
\newcommand{\hf}{\frac12}

\newcommand{\vf}{\varphi}

\newcommand{\be}{\begin{equation}}
\newcommand{\ee}{\end{equation}}
\newcommand{\bea}{\begin{eqnarray}}
\newcommand{\eea}{\end{eqnarray}}
\newcommand{\non}{\nonumber}
\newcommand{\ba}{\begin{array}}
\newcommand{\ea}{\end{array}}

\newcommand{\1}{\underline{1}}

\newcommand{\iu}{\underline{i}}
\newcommand{\ju}{\underline{j}}

\newcommand{\gu}{\underline{g}}
\newcommand{\Tu}{\underline{T}}
\newcommand{\Su}{\underline{S}}
\newcommand{\Zu}{\underline{Z}}
\newcommand{\cPu}{\underline{\cP}}
\newcommand{\cFu}{\underline{\cF}}
\newcommand{\Fu}{\underline{F}}
\newcommand{\Gu}{\underline{G}}
\newcommand{\Lau}{\underline{\L}}
\newcommand{\Siu}{\underline{\S}}
\newcommand{\bm}[1]{\mbox{\boldmath$#1$}}

\def\double #1{#1{\hbox{\kern-2pt $#1$}}}

\newcommand{\hal}{{\hat{\a}}}

\newcommand{\sSU}{\mathsf{SU}}
\newcommand{\sSL}{\mathsf{SL}}
\newcommand{\sGL}{\mathsf{GL}}
\newcommand{\sSO}{\mathsf{SO}}
\newcommand{\sO}{\mathsf{O}}
\newcommand{\sU}{\mathsf{U}}

\newcommand{\sPSU}{\mathsf{PSU}}

\newcommand{\sOSp}{\mathsf{OSp}}

\newcommand{\bsubeq}{\begin{subequations}}
\newcommand{\esubeq}{\end{subequations}}

\numberwithin{equation}{section}

\begin{document}

\begin{titlepage}
\begin{flushright}
May, 2015\\
\end{flushright}
\vspace{2mm}

\begin{center}
{\Large \bf Superconformal field theory in three dimensions:\\
Correlation functions of conserved currents
}\\
\end{center}

\begin{center}

{\bf Evgeny I. Buchbinder, Sergei M. Kuzenko and Igor B. Samsonov%
\footnote{On leave from Tomsk Polytechnic University, 634050
Tomsk, Russia.} }

{\footnotesize{
{\it School of Physics M013, The University of Western Australia\\
35 Stirling Highway, Crawley W.A. 6009, Australia}} ~\\
}

\end{center}

\begin{abstract}
\baselineskip=14pt
For $\cN$-extended superconformal field theories in three spacetime dimensions (3D),
with $1\leq \cN \leq 3$, we compute the two- and three-point correlation functions of
the supercurrent and the flavour current multiplets.
We demonstrate that supersymmetry imposes additional restrictions on the correlators
of conserved currents as compared with the non-supersymmetric case
studied by Osborn and Petkou in hep-th/9307010.
It is shown that
the three-point function of the supercurrent is determined by a single functional form
 consistent with the conservation equation and all the symmetry properties.
Similarly, the  three-point function of the
flavour current multiplets is also determined
by a single functional form in the $\cN=1$ and $\cN=3$ cases.
The specific feature of the $\cN=2$ case is that two independent structures are allowed
for the three-point function of flavour current multiplets, but only one of them
contributes to the three-point function of the conserved  currents contained
in these multiplets. Since the supergravity and super-Yang-Mills Ward identities
are expected to relate the coefficients of the two- and three-point functions under consideration, the results obtained for 3D superconformal field theory are analogous
to those in 2D conformal field theory.

In addition,
we present a new supertwistor
construction for compactified Min\-kowski superspace.
It is
suitable for developing superconformal field theory on 3D spacetimes
other
than Minkowski space, such as $S^1 \times S^2$ and
its universal covering space ${\mathbb R} \times S^2$.
\end{abstract}

\vfill
\end{titlepage}

\newpage
\renewcommand{\thefootnote}{\arabic{footnote}}
\setcounter{footnote}{0}

\tableofcontents{}
\vspace{1cm}
\bigskip\hrule

\section{Introduction}\label{Introduction}
\setcounter{equation}{0}

One of the well-known implications of conformal invariance in $d\geq 2$ dimensions
\cite{Polyakov,Schreier,Migdal}
is that  the functional form of the two- and three-point correlation functions
of primary fields is fixed up to a finite number of parameters.\footnote{In the Euclidean case, these results follow from the following well-known mathematical observation.
For any  three distinct points
$p_1 $, $ p_2 $ and $p_3$ on
the $d$-sphere $S^d = {\mathbb R}^d \bigcup \{\infty\}$, there exists a conformal
transformation  $g \in \sSO_0(d+1,1)$ that maps these points to $\vec{0}$,
$(1, 0, \dots, 0)$ and $\infty$, respectively.
Here $S^d$ is understood to be the conformal compactification of Euclidean space
${\mathbb R}^d$, i.e.
the set of all null straight lines through the origin in ${\mathbb R}^{d+1,1}$. The above observation can be rephrased as the statement that there is no conformal invariant of three points.}
 In general, however, it is a nontrivial technical exercise to determine
explicitly the three-point functions of constrained tensor  operators,
examples of which are the energy momentum-tensor or conserved vector currents.
The point is that such operators obey certain differential constraints
and some work is required in order to classify those functional contributions to the  given three-point  function, which are consistent with all the constraints.
Building on the theoretical ideas and results that may be traced back
as early as the 1970s (see, e.g.,  \cite{FGG,Koller,Mack,TMP,FP}
and references therein), Osborn and Petkou
\cite{OP} presented the group-theoretic formalism to construct the three-point functions
for primary fields of arbitrary spin in $d$ dimensions.\footnote{In four dimensions, the
three-point function of the energy-momentum tensor was first derived
by Stanev \cite{Stanev}.}
They analysed in detail the restrictions
on the correlation functions imposed by the conservation equations for the energy-momentum tensor and conserved currents,
and the outcomes of their study include the following conclusions:
The three-point function of the energy-momentum tensor
contains three
linearly independent functional forms
for $d>3$, while for $d=3$ there are two and for $d=2$ only one.\footnote{In three
dimensions, conformal invariance also allows
parity violating structures for the three-point functions involving either
the energy-momentum tensor, flavour currents
or higher spin currents \cite{Giombi:2011rz, Giombi:2011kc}
(see also \cite{Costa:2011mg}).}
The three-point function of vector currents contains
two linearly independent completely antisymmetric functional forms,
which is in accord with the result obtained in 1971 by Schreier  \cite{Schreier}.
In the $d=4$ case, an additional completely symmetric structure is allowed, which
reflects the presence of anomalies \cite{EO}. In the same $d=4$ case,
the three parameters describing the three-point function of the energy-momentum tensor were demonstrated \cite{OP,EO} to be related to two coefficients in the trace anomaly of a conformal field theory in curved space.

When conformal symmetry is combined with supersymmetry,
the story of the two- and three-point  functions
of conserved currents described in \cite{OP,EO}
has to be supplemented with a sequel, for there appear
conceptually new fermionic conserved currents.
In the realm of supersymmetric field theories,
the energy-momentum tensor is replaced
with the supercurrent \cite{FZ}.
The latter is a supermultiplet
 containing
 the energy-momentum tensor and the supersymmetry current, along with
 some additional components such as the $R$-symmetry current.
Thus the supercurrent
contains fundamental information about the symmetries of
a given supersymmetric field theory.

The supercurrent is the source of supergravity  \cite{OS,FZ2,Siegel},
in the same way as the  energy-momentum tensor is the source of gravity.
For every superconformal field theory, the supercurrent
is an irreducible multiplet that may be coupled to the Weyl multiplet of conformal supergravity.  For non-superconformal theories, the supercurrent is reducible
and contains the so-called trace multiplet which includes
the trace of the energy-momentum tensor. Different supersymmetric
theories may possess different trace multiplets that correspond to different off-shell
formulations for supergravity.

For completeness, it is pertinent to recall the structure of (non-)conformal supercurrents
in four spacetime dimensions (4D). The  $\cN=1$ conformal supercurrent \cite{FZ} is a real vector superfield $J_{\a\bd}$
constrained by $\bar D^\bd J_{\a\bd} =0$ or, equivalently, $D^\a J_{\a \bd} =0$.
The simplest non-conformal supercurrent was given by Ferrara and Zumino
\cite{FZ}; the corresponding conservation equation is $\bar D^\bd J_{\a\bd} = D_\a T$,
where the trace multiplet $T$ is chiral (see \cite{BK} for a review). The Ferrara-Zumino supercurrent proves to be well defined on a dense set in the space of $\cN=1$ supersymmetric field theories.
For recent discussions of the most general 4D $\cN=1$ non-conformal supercurrents,
see \cite{MSW,KS2,K-var}.
The $\cN=2$ conformal supercurrent \cite{Sohnius,HST} is
a real scalar superfield $J$ constrained by $\bar D^i_\ad \bar D^{\ad j} J =0$ or,
equivalently, $D^{\a i }D_\a^j J =0$, see \cite{KT} for more details.  Numerous
$\cN=2$ supersymmetric theories are characterised by the  following conservation
equation \cite{KT,Butter:2010sc}:
$ \bar D^i_\ad \bar D^{\ad j} J = {\rm i} \,T^{ij} +g^{ij} Y$.
Here the trace multiplets
$T^{ij} = T^{ji} = \overline{T_{ij}}$ and $Y$ are linear and reduced chiral
superfields\footnote{The linear superfield is constrained by
$D_\a^{(i}T^{jk)} =\bar D_\ad^{(i}T^{jk)}=0$,
while the reduced chiral superfield obeys the chirality constraint
 $\bar D^i_\ad Y =0$ and the reality condition
 $D^{\a i }D_\a^j Y = \bar D^i_\ad \bar D^{\ad j} \bar Y$.}
respectively;
$g^{ij} = g^{ji}= \overline{g_{ij}}$ is a constant iso-triplet that might be thought of
as an expectation value of the tensor multiplet,
one of the two supergravity compensators, see
 \cite{Butter:2010sc} for the details.

Superconformal symmetry imposes additional restrictions on the structure of
three-point functions of conserved currents as compared with the non-supersymmetric case studied in
\cite{OP,EO}. In 4D $\cN=1$ superconformal theories, the three-point function
of the supercurrent is the sum of two linearly independent functional structures
\cite{Osborn} as compared with the three functional forms in the non-supersymmetric case
\cite{OP}.
The corresponding coefficients $a$ and $c$ were shown \cite{Osborn}  to be
related to those constituting the super-Weyl anomaly in curved superspace
 studied theoretically in \cite{BPT}
 and computed explicitly in \cite{BK86} (see \cite{BK} for a review).\footnote{In his analysis \cite{Osborn},
Osborn used the realisation of superconformal
transformations in 4D $\cN=1$ Minkowski superspace described in \cite{BK}.
Earlier works on superconformal transformations in superspace include
\cite{Sohnius77,Lang,Shizuya}.}
 The same conclusion holds for the three-point function of the supercurrent in 4D $\cN=2$ superconformal field theories \cite{KT}, while
 $\cN=4$ superconformal symmetry is known to demand $a=c$.
 As concerns the three-point function of the flavour current multiplets, there exist two
 independent structures in the 4D $\cN=1$ case \cite{Osborn}
 (as compared with three in the non-supersymmetric case \cite{OP,EO}),
 while $\cN=2$ superconformal symmetry allows only one \cite{KT}.

The present paper is the first in a series
devoted to the correlation functions of conserved currents in $\cN$-extended superconformal field theories in three spacetime dimensions. We start with
a brief discussion of 3D conformal supercurrents and flavour current multiplets.
The 3D $\cN$-extended conformal supercurrents have been described in \cite{KNT-M}
using the conformal superspace formulation for $\cN$-extended conformal supergravity
given in \cite{BKNT-M}.
For every  $\cN=1,2\dots$, the supercurrent is a primary real superfield of certain tensor type and dimension, which obeys some conservation equation formulated in terms of covariant derivatives.
Denoting by $D_\a^I$ the spinor covariant derivatives
of $\cN$-extended Minkowski superspace ${\mathbb R}^{3|2\cN}$,
the conformal supercurrents\footnote{The 3D $\cN=2$ supercurrents were studied in
\cite{DS,KT-M11}.}
for $\cN\leq 4$ are specified by the following properties:
\bea
&&  \left|
\begin{array}{c || c |c|c}
\hline
\mbox{SUSY Type}  ~& ~\mbox{Supercurrent} ~&~\mbox{Dimension}~ &
~\mbox{Conservation Equation}  \\
\hline
\cN=1  ~&  J_{\a\b\g}~& 5/2& ~ D^\a J_{\a\b\g} =0
\\
\hline
\cN=2 ~&  J_{\a\b} ~ &2& ~ D^{I \a} J_{\a\b} =0 \\
\hline
\cN=3 ~&  J_{\a} ~ &3/2& ~ D^{I \a} J_{\a} =0 \\
\hline
\cN=4 ~&  J ~ &1&~ (D^{I \a }D_{\a}^J - \frac{1}{4}\d^{IJ}D^{L \a}D _{\a}^L )J =0\\
\hline
\end{array}
\right| ~~~~~
\label{1}
\eea
For $\cN>4$, the conformal supercurrent is a completely antisymmetric
dimension-1 superfield, $J^{IJKL} = J^{[IJKL]} $, subject to the conservation equation
\bea
D_{\a}^I J^{JKLP} = D_\a^{[I} J^{JKLP]}
- \frac{4}{\cN - 3} D_{\a}^Q J^{Q [JKL} \d^{P] I} ~ .
\eea
The above results follow from the analyses carried out in  \cite{BKNT-M,KNT-M}.
Given an $\cN$-extended superconformal field theory,
it may be coupled to the Weyl multiplet of $\cN$-extended conformal supergravity.
In curved superspace, the supercurrent $J$ (with its indices suppressed)
of the matter model with action $S_{\rm matter}$ is
\bea
J \propto \frac{\d S_{\rm matter} }{\d H}~,
\eea
with  $H$ being an unconstrained  prepotential for conformal supergravity.
The latter has the following index structure:
 $H_{\a\b\g} $
for $\cN=1$ \cite{GGRS}, $H_{\a\b}$ for $\cN=2$ \cite{ZP,Kuzenko12},
$H_\a$ for $\cN=3$ and $H$ for $\cN=4$ \cite{BKNT-M,KNT-M}.\footnote{Using the harmonic superspace techniques
\cite{GIOS}, one may derive the $\cN=3$ and $\cN=4$ prepotentials by generalising
the 4D $\cN=2$ analysis of \cite{KT}.}

The 3D $\cN$-extended flavour current multiplets constitute another family of primary real superfields obeying certain conservation equations.
For $\cN\leq 3$, they have the following structure:
\bea
&&  \left|
\begin{array}{c || c |c|c}
\hline
\mbox{SUSY Type}  ~& ~\mbox{Flavour Current} ~&~\mbox{Dimension}~ &
~\mbox{Conservation Equation}  \\
\hline
\cN=1  ~&  L_{\a}~& 3/2& ~ D^\a L_{\a} =0
\\
\hline
\cN=2 ~&  L ~ &1& ~ (D^{\a I}D_{\a}^J - \frac{1}{2}\d^{IJ}D^{K \a}D _{\a}^K )L =0 \\
\hline
\cN=3 ~&  L^I ~ &1 & ~ D_\a^{ (I}L^{J)} - \frac{1}{3}\d^{IJ}D^{K }_\a L^K  =0 \\
\hline
\end{array}
\right| ~~~~~
\label{2}
\eea
In the $\cN=4$ case, there are two inequivalent flavour current multiplets,
$L^{IJ}_+$ and $L^{IJ}_-$.
Each of them  is described by
a primary antisymmetric dimension-1 superfield,
$L^{I J} = - L^{JI}$, which obeys the conservation equation
\bea
D_{\a}^{I} L^{ J K}&=&
D_{\a}^{[I} L^{ J K]}
- \frac{2}{3}  D_{\a }^L L^{ L[J} \d^{K] I}
~.
\eea
What differs between the two flavour current multiplets, $L^{IJ}_+$ and $L^{IJ}_-$,
is that they are subject to different  self-duality constraints
\bea
 \hf \ve^{IJKL}L^{KL}_\pm  = \pm L^{IJ}_\pm~.
 \eea
The above results naturally follow from the known structure of unconstrained prepotentials
for the $\cN$-extended vector multiplets given in the following publications:
\cite{Siegel79,GGRS} for $\cN=1$,
\cite{Siegel79,HitchinKLR,ZP} for $\cN=2$, \cite{ZH} for $\cN=3$ and
\cite{Zupnik99,Zupnik2009} for $\cN=4$.

The general group-theoretic formalism to construct the two- and three-point functions
of primary superfields in 3D $\cN$-extended Minkowski superspace was
developed by Park \cite{Park3}, as a natural extension of earlier
4D \cite{Osborn,Park4} and 6D \cite{Park6} constructions.
 Instead, we will re-derive the two- and three-point superconformal building blocks,
originally given in \cite{Park3}, by making use of  the 3D $\cN$-extended supertwistor construction
of \cite{KPT-MvU}. Such a re-formulation makes it possible to apply the formalism for computing
correlation functions on more general (conformally flat) superspaces than the standard
Minkowski superspace used in \cite{Park3}.

In this paper we study the correlation functions
of conserved current multiplets in superconformal field theories with $\cN\leq 3$,
while the case $\cN>3$ will be considered elsewhere.
 The main outcomes of our study are as follows: For $\cN\leq3$,
 the three-point function of the supercurrent is determined by a single functional form
 consistent with the conservation equation and all the symmetry properties.
 The same conclusion holds for the three-point function of flavour current multiplets
 in the $\cN=1$ and $\cN=3$ cases.
 As concerns the $\cN=2$ case,  two independent structures are allowed
for the three-point function of flavour current multiplets, but only one of them
contributes to the three-point function of the conserved  currents contained
in these multiplets.
 Thus the 3D superconformal story is analogous to that in 2D conformal field theory.

 In 3D $\cN=2$ superconformal field theories,
of special importance are contact terms of the supercurrent and conserved current
multiplets \cite{CDFKS}. Such contributions to correlation functions
are associated with certain Chern-Simons terms for background fields.
In this paper, we will concentrate on studying the correlation
functions at distinct points where the contact terms do not contribute.

 Before we turn to the technical aspects of this paper, it is worth discussing one more conceptual issue: the symmetry structure of extended supersymmetric field theories from
 the point of view of ``less extended'' supersymmetry.
Every $\cN$-extended superconformal field theory is a special theory with
$(\cN-1)$-extended superconformal symmetry. It is worth elucidating
the structure
of $(\cN-1)$-extended supermultiplets contained in the $\cN$-extended supercurrent or
flavour current multiplet. To uncover this, we split the Grassmann coordinates $\q^\a_I $
of $\cN$-extended Minkowski superspace ${\mathbb M}^{3|2\cN}$ onto two subsets:
(i) the coordinated $\q^\a_{\hat I}$, with $\hat I = 1, \dots, \cN-1$,
corresponding to $(\cN-1)$-extended Minkowski superspace
${\mathbb M}^{3|2(\cN -1)}$; and (ii) two additional coordinates $\q^\a_\cN$.
The corresponding splitting of the spinor derivatives $D_\a^I$  is
$D_\a^{\hat I} $ and $D_\a^\cN$. Given a superfield $V$ on ${\mathbb M}^{3|2\cN}$,
its bar-projection onto ${\mathbb M}^{3|2(\cN-1)}$  is defined by
$V| := V|_{\q_\cN =0} $.

Consider the $\cN=2$ case. The spinor covariant derivatives
$D_\a^{\hat I} $ and $D_\a^\cN$ introduced above, now become $D_\a $ and $D_\a^2$ respectively. The  supercurrent $J_{\a\b}$ leads to the following $\cN=1$
 supermultiplets:
\begin{subequations}
\bea
S_{\a\b} &:=& J_{\a\b}|~, \qquad \qquad D^\a S_{\a\b} =0~; \\
J_{\a\b\g} &:=& \ri D^2_{(\a}J_{\b\g)} |~; \qquad D^\a J_{\a\b\g} =0~.
\label{1.7b}
\eea
\end{subequations}
Here $J_{\a\b\g}$ is the $\cN=1$ supercurrent, while the additional superfield
$S_{\a\b}$  contains the $\sU(1) $
$R$-symmetry current (the $\q$-independent component of $S_{\a\b}$)
and the second supersymmetry current
(the top component of $S_{\a\b}$).

The  $\cN=3$ supercurrent $J_{\a}$
leads to the following $\cN=2$ supermultiplets:
\begin{subequations}
\bea
R_\a &:=& J_\a|~, \qquad \qquad D^{\hat I \a} R_\a = 0 ~; \label{1.6a}\\
J_{\a\b} &:= &  D^3_{(\a} J_{\b)} |~, \qquad D^{\hat I \a} J_{\a\b} =0~.
\eea
\end{subequations}
Here $J_{\a\b}$ is the $\cN=2$ supercurrent, while
$R_\a$ contains the third supersymmetry current and two
$R$-symmetry currents corresponding to $\sSO(3)/\sSO(2)$.

Next, the $\cN=4$ supercurrent $J$
contains the following $\cN=3$
supermultiplets:
\begin{subequations}
\bea
S &:=& J|~, \qquad (D^{\hat I\a }D_{\a}^{\hat J} - \frac{1}{3}\d^{\hat I \hat J}
D^{\hat K \a}D _{\a}^{\hat K} )S =0~; \\
J_\a&:=& \ri D^4_\a J |~, \qquad D^{\hat I \a} J_\a =0~.
\eea
\end{subequations}
Here $J_{\a}$ is the $\cN=3$ supercurrent, while
$S$ contains the fourth supersymmetry current and three
$R$-symmetry currents corresponding to $\sSO(4)/\sSO(3)$.
Upon reduction to $\cN=2$ superspace, the scalar $S$ generates two
primary $\cN=2$ superfields: (i) the scalar $S|_{\q_2=0}$, which is an
$\cN=2$ flavour current multiplet; and (ii) the spinor  $D^3_\a S|_{\q_2=0}$,
which is of the type \eqref{1.6a}.

Finally, we just mention the $\cN \to (\cN-1)$ decomposition of flavour current multiplets.
The $\cN=2$ multiplet $L$ leads to the following $\cN=1$
real supermultiplets:
\begin{subequations}
\bea
S &:=& L|~; \\
L_{\a} &:=& \ri D^2_{\a}L |~; \qquad D^\a L_{\a} =0~.
\label{1.10b}
\eea
\end{subequations}
Here $L_\a$ is an $\cN=1$ flavour current multiplet, and
the real scalar $S$ is unconstrained. The $\cN=3$ multiplet $L^I$
leads to an $\cN=2$ flavour current multiplet $L$ and a chiral scalar.

Therefore, if one studies $\cN$-extended superconformal field theories in
$(\cN-1)$-extended superspace, it is not sufficient to analyse the correlation
functions of those currents which correspond to the manifestly realised symmetries.

There is a remarkable difference between superconformal field theories and ordinary  conformal ones in diverse dimensions. For the action of the conformal group on (compactified) Minkowski
 space, there is no conformal invariant of three points. The situation is different
 in superspace. On (compactified) Minkowski superspace, the superconformal group does not act transitively on the set consisting of triples of distinct superspace points.
 As a result, there exist
{\it nilpotent} superconformal invariants of three points. Such invariants have been
constructed by Park in diverse dimensions \cite{Park:1997bq,Park3,Park4,Park6}.

This paper is organised as follows.
Following \cite{KPT-MvU}, in section 2
we review the supertwistor construction of $\cN$-extended
compactified Minkowski superspace  $\overline{\mathbb M}{}^{3|2\cN}$.
Minkowski superspace  ${\mathbb M}{}^{3|2\cN}$ originates as a dense open subset
of  $\overline{\mathbb M}{}^{3|2\cN}$.
It is shown that  $\overline{\mathbb M}{}^{3|2\cN}$
is a homogeneous space for the superconformal group $\sOSp(\cN|4, {\mathbb R})$,
while only the infinitesimal superconformal transformations are well defined on
${\mathbb M}{}^{3|2\cN}$.
Section 3 describes  a different isomorphic realisation for $\sOSp(\cN|4, {\mathbb R})$.
Using this realisation, we
 construct a global supermatrix parametrisation of $\overline{\mathbb M}{}^{3|2\cN}$
as well as a smooth metric
on $\overline{\mathbb M}{}^{3|2\cN}$, which  only scales under the superconformal transformations.
The supertwistor formalism is used in section 4 to derive all building blocks
in terms of which the two- and three-point functions of primary superfields are
constructed. The general structure of the two- and three-point functions of primary superfields
is described in section 5 following \cite{Park3}. The two- and three-point functions for
the supercurrent and flavour current multiplets in superconformal field theories
with $\cN=1$,  $\cN= 2$ and $\cN=3$ are computed in sections 6, 7 and 8 respectively.
Concluding comments are given in section 9.

We have also included several technical appendices.
Appendix A gives a summary of our 3D notation and conventions.
Appendix \ref{AppendixB} is devoted to
the correlation functions involving (anti)chiral superfields.
Appendix C is concerned with
the $\cN=2\to\cN=1$ superspace reduction of
the three-point functions for the $\cN=2$ supercurrent and flavour current multiplets.
In Appendix D
we reduce to components the three-point function for $\cN=1$ flavour current multiplets.


\section{Supertwistor construction}

In this section we describe the supertwistor construction of $\cN$-extended
compactified Minkowski superspace.
Our presentation mostly follows the construction given in \cite{KPT-MvU}
and inspired by \cite{K-compactified06} (see also
\cite{K-compactified12}).

\subsection{Supertwistors and the superconformal group}\label{section2}

In three spacetime dimensions, the $\cN$-extended superconformal group\footnote{This
supergroup was denoted  $\sOSp(\cN|2, {\mathbb R})$ in \cite{KPT-MvU}.}
is  $\sOSp(\cN|4; {\mathbb R})$.
It naturally acts on the space of {\it real even} supertwistors and on
the space of {\it real odd} supertwistors.

An arbitrary supertwistor is a column vector
\bea
T = (T_A) =\left(
\begin{array}{c}
T_\hal \\
\hline \hline
 T_I
\end{array}
\right)~, \qquad
(T_\hal ) = \left(
\begin{array}{c}
 f_\a \\
  g^\b
  \end{array}
\right)~, \qquad \a, \b = 1,2 ~, \quad I = 1, \dots, \cN ~.
\eea
In the case of even supertwistors, $ T_\hal$ is bosonic
and $T_I$ is fermionic.
In the case of odd supertwistors, $ T_\hal$ is fermionic while  $T_I$ is bosonic.
The even and odd supertwistors are called pure.
We introduce the parity function $\ve ( T )$ defined as:
$\ve ( T ) = 0$ if $ T$ is even, and $\ve ( T ) =1$ if $T $ is odd.
It is also useful to define
\bea
 \ve_A = \left\{
\begin{array}{c}
 0 \qquad A=\hal \\
 1 \qquad A=I
\end{array}
\right.{}~.
\non
\eea
Then the components $T_A$ of a pure supertwistor
 have the following  Grassmann parities
\bea
\ve ( T_A) = \ve ( T ) + \ve_A \quad (\mbox{mod 2})~.
\eea
A pure supertwistor is said to be real if its components obey the reality condition
\bea
\overline{T_A} = (-1)^{\ve(T) \ve_A + \ve_A} T_A~.
\label{realcon}
\eea
The space of complex (real) even supertwistors is naturally identified with
${\mathbb C}^{4|\cN}$ (${\mathbb R}^{4|\cN}$),
while the space of complex (real) odd supertwistors may be identified with
${\mathbb C}^{\cN |4}$
(${\mathbb R}^{\cN |4}$).

Introduce a graded antisymmetric supermatrix
\bea
{\mathbb J} = ({\mathbb J}^{AB}) = \left(
\begin{array}{c ||c}
J ~&~ 0 \\\hline \hline
0 ~& ~{\rm i} \,{\mathbbm 1}_\cN
\end{array} \right) ~, \qquad
J
=\big(J^{\hat \a \hat \b} \big)
=\left(
\begin{array}{cc}
0  & {\mathbbm 1}_2\\
 -{\mathbbm 1}_2  &    0
\end{array}
\right) ~,
\label{supermetric}
\eea
where ${\mathbbm 1}_\cN $ denotes the unit  $\cN \times \cN $ matrix.
With the aid of $\mathbb J$, we may define a graded symplectic inner product on the space of pure supertwistors by the rule: for arbitrary pure supertwistors $T$ and $S$,
the inner product is
\bea
\langle { T}| { S} \rangle_{\mathbb J}: = {T}^{\rm sT}{\mathbb J} \, {S}
~,
\label{innerp}
\eea
where the row vector  ${ T}^{\rm sT} $ is defined by
\bea
{ T}^{\rm sT} := \big( T_\hal , - (-1)^{\ve(T)}  T_I \big)
= (  T_A (-1)^{\ve(T)\ve_A +\ve_A} )
\eea
and is called the super-transpose of $T$.
The above inner product has  the following symmetry property
\bea
\langle { T}_1 | { T}_2  \rangle_{\mathbb J}
= -(-1)^{\ve_1 \ve_2} \langle { T}_2 | { T}_1  \rangle_{\mathbb J} ~,
\eea
where $\ve_i$ stands for the Grassmann parity of ${ T}_i$.
If $T$ and $S$ are real supertwistors, then applying the complex conjugation
gives
\bea
\overline{
\langle { T}| { S} \rangle_{\mathbb J} } =
-\langle {S}| { T} \rangle_{\mathbb J}~.
\eea

By definition, the supergroup   $\sOSp(\cN|4; {\mathbb C})$
consists of those even $(4|\cN) \times (4|\cN)$ supermatrices
\bea
g = (g_A{}^B) ~, \qquad \ve(g_A{}^B) = \ve_A + \ve_B ~,
\eea
which preserve the inner product \eqref{innerp} under the action
\bea
T =(T_A) ~\to ~ gT = (g_A{}^B T_B)~. \label{2.9}
\eea
Such a transformation maps the space of even (odd) supertwistors onto itself.
The condition of invariance of the inner  product \eqref{innerp}
under \eqref{2.9} is
\bea
g^{\rm sT} {\mathbb J}\, g = {\mathbb J} ~, \qquad
(g^{\rm sT})^A{}_B := (-1)^{\ve_A \ve_B + \ve_B} g_B{}^A~.
\label{groupcond}
\eea
The subgroup $\sOSp(\cN|4; {\mathbb R}) \subset \sOSp(\cN|4; {\mathbb C})$
consists of those transformations which preserve the reality condition
\eqref{realcon},
\begin{subequations}
\bea
\overline{T_A} = (-1)^{\ve(T) \ve_A + \ve_A} T_A \quad \longrightarrow \quad
\overline{(gT)_A} = (-1)^{\ve(T) \ve_A + \ve_A} (gT)_A~.
\eea
This is equivalent  to
\bea
\overline{ g_A{}^B} = (-1)^{\ve_A \ve_B + \ve_A} g_A{}^B \quad
\Longleftrightarrow \quad g^\dagger = g^{\rm sT}~.
\eea
In conjunction with \eqref{groupcond}, this reality condition is equivalent to
\bea
g^\dagger {\mathbb J}\, g = {\mathbb J} ~.
\eea
\end{subequations}

A dual supertwistor
\bea
{ Z} = ({Z}^A) = \left(
{Z}^{\hal}, Z^I
\right)
\eea
is a row vector that transforms under  $\sOSp(\cN|4; {\mathbb R}) $ such that $Z^A T_A$ is invariant
for every supertwistor $T$,
\bea
Z~\to ~Z' = Zg^{-1}~,  \qquad g \in\sOSp(\cN|4; {\mathbb R}) ~.
\eea
A dual supertwistor $Z$ is even (odd) if $Z^A T_A$ is a $c$-number
for every even (odd) supertwistor $T$. Given a pure dual supertwistor $Z$,
its super-transpose $Z^{\rm sT}$ will be defined to be the following column vector
\bea
(Z^{\rm sT})^A := (-1)^{\ve(Z) \ve_A +\ve (Z) } Z^A~,
\eea
such that $Z^A T_A = (T^{\rm sT})_A (Z^{\rm sT} )^A$.

The superconformal algebra $\mathfrak{osp}(\cN |4; {\mathbb R})$ consists of real supermatrices $\O$
obeying the master equation
\bea
{ \O}^{\rm sT} {\mathbb J} + {\mathbb J} \,{ \O}=0~.
\eea
The general solution of this equation is
\bea
{ \O}
&=& \left(
\begin{array}{c | c ||c}
  \l -\hf \s {\mathbbm 1}_2  ~& ~ \check{b} &~\sqrt{2} \check{\eta}^{\rm T}   \\
\hline
-\hat{a}  ~&   -  \l^{\rm T} +\hf \s {\mathbbm 1}_2~&~ -\sqrt{2}\hat{\e}^{\rm T}
\\
\hline
\hline
{\rm i}\sqrt{2}\,\hat \e ~& ~{\rm i}\sqrt{2}\, \check \eta &~\L
\end{array}
\right) \non \\
&\equiv& \left(
\begin{array}{c | c ||c}
  \l_\a{}^\b -\hf \s \d_\a{}^\b  ~& ~ b_{\a \b} &~\sqrt{2} \eta_{\a J}   \\
\hline
-a^{\a \b}  ~&   -  \l^\a{}_\b +\hf \s \d^\a{}_\b~&~ -\sqrt{2}\e^\a{}_J
\\
\hline
\hline
{\rm i}\sqrt{2}\,{\e_I{}^\b} ~& ~{\rm i}\sqrt{2}\, \eta_{I\b}&~\L_{IJ}
\end{array}
\right) ~,
\label{SP-g} \\
&&  \l_\a{}^\a =0~, \qquad {a}^{\a\b} = {a}^{\b \a} ~, \qquad {b}_{\a \b} = {b}_{\b \a}~, \qquad
\L_{IJ}=-\L_{JI}~.
\non
\eea
The bosonic  parameters $\l_\a{}^\b$, $\s$, $a_{\a\b}$, $b^{\a\b}$ and $\L_{IJ}$, as well as
the fermionic parameters $\e^\a{}_I\equiv \e_I{}^\a $ and $\eta_{\a I} \equiv \eta_{I\a}$
in (\ref{SP-g}) are real.

\subsection{Compactified Minkowski superspace}

In accordance with \cite{KPT-MvU}, the compactified $\cN$-extended Minkowski
superspace $\overline{\mathbb M}{}^{3|2\cN}$ is defined to be the set of all
Lagrangian subspaces of ${\mathbb R}^{4|\cN}$, the space of real even supertwistors.
We recall that a Lagrangian subspace of ${\mathbb R}^{4|\cN}$ is a maximal isotropic subspace of ${\mathbb R}^{4|\cN}$.
By definition, such a subspace is spanned by two even supertwistors
${ T}^\m$ with the properties
that (i) the bodies of ${T}^1$ and $ {T}^2$ are linearly independent;
(ii) they obey the null condition
\bea
\langle { T}^1 | { T}^2  \rangle_{\mathbb J}=0~;
\label{null3.1}
\eea
(iii) they are defined only modulo
the equivalence relation
\bea
\{ { T}^\m \} ~ \sim ~ \{ \tilde{ T}^\m \} ~, \qquad
\tilde{ T}^\m = { T}^\n\,\X_\n{}^\m~,
\qquad \X \in \sGL(2,{\mathbb R}) ~.
\label{super-nullplane2}
\eea
Equivalently, the space  $\overline{\mathbb M}{}^{3|2\cN}$ can be defined
to consist of rank-two supermatrices of the form
\bea
\cP = \big( { T}^1 ~ { T}^2 \big)=\left(
\begin{array}{c}
{ F}\\ { G} \\ \hline \hline
{\rm i} \U
\end{array}
\right) ~, \qquad
G^{\rm T}F =F^{\rm T}G + {\rm i}\U^{\rm T} \,\U~,
\label{super-two-plane}
\eea
which are defined modulo the equivalence relation
\bea
\cP =\left(
\begin{array}{c}
 F\\  G \\ \hline \hline
 {\rm i} \U
\end{array}
\right) ~ \sim ~
\left(
\begin{array}{r}
 F\, \X\\  G\,\X \\ \hline \hline
 {\rm i}\U\,\X
\end{array}
\right) = \cP \X
~, \qquad \X \in \sGL(2,{\mathbb R}) ~.
\eea
Here $F$ and $G$ are $2\times 2$
real bosonic matrices, and $\U$ is a $\cN \times 2$
real fermionic matrix.
The null condition \eqref{null3.1} can be rewritten as
\bea
\cP^{\rm sT} {\mathbb J} \cP =0~.
\eea

It may be shown that
the superconformal group $ \sOSp(\cN|4; {\mathbb R})$ acts transitively
on the compactified Minkowski superspace.
Thus  $\overline{\mathbb M}{}^{3|2\cN}$ can be identified with the coset space
 $ \sOSp(\cN|4; {\mathbb R}) / G_{\cP_0}$, where $G_{\cP_0} $ denotes
 the isotropy group at a given two-plane $\cP_0 \in \overline{\mathbb M}{}^{3|2\cN}$.

\subsection{Minkowski superspace}

As discussed in \cite{KPT-MvU},
Minkowski superspace ${\mathbb M}{}^{3|2\cN}$ is identified with a dense open subset
$U_F$ of $\overline{\mathbb M}{}^{3|2\cN}$ spanned by supermatrices (\ref{super-two-plane})
under the condition
\bea
\det F \neq 0
~.
\eea
Every null two-plane in
${\mathbb M}{}^{3|2\cN}$
may be described by a supermatrix
of the form
\bea
\cP  \sim\left(
\begin{array}{c}
 {\mathbbm 1}_2 \\ - \hat{\bm x} \\ \hline \hline
 {\rm i} \sqrt{2} \hat \q \phantom{\Big|}
\end{array} \right)
= \left(
\begin{array}{c}
 \d_\a{}^\b  \\ - {\bm x}^{\a\b} \\ \hline \hline \phantom{\Big|}
 {\rm i} \sqrt{2}  \q_I{}^\b
\end{array} \right) \equiv \cP(z) ~,
\label{2.23}
\eea
where the real matrix $\hat{\bm x}$ is constrained by
\bea
\hat{\bm x} - \hat{\bm x}^{\rm T} = 2\ri \hat{\q}^{\rm T} \hat \q \quad
\Longrightarrow \quad {\bm x}^{\a\b} = x^{\a\b} +\ri \q_I{}^\a \q_I{}^\b~,
\quad x^{\a\b } = x^{\b\a}~.
\label{2.24}
\eea
The points of ${\mathbb M}^{3|2\cN}$
are naturally
parametrised by the variables
$ z^A = (x^a, \q^\a_I)$.

Given a group element $g \in  \sOSp(\cN|4; {\mathbb R})$, its action $\cP \to g \cP$
on the two-plane $\cP (z)  \in {\mathbb M}{}^{3|2\cN}$
can be represented as
\bea
g \,
\left(
\begin{array}{c}
{\mathbbm 1}_2 \\  -\hat{\bm x} \\
\hline \hline \phantom{\Big|}
{\rm i}\sqrt{2}\,\hat \q
\end{array}
\right)
= \left(
\begin{array}{c}
{\mathbbm 1}_2 \\ -\hat{\bm x}' \\ \hline \hline \phantom{\Big|}
{\rm i}\sqrt{2}\,\hat{\q}'
\end{array}
\right) \X(g;z) ~, \qquad  \X(g;z) \in \sGL(2,{\mathbb R})
\label{3.8}
\eea
provided the transformed two-plane, $g \cP (z)$,  belongs to ${\mathbb M}{}^{3|2\cN}$.
In general, this property holds only locally, since $\X(g;z) $ may become singular for
certain group elements $g$ (special conformal transformations)
and some spacetime points $x$.

Let us consider an infinitesimal superconformal transformation,
$g= {\mathbbm 1} + \k \O $,
where $\k$ is an infinitesimal parameter and $\O$ is given by  (\ref{SP-g}).
Then from (\ref{3.8})
we derive
\begin{subequations} \label{3.10}
\bea
\d \hat{\bm x} &=& \hat{a} - \l^{\rm T} \hat{\bm x} - \hat{\bm x} \l +\s \hat{\bm x}
+\hat{\bm x}\check{b} \,\hat{\bm x} + 2{\rm i}\, \hat{\e}^{\rm T}\hat \q
- 2{\rm i} \,\hat{\bm x} \check{\eta}^{\rm T}\hat \q~, \\
\d \hat\q &=&\hat \e -\hat \q \l +\hf \s  \hat \q +\L \hat \q +  \hat \q \check{b}\hat{\bm x}
-\check \eta \hat{\bm x}-2{\rm i}\, \hat \q\, \check \eta^{\rm T} \hat \q~.
\eea
\end{subequations}
We see that the
matrix elements in \eqref{SP-g} correspond to a
Lorentz transformation  ($\l_\a{}^\b$),  spacetime translation  ($a^{\a\b}$),
special conformal transformation ($b_{\a\b}$), dilatation ($\s$),
$Q$-supersymmetry ($\e_I{}^\b$), $S$-supersymmetry ($\eta_{I\b}$) and
$R$-symmetry transformation ($\L_{IJ}$).

As pointed out in the previous subsection,
  $\overline{\mathbb M}{}^{3|2\cN}$ can be identified with  the homogeneous space
 $ \sOSp(\cN|4; {\mathbb R}) / G_{\cP_0}$, where $G_{\cP_0} $ denotes
 the isotropy group at a given two-plane $\cP_0 \in \overline{\mathbb M}{}^{3|2\cN}$.
Consider a special null two-plane $\cP_0 \in\overline{\mathbb M}{}^{3|2\cN}$
which corresponds to the origin of ${\mathbb M}{}^{3|2\cN}$,
that is  $\cP_0 = \cP(z=0)$.
Its isotropy group  $G_{\cP_0}$
 is the subgroup of $ \sOSp(\cN|4; {\mathbb R})$ generated by
supermatrices \eqref{SP-g} of the form:
\bea
\o &=& \left(
\begin{array}{c | c ||c}
  \l -\hf \s {\mathbbm 1}_2  ~& ~ \check{b} &\sqrt{2} \check{\eta}^{\rm T}   \\
\hline
0  ~&   -  \l^{\rm T} +\hf \s {\mathbbm 1}_2~&~ 0\\
\hline \hline
0 ~& ~{\rm i}\sqrt{2}\, \check \eta &~\L
\end{array}
\right) ~.
\label{isotropy_subalgebra}
\eea
The isotropy group $G_{\cP_0}$ consists of the followings supermatrices
\begin{subequations} \label{isotropy_group}
\bea
\left(
\begin{array}{c | c ||c}
 A  ~& ~ A\check{\bm b} &\sqrt{2} A\check{\eta}^{\rm T}   \\
\hline
0  ~&   (A^{-1})^{\rm T}~&~ 0\\
\hline \hline
0 ~& ~{\rm i}\sqrt{2}\, R\check \eta &~R
\end{array}
\right) ~, \qquad A \in \sGL(2, {\mathbb R})~,  \quad R \in \sO(\cN)~,
\eea
where
the $2\times 2$ matrix $\check{\bm b}$
is constrained by
\bea
\check{\bm b} - \check{\bm b}^{\rm T} = 2\ri \check{\eta}^{\rm T} \check \eta \quad
\Longrightarrow \quad {\bm b}_{\a\b} = b_{\a\b} +\ri \eta_{I \a }\eta_{I\b}~,
\quad b_{\a\b } = b_{\b\a}~.
\eea
\end{subequations}
As follows from \eqref{isotropy_group},
 $G_{\cP_0}$ includes space reflections.
Choosing $i =1$ or $i=3$,
let us consider the following element of  $G_{\cP_0}$:
\bea
g_i = \left(
\begin{array}{c | c ||c}
 \s_i  ~& ~ 0 & 0   \\
\hline
0  ~&   \s_i~&~ 0\\
\hline \hline
0 ~& ~0 &~{\mathbbm 1}_\cN
\end{array}
\right) ~,
\eea
with $\s_i$ being the $i$-th
Pauli matrix,
$\det \s_i = -1$.
Associated with this group element is the transformation on ${\mathbb M}{}^{3|2\cN}$
\bea
P_i: \qquad \hat x' = \s_i \hat x \s_i~, \qquad \hat{\q}'= \hat \q \s_i~,
\eea
which is a reflection about one of the coordinate axes in two-space.

It is also seen from \eqref{isotropy_group} that
 $G_{\cP_0}$ includes arbitrary $R$-symmetry transformations from the group
 $\sO(\cN)$
and  not necessarily from its connected component of the identity, $\sSO(\cN)$, as discussed by  \cite{Park3}.

A complement of the  subalgebra
\eqref{isotropy_subalgebra}
in $ \frak{osp}(\cN|4; {\mathbb R})$ generates
a subgroup of the superconformal group consisting of all supermatrices of the form:
\bea \label{232}
s(a, \e) &=& \left(
\begin{array}{c | c ||c}
  {\mathbbm 1}_2 ~& ~ 0 ~&~0  \\
\hline
-\hat{\bm a}  ~& ~ {\mathbbm 1}_2~& \, -\sqrt{2}\hat{\e}^{\rm T}
\\
\hline
\hline
{\rm i}\sqrt{2}\,\hat \e ~& ~0 ~&~{\mathbbm 1}_\cN
\end{array}
\right) ~, \qquad \hat{\bm a} = \hat{a}+ \ri \hat{\e}^{\rm T} \hat \e ~.
\eea
Such a supermatrix describes a spacetime translation ($\e=0$) and a $Q$-supersymmetry transformation ($a=0$) when acting on ${\mathbb M}{}^{3|2\cN}$.

The {$\cN$-extended Minkowski superspace} can be also
realised as a homogeneous space. The standard realisation is
\bea
{\mathbb M}^{3|2\cN}
={\mathfrak P}(3|\cN)\big/ {\sSL}(2,{\mathbb R})
~,
\eea
where ${\mathfrak P}(3|\cN)$ denotes the $\cN$-extended super-Poincar\'e group
and
${\sSL}(2,{\mathbb R})$ the spin group in three spacetime dimensions.
Every group element $g \in {\mathfrak P}(3|\cN)$ can uniquely be represented
in the form $g = s(a, \e) \, h( M)$, where
\bea
h(M) &=&  \left(
\begin{array}{c | c ||c}
  M  ~& ~ 0~ &~0   \\
\hline
0  ~& ~  (M^{-1})^{\rm T}~&~ 0
\\
\hline
\hline
0 ~& ~0~&~{\mathbbm 1}_\cN
\end{array}
\right) ~, \qquad
M \in \sSL(2, {\mathbb R})~.
\eea
Here the supermatrix $h(M)$ describes a Lorentz transformation.
The points of ${\mathbb M}^{3|2\cN}$
can be parametrised by
the following
coset representative
\bea
s(z) :=s(x, \q) &=& \left(
\begin{array}{c | c ||c}
  {\mathbbm 1}_2 ~& ~ 0 ~&~0  \\
\hline
-\hat{\bm x}  ~& ~ {\mathbbm 1}_2~& \, -\sqrt{2}\hat{\q}^{\rm T}
\\
\hline
\hline
{\rm i}\sqrt{2}\,\hat \q ~& ~0 ~&~{\mathbbm 1}_\cN
\end{array}
\right) ~.
\label{s(z)}
\eea

The $Q$-supersymmetry transformation $s(0, \e) $ acts on ${\mathbb M}^{3|2\cN}$
according to the law $s(z)  \to s(z') = s(0,\e) s(z) $, and thus
\bea
x'^{\a\b} = x^{\a \b} + {\rm i} ( \e^\a_I  \q^\b_I  +   \e^\b_I \q^\a_I)~, \qquad
\q'^\a_I = \q^\a_I + \e^\a_I~.
\label{Q2.34}
\eea
These results can be rewritten
as
\bea
z'^A =z^A -{\rm i} \, \e^\b_J Q^J_\b \,z^A~,
\eea
where we have introduced
the supersymmetry generators
\bea
Q^I_\a ={\rm i}\, \frac{\pa}{\pa \q^\a_I} + (\g^m)_{\a\b}\, \q^\b_I \pa_m
= {\rm i}\, \frac{\pa}{\pa \q^\a_I} +   \q^\b_I \pa_{\b\a}~.
\eea
From here we immediately  read off the spinor covariant derivatives
\begin{subequations} \label{D}
\bea
D^I_\a =\frac{\pa}{\pa \q^\a_I} + {\rm i}  (\g^m)_{\a\b}\, \q^\b_I \pa_m
=  \frac{\pa}{\pa \q^\a_I} + {\rm i}  \q^\b_I \pa_{\b\a}~,
\eea
which anti-commute with the supercharges, $\{D^I_\a, Q^J_\b\} =0$,
and obey  the anti-commutation relations
\bea
\big\{ D^I_\a , D^J_\b \big\} = 2{\rm i}\, \d^{IJ}  (\g^m)_{\a\b}\,\pa_m~.
\eea
\end{subequations}

We introduce the 3D extension of the Volkov-Akulov supersymmetric one-form
\cite{VA,AV}
\bea
\hat e = \rd \hat{\bm x} - 2\ri \hat{\q}^{\rm T} \rd \hat \q = \rd \hat x +\ri \rd \hat{\q}^{\rm T} \hat \q
-\ri \hat{\q}^{\rm T} \rd \hat \q~,
\qquad \hat{e}^{\rm T} = \hat e~.
\label{VA}
\eea
It is obviously invariant under the $Q$-supersymmetry transformation
\eqref{Q2.34}.

\subsection{Twin Minkowski superspace}\label{subsection3.3}

The chart $U_F$, which we have identified with Minkowski superspace,
does not cover $\overline{\mathbb M}{}^{3|2\cN}$. Another
 dense open subset
$U_G$ of $\overline{\mathbb M}{}^{3|2\cN}$ consists  of
those supermatrices (\ref{super-two-plane}) which are characterised by
\bea
\det G \neq 0
~.
\eea
Every null two-plane in $U_G$
may be described by a supermatrix of the form
\bea
\cP  \sim\left(
\begin{array}{c}
\check{\bm y}
 \\ {\mathbbm 1}_2 \\
 \hline \hline
 {\rm i} \sqrt{2} \check \r
\end{array} \right)
= \left(
\begin{array}{c}
{\bm y}_{\a\b}
  \\   \d^\a{}_\b\\  \hline \hline \phantom{\Big|}
  {\rm i} \sqrt{2}  \r_{I\b}
\end{array} \right) ~,
\eea
where the real $2\times 2$ matrix $\check{\bm y}$ is constrained by
\bea
\check{\bm y} - \check{\bm y}^{\rm T} = 2\ri \check{\r}^{\rm T} \check \r \quad
\Longleftrightarrow \quad {\bm y}_{\a\b} = y_{\a\b} +\ri \r_{I\a} \r_{I\b}~,
\quad y_{\a\b } = y_{\b\a}~.
\eea

One may think of $U_G$ as a twin of $U_F$
obtained by replacing the spacetime translations and $Q$-supersymmetry
transformations with the special conformal boosts and $S$-supersymmetry
transformations, respectively. The two-plane $\cP_0$, which is the origin of $U_F$,
 is replaced with $\cP_\infty$ corresponding to $y_{\a\b}=0$ and $\r_{I\a}=0$,
the origin of $U_G$.
The two-plane
$\cP_{\infty} $ is an infinitely separated point from the viewpoint of $U_F$.
The isotropy group $\cP_{\infty} $, denoted $G_{\cP_\infty}$,
consists of the following supermatrices
\bea
\left(
\begin{array}{c | c ||c}
 A  ~& ~ 0 & 0   \\
\hline
- (A^{-1})^{\rm T}\hat{\bm a}  ~&   (A^{-1})^{\rm T}~&~  -\sqrt{2}(A^{-1})^{\rm T}\hat{\q}^{\rm T}\\
\hline \hline
{\rm i}\sqrt{2}\,R\hat \q ~& ~0 &~R
\end{array}
\right) ~, \qquad A \in \sGL(2, {\mathbb R})~,  \quad R \in \sO(\cN)~,~~~
\eea
where $\hat{\bm a} $ is defined in \eqref{232}.
The following one-form
\bea
\check e_{\rm M} = \rd \check{\bm y} - 2\ri \check{\r}^{\rm T} \rd \check \r
~,\qquad \check{e}^{\rm T}_{\rm M} = \check e_{\rm M}
\label{VA-M}
\eea
is invariant under the $S$-supersymmetry transformations.

In the intersection of the two charts introduced, $U_F \bigcap U_G $,
the transition functions are
\bea
\check{\bm y} = -\hat{\bm x}^{-1} ~, \qquad \check{\r} = - \hat{\q} \hat{\bm x}^{-1}~.
\eea
The one-forms \eqref{VA} and \eqref{VA-M} are related to each other
by the rule
\bea
\hat e = (\check{\bm y}^{\rm T})^{-1} \check{e}_{\rm M} \check{\bm y}^{-1}~.
\eea
The charts $U_F$ and $U_G$ are mapped onto each other by  the superconformal
transformation
\bea
 \left(
\begin{array}{c ||c}
-J ~&~ 0 \\\hline \hline
0 ~& ~{\mathbbm 1}_\cN
\end{array} \right) \in \sOSp(\cN|4; {\mathbb R})~,
\eea
where the matrix $J$ is defined by \eqref{supermetric}.

The two charts $U_F$ and $U_G$ do not cover the compactified Minkowski superspace.
It may be shown  \cite{KPT-MvU}
that the bosonic body of $\overline{\mathbb M}{}^{3|2\cN} \setminus
\{ U_F \bigcup U_G\} $ is topologically $S^1$.
Additional charts are required if we are interested in the global description of
$\overline{\mathbb M}{}^{3|2\cN} $. Instead of introducing such additional charts,
there is actually a better way out. It turns out that there exists an isomorphic realisation for
$\sOSp(\cN|4; {\mathbb R}) $ that is ideally suited for a global description of
$\overline{\mathbb M}{}^{3|2\cN} $. It will be presented in the next section.

\subsection{Alternative definition of compactified Minkowski superspace}
\label{subsection2.5}

We would like to introduce
one more refinement of the formalism that will be rather useful for the discussion in next sections. Following  \cite{KPT-MvU},
 we have defined the compactified $\cN$-extended Minkowski
superspace to be the space of null two-planes (through the origin) in ${\mathbb R}^{4|\cN}$.
However, every two-plane in ${\mathbb R}^{4|\cN}$ is a real two-plane
in ${\mathbb C}^{4|\cN}$,  the space of complex even supertwistors.
A two-plane in ${\mathbb C}^{4|\cN}$ is described by a supermatrix
$\cP =(T_1 , T_2)$, where the supertwistors $T_1$ and $T_2 $ constitute a basis
of the two-plane. This supermatrix is defined modulo the equivalence relation
\bea
\cP~\sim ~\cP \X
~, \qquad \X \in \sGL(2,{\mathbb C}) ~.
\eea
The equivalent supermatrices define the same two-plane.
A two-plane $\cP$  in ${\mathbb C}^{4|\cN}$ is said to be real if it possesses
a basis $\cP_0$
consisting of real supertwistors, which means $\cP_0^\dagger =\cP_0^{\rm sT}$.
 Given an arbitrary basis $\cP$ of the real two-plane, it holds that
 $\cP = \cP_0 \X$, for some nonsingular matrix $\X$, and hence
$\cP^\dagger \sim \cP^{\rm sT}$.
We will adopt this new point of view in what follows.
It allows us to define
the compactified $\cN$-extended Minkowski
superspace $\overline{\mathbb M}{}^{3|2\cN}$  to be the space of all
real Lagrangian subspaces of ${\mathbb C}^{4|\cN}$, the space of  even supertwistors.

\section{Pseudo-unitary realisation of $\sOSp(\cN|4; {\mathbb R}) $}

In this section we present a different isomorphic
realisation for the superconformal group, which allows
us to construct (i) a global supermatrix parametrisation of $\overline{\mathbb M}{}^{3|2\cN}$;
and (ii) a globally defined smooth metric, $\rd s^2$,
on $\overline{\mathbb M}{}^{3|2\cN}$
with the property that $\rd s^2$  only scales under the superconformal transformations.
The crucial feature of this realisation is that it is
suitable for developing superconformal field theory on 3D spacetimes
more general than Minkowski space, such as $S^1 \times S^2$ and
its universal covering space ${\mathbb R} \times S^2$.
Our results in this section are analogous to those for
the supersphere $S^{3|4n}$ \cite{KS}.

\subsection{Algebraic aspects}

The superconformal group possesses an alternative realisation based on the isomorphism
\bea
\sOSp(\cN|4; {\mathbb R}) \cong \sU(2,2 |\cN) \bigcap \sOSp (\cN| 4; {\mathbb C})~.
\label{4.1}
\eea
Here the supergroup on the right
consists of all even $(4|\cN) \times (4|\cN)$ supermatrices $\gu$ constrained by
\begin{subequations}\label{3.2ab}
\bea
\gu^\dagger {\mathbb I} \gu &=& {\mathbb I}~, \label{3.2a}\\
\gu^{\rm sT} {\mathbb J}\, \gu &=&{\mathbb J}~, \label{3.2b}
\eea
\end{subequations}
where we have introduced
\bea
{\mathbb I}  = \left(
\begin{array}{c||c}
I ~&~ 0 \\ \hline \hline
0 ~& - {\mathbbm 1}_\cN
\end{array} \right) ~, \qquad
I
=\left(
\begin{array}{c||c}
  {\mathbbm 1}_2 & 0 \\ \hline \hline
0  &     -{\mathbbm 1}_2
\end{array}
\right) ~.
\eea
The condition \eqref{3.2a} defines the supergroup $\sU(2,2 |\cN)$.
It should be pointed out that for $\cN\neq 4$ the supergroup $\sSU(2,2 |\cN)$ is
the $\cN$-extended
superconformal group in four spacetime dimensions, as defined in \cite{KT}
($\sPSU(2,2|4)$ in the $\cN=4$ case).
In what follows, the supergroup on the right of \eqref{4.1} will be denoted
$\sOSp(\cN|4; {\mathbb R})_U$.

The proof of \eqref{4.1} is based on considering the following correspondence
\begin{subequations}
\bea
g ~&\to & ~\gu := {\mathbb U} g {\mathbb U}^{-1} ~, \qquad
g \in \sOSp(\cN|4; {\mathbb R}) ~, \label{3.4a}\\
T ~&\to & ~\Tu := {\mathbb U} T~, \\
Z ~&\to & ~\Zu := Z{\mathbb U}^{-1} ~,
\eea
\end{subequations}
for every supertwistor  $T$ and dual supertwistor $ Z$.
Here the supermatrix $\mathbb U$ is defined by
\bea
{\mathbb U}  = \left(
\begin{array}{c||c}
U ~&~ 0 \\\hline \hline
0 ~& ~ {\mathbbm 1}_\cN
\end{array} \right) ~, \qquad
U=  \frac{1}{  \sqrt{2} }
\left(
\begin{array}{c||r}
 {\mathbbm  1}_2   & \ri {\mathbbm 1}_2\\ \hline \hline
\ri {\mathbbm  1}_2 &   ~ {\mathbbm 1}_2
\end{array}
\right) = U^{\rm T}~.
\eea
The symmetric $4\times 4$ matrix $U$ is unitary,
$U^\dagger U = \overline{U}U={\mathbbm 1}_4$,
and symplectic, $U J U = J$. Another important property is
$U J U^{-1} = - \ri I$. These properties have obvious counterparts in terms
of $\mathbb U$:
\bea
{\mathbb U}^\dagger {\mathbb U} ={\mathbbm 1}_{4|\cN}~, \qquad
{\mathbb U}{\mathbb J}{\mathbb U}= {\mathbb J}~, \qquad
{\mathbb U}{\mathbb J}{\mathbb U}^{-1} = -\ri {\mathbb I}~.
\label{4.6}
\eea
Of special importance for us will be the identity
\bea
{\mathbb U}^{-2} {\mathbb I}= \ri {\mathbb J}~.
\label{4.7}
\eea
The above properties of $\mathbb U$ imply that $\gu$ defined by \eqref{3.4a}
obeys the conditions
\eqref{3.2ab},  and hence $\gu \in \sOSp(\cN|4; {\mathbb R})_U$,
for every $g \in \sOSp(\cN|4; {\mathbb R})$, and vice versa.

Associated with the supergroup $\sOSp(\cN|4; {\mathbb R})_U$
are two invariant inner products defined as follows:
\begin{subequations}
\bea
\langle { \Tu}| { \Su } \rangle_{\mathbb J}&: =& {\Tu}^{\rm sT}{\mathbb J} \, {\Su}~,
\label{4.8a} \\
\langle { \Tu}| { \Su } \rangle_{\mathbb I}&: =& \Tu^\dagger {\mathbb I} \,\Su~,
\label{4.8b}
\eea
\end{subequations}
for arbitrary pure supertwistors $\Tu$ and $\Su$.

An important feature of the supergroup $\sOSp(\cN|4; {\mathbb R})_U$ is that
one can define an involution $\star$ that acts on the space of  supertwistors
and commutes with the superconformal transformations.
Associated with  a pure supertwistor $\Tu$ is its star-image, $\star \Tu$, defined by
\bea
\Tu^\dagger {\mathbb I} = (\star \Tu)^{\rm sT} {\mathbb J}~.
\eea
Explicitly the map $\star$ acts as follows:
\bea
\Tu  = \left(
\begin{array}{c}
f \\
g\\ \hline \hline
\j
\end{array}
\right)
\quad \longrightarrow \quad
\star \Tu = - \left(
\begin{array}{c}
\bar g  \\
\bar f \\ \hline \hline
(-1)^{\ve (\Tu)} \bar \j \phantom{\Big|}
\end{array} \right) ~.
\eea
This shows that $\star (\star \Tu )= \Tu$, for every supertwistor $\Tu$.

\subsection{Compactified Minkowski superspace} \label{subsection4.2}

Let us see how the compactified Minkowski superspace
is described within the supergroup realisation introduced above.
The null two-plane $\cP \in \overline{\mathbb M}^{3|2\cN}$  turns into
$\cPu= {\mathbb U} \cP$. Since $\cP$ obeys the null condition
$\cP^{\rm sT} {\mathbb J} \cP =0$ and is real,
$\cP^\dagger =\cP^{\rm sT}$,
the two-plane $\cPu$ enjoys the two null conditions
\begin{subequations} \label{4.11ab}
\bea
\cPu^\dagger {\mathbb I} \cPu &=& 0~, \label{4.11a}\\
\cPu^{\rm sT} {\mathbb J} \cPu &=&0 ~. \label{4.11b}
\eea
\end{subequations}
The reality condition $\cP^\dagger =\cP^{\rm sT}$ turns into
\bea
\cPu^\dagger = \cPu^{\rm sT} {\mathbb U}^{-2} ~.
\eea
It may be seen that
this reality condition preserves its form only under the {\it real} equivalence transformations
\bea
\cPu~ \sim ~\cPu \X~, \qquad \X \in \sGL (2, {\mathbb R}) ~.
\eea
However, making use of the identities \eqref{4.6} and \eqref{4.7},
it may be rewritten in a different but equivalent form
\bea
\star \cPu~ \sim ~\cPu ~, \label{4.14}
\eea
which does not change its form under arbitrary {\it complex} equivalent transformations
\bea
\cPu~ \sim ~\cPu \X~, \qquad \X \in \sGL (2, {\mathbb C}) ~,
\label{4.15}
\eea
see subsection \ref{subsection2.5}.
Thus, the compactified $\cN$-extended Minkowski
superspace $\overline{\mathbb M}{}^{3|2\cN}$  is equivalently defined as the set of all
two-planes in the space of even supertwistors
${\mathbb C}^{4|\cN}$ which obey (i) the null conditions \eqref{4.11ab};
and (ii) the reality condition \eqref{4.14}.

We can represent
\bea
\cPu =\left(
\begin{array}{c}
\Fu \\
\Gu \\ \hline \hline
\Lau
\end{array}
\right)~,
\eea
where $\Fu$ and $\Gu$ are bosonic $2\times 2$ matrices, and the remaining
$\cN\times 2$ matrix
 $\Lau$ is fermionic. Then the null condition \eqref{4.11a} tells us that
 \bea
 \Fu^\dagger \Fu - \Gu^\dagger \Gu = \Lau^\dagger \Lau~.
 \eea
 In conjunction with the fact that the supermatrix $\cPu$ has rank two,
 this condition implies that $\det \Fu \neq 0$ and $\det \Gu \neq 0$,
 see \cite{K-compactified06} for the proof.
 As a result, making use of the  equivalence relation \eqref{4.15}
 allows us to bring every two-plane $\cPu \in \overline{\mathbb M}^{3|2\cN}$  to the form
 \bea
\cPu \sim \left(
\begin{array}{c}
\bm h \\
{\mathbbm 1}_2
\\
\hline \hline
\z
\end{array}
\right)~.
\label{4.18}
\eea
Now, the null conditions \eqref{4.11a} and \eqref{4.11b}
 turn into
 \begin{subequations} \label{4.19}
 \bea
 {\bm h}^\dagger \bm h  -{\mathbbm 1}_2 &=&  \z^\dagger \z~, \\
 {\bm h}^{\rm T} - \bm h &=& \ri \z^{\rm T} \z~.
 \eea
 \end{subequations}
Moreover, the reality condition \eqref{4.14} gives
\begin{subequations}\label{4.20}
\bea
\bar{\bm h}  &=& {\bm h}^{-1}~, \\
\bar \z  &=& -\ri \z  {\bm h}^{-1}~.
 \eea
 \end{subequations}
The relations (\ref{4.18} -- \ref{4.20}) provide a global supermatrix parametrisation of $ \overline{\mathbb M}^{3|2\cN}$.

 Consider the bosonic body $ \overline{\mathbb M}^{3}$ of compactified Minkowski superspace
$ \overline{\mathbb M}^{3|2\cN}$, which is obtained by switching off the Grassmann variables
 $\z$ and is described by a $2\times 2$ matrix $h$ defined by
 $
 h := {\bm h} |_{\z=0}$.
As follows from \eqref{4.19}, its properties are
$h^\dagger h ={\mathbbm 1}_2$ and $h^{\rm T} = h$. The general solution of these
constraints is
\bea
h = \re^{\ri \vf} \Big( a {\mathbbm 1}_2 + \ri  b \s_1 +  \ri c\s_3\Big)
~, \quad a^2 +b^2 +c^2 =1~,
\eea
for real parameters $\vf$ and  $a, \,b, \,c$ parametrising, respectively,
$S^1$ and  $S^2$. We also have
$$
 \re^{\ri \vf} \Big( a {\mathbbm 1}_2 + \ri  b \s_1 +  \ri c\s_3\Big)
= \re^{\ri (\vf +\p) } \Big( -a {\mathbbm 1}_2 - \ri b \s_1 -\ri   c\s_3\Big)~,
$$
and thus  compactified Minkowski space $ \overline{\mathbb M}^3 $
is $ (S^1 \times S^2)/ {\mathbb Z}_2$.

\subsection{Superconformal metric}
As shown in the previous subsection,
every null two-plane $\cPu \in \overline{\mathbb M}^{3|2\cN}$
is uniquely represented in the form \eqref{4.18} for some
matrices  $\bm h$ and $\z$  constrained
according to \eqref{4.19} and \eqref{4.20}. This means that,
given a group element $\gu \in \sOSp(\cN|4; {\mathbb R})_U $,
it acts on $\cPu$ as
\bea
\gu \left(
\begin{array}{c}
\bm h \\
{\mathbbm 1}_2 \\ \hline \hline
 \z
\end{array}
\right) =  \left(
\begin{array}{c}
{\bm h}' \\
{\mathbbm 1}_2 \\  \hline \hline
\z'
\end{array}
\right) \vf(\gu, \bm h, \z)~, \qquad \vf(\gu, \bm h, \z) \in \sGL (2, {\mathbb C})~,
\label{4.22}
\eea
for some nonsingular matrix $\vf(\gu, \bm h, \z)$.
Explicitly, if we represent $\gu$ in the block form
\bea
\gu
&=& \left(
\begin{array}{c|
c ||c}
A ~&  B & \g   \\ \hline
C  ~&   D~&~ \d
\\
\hline
\hline
\l ~& \r &~R
\end{array}
\right) \in \sOSp(\cN|4; {\mathbb R})_U~,
\eea
then $\bm h'$ and $\z'$ are seen to be fractional linear functions of $\bm h$ and $\z$,
\begin{subequations}
\bea
\bm h' &=& (A \bm h + B + \g \z)(C \bm h + D + \d \z)^{-1}~,\\
\z' &=& (\l \bm h + \r + R \z)(C \bm h + D + \d \z)^{-1}~,
\eea
\end{subequations}
and $\vf(\gu, \bm h, \z)= C \bm h + D + \d \z$. By construction,
$\vf (\gu, \bm h, \z)$ is nonsingular for every
group element $\gu  \in \sOSp(\cN|4; {\mathbb R})_U$.

Cartan's one-form
\bea
\cE := \cPu^\dagger {\mathbb I} \rd \cPu = {\bm h}^\dagger \rd \bm h
- \z^\dagger \rd \z ~, \qquad \cE^\dagger = - \cE
\eea
takes its values in the superalgebra $\frak{osp}(\cN|4; {\mathbb R})_U$
and possesses the superconformal transformation law
\bea
\cE ~\to ~ \cE' = (\vf^\dagger )^{-1} \cE \vf^{-1}~,  \label{4.26}
\eea
where the shorthand notation $\vf = \vf(\gu, \bm h, \z) $ has been used.
We can introduce  a super-interval
\bea
\rd s^2 := \frac{1}{4} \det \cE~,
\eea
which is a globally defined tensor field over $\overline{\mathbb M}^{3|2\cN}$.
It follows from \eqref{4.26}
that $\rd s^2$ only scales under the superconformal transformations,
\bea
\rd s^2 ~\to ~ \rd s^2 | \det \vf |^{-2}~.
\eea

In the Minkowski chart, it may be seen that
the variables $\bm h$ and $\z$ are expressed in terms of the superspace coordinates
as
\bea
\bm h = -\ri ( {\mathbbm 1}_2 - \ri \hat{\bm x} )( {\mathbbm 1}_2 + \ri \hat{\bm x} )^{-1} ~,
\qquad \z = 2 \q ( {\mathbbm 1}_2 + \ri \hat{\bm x} )^{-1}~.
\eea
A direct calculation of $\cE$ gives the following expression:
\bea
\cE=  -2 \ri ({\mathbbm 1}_2 -\ri \hat{\bm x}^{\rm T})^{-1} \hat e \,
({\mathbbm 1}_2 +\ri \hat{\bm x})^{-1}~,
\eea
where $\hat e$ is the supersymmetric one-form \eqref{VA}.

\subsection{Pseudo inversion}
Consider a particular superconformal transformation
\bea
\cFu
&=& \left(
\begin{array}{c |
c ||c}
\s_2 ~&  0 & 0   \\ \hline
0  ~&   -\s_2&~ 0
\\
\hline
\hline
0 ~&  0 &~{\mathbbm 1}_\cN
\end{array}
\right) \in \sOSp(\cN|4; {\mathbb R})_U~, \qquad \cFu^2 = {\mathbbm 1}_{4|\cN}~,
\label{pseudoin}
\eea
where $\s_2$ is the second Pauli matrix.
It acts on $\overline{\mathbb M}^{3|2\cN}$ as follows
\bea
{\bm h}' = -\s_2 \bm h \s_2~, \qquad \z' = - \z\s_2~.
\eea

In the real realisation of the superconformal group,
the supermatrix \eqref{pseudoin} turns into
\bea
\cF = {\mathbb U}^{-1} \cFu {\mathbb U}
&=& \left(
\begin{array}{c|
c ||c}
0 ~&
-\ve
& 0   \\ \hline
-\ve^{-1}
~&   0 &~ 0
\\
\hline
\hline
0 ~&  0 &~{\mathbbm 1}_\cN
\end{array}
\right) \in \sOSp(\cN|4; {\mathbb R})~, \qquad \ve:= (\ve_{\a\b} ) = - \ri \s_2~.
\eea
Its action on the two-plane $\cP (z) $ defined by \eqref{2.23} is
\bea
\cF \cP(z) =  \left(
\begin{array}{c}
 {\mathbbm 1}_2 \\ \ve^{-1} \hat{\bm x}^{-1} \ve^{-1}  \\ \hline \hline \phantom {\Big|}
 {\rm i} \sqrt{2} \hat \q \hat{\bm x}^{-1} \ve^{-1}
\end{array} \right) \X(z)
=
\cP (z') \X( z) ~, \qquad
\X( z) := \ve \hat{\bm x}  ~.
\label{3.35}
\eea
This leads to
\begin{subequations} \label{pseudo-inversion}
\bea
\hat{\bm x}' = - \check{\bm x}^{-1} ~, \qquad \hat{\q}' = \check{\q} \check{\bm x}^{-1}
\label{4.34b}
\eea
or, equivalently,
\bea
\check{\bm x}' = - \hat{\bm x}^{-1} ~, \qquad \check{\q}' = - \hat{\q} \hat{\bm x}^{-1} ~.
\eea
\end{subequations}

Let $g= {\mathbbm 1} + \k \O $ be an infinitesimal superconformal transformation,
where $\k$ is an infinitesimal parameter and
$\O$ is an arbitrary element of the superconformal algebra $\mathfrak{osp}(\cN |4; {\mathbb R})$
given by \eqref{SP-g}. In Minkowski superspace, its action is given by eq.  \eqref{3.10}.
It is an instructive exercise to check that the transformation
$\cF ({\mathbbm 1} + \k \O ) \cF$ generates the following infinitesimal transformation:
\begin{subequations}
\bea
\d \check{\bm x} &=& \check{b}
+\l \check{\bm x}  + \check{\bm x} \l^{\rm T} - \s \check{\bm x}
+\check{\bm x}\hat{a} \,\check{\bm x} + 2{\rm i}\, \check{\eta}^{\rm T}\check \q
+ 2{\rm i} \,\check{\bm x} \hat{\e}^{\rm T}\check \q~, \\
\d \check \q &=&\check \eta +\check \q \l^{\rm T} -\hf \s  \check \q +\L \check \q
+ \check \q \hat{a}\check{\bm x}
+\hat \e \check{\bm x} + 2{\rm i}\, \check \q\, \hat \e^{\rm T} \check \q~.
\eea
\end{subequations}
As compared with $g= {\mathbbm 1} + \k \O $,
the transformation $\cF ({\mathbbm 1} + \k \O ) \cF$ swaps the spacetime
translations and special conformal boosts, as well as
the $Q$-supersymmetry and $S$-supersymmetry transformations. It also changes the
sign of the scale parameter $\s$.

The transformation $\cF$ has properties analogous to those of the  superinversion
(i.e., a supersymmetric extension of the conformal inversion), see e.g. \cite{BK}
for the 4D case. However, the restriction of
$\cF$ to compactified Minkowski space is a transformation that belongs to the connected
component of the identity of the conformal group,
and thus it differs from the 3D conformal inversion
\bea
\hat x' = \check{x}^{-1}~,
\label{conf_inversion}
\eea
which belongs to the other component of the conformal group.  This is why it is
appropriate to call $\cF$ ``pseudo inversion.'' The transformation
\eqref{pseudo-inversion} was called ``superinversion'' in \cite{Park3}.
Our consideration shows that this terminology is somewhat misleading. An extension of
conformal inversion \eqref{conf_inversion} is unclear to us.

\subsection{Fibre bundles over compactified Minkowski superspace}

Fibre bundles over $\overline{\mathbb M}^{3|2\cN}$, such as compactified
harmonic/projective superspaces  in three spacetime dimensions \cite{KPT-MvU},
can be obtained by generalising the construction of subsection \ref{subsection4.2}
to include odd supertwistors.\footnote{Our approach here is inspired
 by the construction of compactified harmonic/projective superspaces
 with Lorentzian signature given in
 \cite{K-compactified06,KPT-MvU,K-compactified12,KS}. These papers built on
earlier works \cite{Rosly,LN2,HH}. }
Odd supertwistors are destined to parametrise fibres over $\overline{\mathbb M}^{3|2\cN}$.
In the unitary realisation of the superconformal group,
given  an odd supertwistor $\Siu$, it is defined
by the following two conditions:
\begin{itemize}

\item it is orthogonal
to the even supertwistors $\Tu^\m$ which form a basis of the  null two-plane
$\cPu \in \overline{\mathbb M}^{3|2\cN}$,
with respect to the inner products \eqref{4.8a} and \eqref{4.8b},
\bea
\langle \Tu^\m | \Siu \rangle_{\mathbb J} = 0~,\qquad
\langle \Tu^\m | \Siu \rangle_{\mathbb I} = 0~;
\label{4.35}
\eea

\item
 it is defined modulo the equivalence relation
\bea
\Siu ~\sim ~ \Siu + \Tu^\m a_\m~,
\label{4.36}
\eea
for arbitrary $a$-numbers $a_\m$ (i.e. odd elements of the Grassmann algebra).
\end{itemize}

If the null two-plane $\cPu$ is chosen in the form \eqref{4.18}, then
imposing the first null condition \eqref{4.35} and making use of
the equivalence relation
\eqref{4.36}, the odd supertwistor
may be brought to the form
\bea
\Siu =  \left(
\begin{array}{c}
-\ri \z^{\rm T} v \\
0
\\
\hline \hline
v
\end{array}
\right) ~,
\label{4.37}
\eea
with $ v= (v_I)$ being a bosonic $\cN$-vector.
Here we have used the reality conditions \eqref{4.20}.
It is important to point out that the second null condition \eqref{4.35}
also leads to the same explicit expression \eqref{4.37}  for $\Siu$. Thus the space of odd supertwistors at $\cPu$ may be identified with ${\mathbb C}^\cN$.

As simple examples of fibre bundles over $\overline{\mathbb M}^{3|2\cN}$,
we can  consider odd supertwistor Grassmannians $\frak G (m, \cN)$,
where $m $ may take values from 1 to $ \cN$. By definition,
the points of $\frak G (m, \cN)$
are described by
 $m $ odd supertwistors ${\Siu}^{\iu}$, with ${\iu}=1,\dots, m$, such that  (i) the bodies of ${\Siu}^{\iu}$ are linearly independent; and (ii) the supertwistors ${\Siu}^{\iu}$ are defined
 modulo the equivalence relation
 \bea
 {\Siu}^{\iu} ~\sim ~{\Siu}^{\ju}
\mathscr D_{\ju}{}^{\iu}~, \qquad
\mathscr D
 \in \sGL(m,{\mathbb C})~.
\eea
In this paper, we are mostly interested in the  $m=\cN$ case, for which
${\Siu}^{\iu}$ may be chosen in the form:
\bea
\underline{\cO} = (\Siu^{1}, \dots , \Siu^\cN) =  \left(
\begin{array}{c}
-\ri \z^{\rm T}  \\
0  \\ \hline \hline
{\mathbbm 1}_\cN
\end{array}
\right) ~.
\label{4.39}
\eea

In the remainder of the paper, we use the real realisation of the superconformal group
described in section \ref{section2}.
In this realisation, only one of the two null conditions \eqref{4.35} remains,
\bea
\langle T^\m | \S \rangle_{\mathbb J} = 0~.
\eea
We recall that for every point $z$ in Minkowski superspace,
 the supermatrix $\cP= (T^1, T^2) $ can be chosen as
\bea
\cP (z)  = \left(
\begin{array}{c}
 {\mathbbm 1}_2 \\ - \hat{\bm x} \\ \hline \hline \phantom{\Big|}
 {\rm i} \sqrt{2} \hat \q
\end{array} \right) ~, \qquad
\hat{\bm x} = \hat{x} + \ri \hat{\q}^{\rm T} \hat \q ~.
\label{5.2}
\eea
Therefore, instead of the representation \eqref{4.37}, now  every
odd supertwistor $\S$ from the fibre at $\cP(z)$ can be brought to the form:
\bea
\S =  \left(
\begin{array}{c}
0 \\
-\sqrt{2} \hat \q^{\rm T} v
\\ \hline \hline \phantom{\Big|}
v
\end{array}
\right) ~.
\eea
Finally, the expression \eqref{4.39} turns into
\bea
{\cO} (z)= (\S^{1}, \dots , \S^\cN)
= \left(
\begin{array}{c}
0 \\
-\sqrt{2} \hat \q^{\rm T}
\\ \hline \hline \phantom{\Big|}
{\mathbbm 1}_\cN
\end{array}
\right) ~.
\label{5.4}
\eea


\section{Two-point and three-point building blocks}

Here we derive those two- and three-point functions of superspace coordinates
which are building blocks for the correlation functions of primary superfields.
An alternative derivation was given by Park \cite{Park3}.

As is seen from \eqref{5.2} and \eqref{5.4}, the coset representative $s(z)$,
defined by \eqref{s(z)},  is built from the supermatrices corresponding to
the even two-plane $\cP (z)$ and the odd $\cN$-plane $\cO (z)$ according to the rule
\bea
s(z) = \Big( \cP(z), ~\cP_{\infty} ,~ \cO(z)\Big)~,
\qquad
\cP_{\infty}  = \left(
\begin{array}{c}
0  \\ {\mathbbm 1}_2 \\ \hline \hline
0
\end{array} \right) ~.
\label{s(z)2}
\eea
Here $\cP_{\infty} $ denotes the null two-plane corresponding to the origin
of the chart $U_G \subset \overline{\mathbb M}{}^{3|2\cN}$, see subsection
\ref{subsection3.3}. The two-plane
$\cP_{\infty} $ is an infinitely separated point from the viewpoint of an observer living in Minkowski  superspace.


\subsection{Infinitesimal superconformal transformations}

For our subsequent analysis, it is advantageous to recast the superconformal
transformation laws of $\cP(z)$ and $\cO(z)$ in terms of the coset representative
\eqref{s(z)2}. Before discussing the  transformation of $s(z)$,
we first point out that the infinitesimal transformation $z^A \to z^A +\d z^A$,
with $\d z^A$ given by \eqref{3.10}, can be rewritten as
\bea
\d z^A = \x z^A \quad \Longleftrightarrow \quad
\d x^a &=& \x^a +{\rm i}\, \x^\a_I (\g^a)_{\a\b} \q^\b_I~, \qquad
\d \q^\a_I = \x^\a_I~,
\eea
where $\x$ denotes a first-order
differential operator
\bea
\x  = \x^A (z) D_A = \x^{a}   \pa_{ a}
+  \x^{\a}_I  D_{\a}^I = -\hf \x^{\a\b} \pa_{\a\b} +  \x^{\a}_I  D_{\a}^I ~.
\label{xi}
\eea
The components of this operator are as follows:
\begin{subequations}\label{explicit-KV}
\bea
\xi^{\alpha\beta}&=&a^{\alpha\beta}
-\lambda^\alpha{}_\gamma x^{\gamma\beta}
-x^{\alpha\gamma}\lambda_\gamma{}^\beta + \s x^{\alpha\beta}
+4\ri \epsilon_I^{(\alpha}\theta_I^{\beta)}
+ 2\ri \L_{IJ}\theta_J^{\alpha}\theta_I^\beta
+ x^{\alpha\gamma}x^{\beta\delta}b_{\gamma\delta}
~~~ \non\\
&&
+\ri b^{(\alpha}_\delta x^{\beta)\delta} \theta^2
-\frac14 b^{\alpha\beta}\theta^2 \theta^2
-4\ri\eta_{I\gamma} x^{\gamma(\alpha}\theta_I^{\beta)}
+2\eta_I^{(\alpha}\theta_I^{\beta)}\theta^2~, \\
\xi_I^\alpha&=&\epsilon_I^\alpha-\lambda^\alpha{}_\beta\theta_I^\beta
+\frac12 \s \theta_I^\alpha
+\L_{IJ}\theta_J^\alpha
+b_{\beta\gamma}{\bm x}^{\beta\alpha}\theta^\gamma_I
+\eta_{J\beta}(2\ri\theta_I^\beta\theta_J^\alpha-\delta_{IJ}{\bm x}^{\beta\alpha})~.
\eea
\end{subequations}
Following the terminology of \cite{BK},
the supervector field $\x$ is called
a conformal Killing supervector field.
It may equivalently be defined \cite{KPT-MvU} as the most
general solution to the equation $[\x , D^I_\a ] \propto D^J_\b$,
and therefore
\bea
[\x, D_\a^I ] = -(D^I_\a \x^\b_J) D^J_\b = \l_\a{}^\b (z)D_\b^I
+ \L^{IJ}(z) D_\a^J -\hf \s (z) D^I_\a~.
\label{master1}
\eea
Here the coefficient functions in the right-hand side read
\bea
\l_{\alpha\beta} (z) =-\frac1{\cN}D^I_{(\alpha}\xi^I_{\beta)}~,\quad
\Lambda^{IJ}(z) =-2D_\alpha^{[I}\xi^{J]\alpha}~,\quad
\sigma (z) =\frac1{\cN}D^I_\alpha\xi_I^\alpha=\frac13\partial_a\xi^a~,
\eea
and may be thought of as the parameters of local Lorentz, $R$-symmetry
and scale transformations, respectively.
The explicit calculation of these parameters gives
\begin{subequations}\label{Local-param}
\bea
\l^{\alpha\beta} (z)&=&\lambda^{\alpha\beta}-x^{\gamma(\alpha}b_\gamma^{\beta)}
-\frac{\ri}2 b^{\alpha\beta}\theta^\gamma_I \theta_{I\gamma}
+2\ri\eta_I^{(\alpha}\theta_I^{\beta)}~,\\
\Lambda_{IJ} (z) &=&\L_{IJ}
+4 \ri\eta_{[I}^\alpha \theta_{J]\alpha}
+2\ri b_{\alpha\beta}\theta_I^\alpha \theta_J^\beta~, \\
\sigma (z)&=&\s+b_{\alpha\beta}x^{\beta\alpha}+2\ri\theta^\alpha_I
\eta_{I\alpha}~.
\eea
\end{subequations}

Under the infinitesimal superconformal transformation associated with $\O$,
eq. \eqref{SP-g},
the even two-plane $\cP (z)$ and the odd $\cN$-plane $\cO (z)$
vary as $\x \cP (z) = \cP (z +\d z) - \cP (z)$ and $\x \cO(z) = \cO(z+\d z ) -\cO(z)$, respectively.
These variations are computed by the rule\footnote{The index structure
of the matrices $\l(z)$, $\L(z)$ and $\check \eta (z)$  in  \eqref{4.8ab} is the same as in \eqref{SP-g}.}
\begin{subequations}\label{4.8ab}
\bea
\O \cP(z) &=& \x \cP (z) + \cP (z) \Big( \l (z) - \hf \s (z)  {\mathbbm 1}_2 \Big)~,
 \\
\O \cO (z) &=& \x \cO(z) + \cO(z) \L(z)
+ \cP(z) \sqrt{2} \check{\eta}^{\rm T} (z)~,
\eea
\end{subequations}
where we have introduced the $z$-dependent $S$-supersymmetry parameter
\bea
\eta_{I\alpha} (z)&:=&\eta_{I\alpha} - b_{\alpha\beta}\theta^\beta_I
=-\frac{\ri}{2} D_\a^I \s(z)~.
\eea
As concerns the coset representative
\eqref{s(z)2},
it follows from first principles that
\bea
\O s(z) = \x s(z) + s(z) \o( z)~,
\label{s(z)-transform}
\eea
for some supermatrix $\o(z)$ belonging to the isotropy subalgebra
\eqref{isotropy_subalgebra}.
Making use of \eqref{4.8ab},
the explicit form of $\o(z)$ is
\bea
\o ( z)= \left(
\begin{array}{c|c||c}
\l_\alpha{}^\beta(z) -\frac12\delta_\alpha{}^\beta\sigma(z) &
b_{\alpha\beta}& \sqrt2 \eta_{I\alpha}(z) \\\hline 0 &
-\l^\alpha{}_\beta(z)+\frac12\delta^\alpha{}_\beta\sigma(z) &0
\\\hline\hline
0 & \ri\sqrt2 \eta_{I\beta}(z) &\Lambda_{IJ}(z)
\end{array}
\right)~.
\label{tomega}
\eea


\subsection{Two-point functions}

By construction, it holds that $\cP^{\rm sT} (z) {\mathbb J} \cP (z)= 0$ and
$\cP^{\rm sT} (z){\mathbb J} \cO (z) =0$. Consider two different superspace points
$z_1$ and $z_2$. Then we can define two-point functions
\begin{subequations} \label{4.12ab}
\bea
\cP^{\rm sT} (z_1) {\mathbb J} \cP (z_2)&=& \hat{\bm x}_1^{\rm T} -\hat{\bm x}_2
+2\ri \hat{\q}_1^{\rm T} \hat{\q}_2\equiv \hat{\bm x}_{12}~, \qquad
\hat{\bm x}_{21} =- \hat{\bm x}^{\rm T}_{12}
\\
\cP^{\rm sT} (z_1) {\mathbb J} \cO (z_2)&=& \sqrt{2} (\hat \q_1 -\hat \q_2 )^{\rm T}
\equiv \sqrt2 \hat\theta^{\rm T}_{12} ~, \qquad ~~~
\hat \theta_{21} = - \hat \q_{12}~.
\eea
\end{subequations}
Making use of (\ref{2.24}),
$\hat{\bm x}_{12}$ and $\hat \q_{12}$
can be rewritten
with explicit spinor indices as follows:\begin{subequations}
\bea
{\bm x}_{12}^{\alpha\beta} &=& (x_{1}-x_{2})^{\alpha\beta}
+2\ri \theta_{1I}^{(\alpha}\theta_{2I}^{\beta)}
-\ri\theta_{12I}^\alpha \theta_{12I}^\beta~,\label{super-interv-X}\\
\theta_{12I}^\alpha &=& (\theta_{1}-\theta_2)^\alpha_{I}~.
\label{super-interv-Theta}
\eea
\label{super-interv}
\end{subequations}
According to \eqref{4.8ab},
the above two-point functions  transform
semi covariantly under the superconformal group
\begin{subequations}\label{2pt-transf}
\bea
\widetilde{\delta} {\bm x}_{12}^{\alpha\beta} &=& \left(\frac12\delta^\alpha{}_\gamma
 \sigma(z_1) -\l^\alpha{}_\gamma(z_1) \right){\bm x}_{12}^{\gamma\beta}
+{\bm x}_{12}^{\alpha\gamma}\left(
\frac12\delta_\gamma{}^\beta \sigma(z_2) -\l_\gamma{}^\beta(z_2)
\right)~, \\
\widetilde\delta\theta^\alpha_{12I}&=&\left(\frac12\delta^\alpha{}_\beta\sigma(z_1)
 -\l^\alpha{}_\beta(z_1)\right)\theta^\beta_{12I}
-{\bm x}_{12}^{\alpha\beta}\eta_{I\beta}(z_2)
+\Lambda_{IJ}(z_2)\theta^\alpha_{12J}~.
\eea
\end{subequations}
Here the variation $\widetilde \d$ is defined by its action on an $n$-point function
$\F(z_1, \dots , z_n) $ to be
\bea
\widetilde \d \F (z_1, \dots, z_n) = \sum_{i=1}^n \x_{z_i} \F (z_1, \dots, z_n) ~.
\eea

We note that the definitions \eqref{4.12ab} can be recast
in terms of the coset representative \eqref{s(z)2} by introducing
the following two-point supermatrix \cite{Park3}
\bea
\cS
(z_1,z_2):= (s(z_1))^{\rm sT}{\mathbb J} s(z_2)~.
\label{Upsilon}
\eea
Using the transformation law of $s(z)$, eq. (\ref{s(z)-transform}),
we read off the superconformal transformation of $\cS(z_1, z_2) $
\bea
\widetilde\delta\cS(z_1,z_2)=-\o(z_1)^{\rm sT}\cS(z_1,z_2)
-\cS(z_1,z_2)\o(z_2)~.
\label{delta-Upsilon}
\eea

Let us introduce the following objects%
\begin{subequations}\label{4.13}
\bea
{\bm x}_{12}{}^2 &:=& -\frac12\tr (\hat {\bm x}_{12}\check {\bm x}_{12}^{\rm T})
= -\frac12{\bm x}_{12}^{\alpha\beta}{\bm
x}_{12\alpha\beta}~,\\
\underline{\hat{\bm x}}_{12}&:=&\frac{\hat{\bm x}_{12}}{
\sqrt{-{\bm x}_{12}{}^2}}~, \qquad
(\ve \underline{\hat{\bm x}}_{12})^2 = {\mathbbm 1}_2~,
\eea
\end{subequations}
with $\ve =(\ve_{\a \b})$.
Using (\ref{2pt-transf}) it is easy to check that ${\bm
x}_{12}{}^2$ transforms only under local scale transformations
while $\underline{\hat{\bm x}}_{12}$ varies only with the local Lorentz
parameters
\begin{subequations}\label{delta-x2}
\bea
\widetilde\delta {\bm x}_{12}{}^2 &=& \left(\sigma(z_1)+ \sigma(z_2) \right) {\bm
x}_{12}{}^2~, \\
\widetilde\delta \underline{\bm x}_{12}^{\alpha\beta}
&=& -\lambda^\alpha{}_\gamma(z_1)\, \underline{\bm
x}_{12}^{\gamma\beta}-\underline{\bm x}_{12}^{\alpha\gamma}\,
\lambda_\gamma{}^\beta(z_2)~.
\eea
\end{subequations}
Thus, they will naturally appear as two-point building blocks in the correlation
functions of primary superfields to be studied in the next
sections.

Since the two-point function ${\bm x}_{12}^{\alpha\beta}$ has the
symmetry property
\be
{\bm x}_{21}^{\alpha\beta} = -{\bm x}_{12}^{\beta\alpha}~,
\label{antisymmetry}
\ee
it can be divided into the symmetric and antisymmetric parts
\be
{\bm x}_{12}^{\alpha\beta} = x_{12}^{\alpha\beta} + \frac{\ri}2\varepsilon^{\alpha\beta} \theta_{12}{}^2~, \qquad
\theta_{12}{}^2 \equiv \theta_{12I}^\alpha
\theta_{12I\alpha}~.
\label{x12-sym}
\ee
The symmetric part is nothing but the
bosonic component of the standard two-point superspace interval
\be
x^{\alpha\beta}_{12} =
(x_{1}-x_{2})^{\alpha\beta}+2\ri\theta_{1I}^{(\alpha}\theta_{2I}^{\beta)}~.
\label{sym-interval}
\ee
We stress that both $x^{\alpha\beta}_{12}$ and ${\bm
x}^{\alpha\beta}_{12}$ are invariant under supersymmetry while
only the latter transforms covariantly under the superconformal
group according to (\ref{2pt-transf}).

To introduce one more important building block,
we point out that the pseudo inversion $\cal F$ acts on $\cP(z)$ by the rule \eqref{3.35}.
One may also work out the action of $\cF $ on $\cO(z)$.
These results allow us to compute the action of $\cF$ on the coset representative
\eqref{s(z)}
\bea
\cF s(z) = s (z') h (\cF ; z) ~,
\eea
where  $h (\cF ; z) $ is a supermatrix from the isotropy group $G_{\cP_0}$.
The latter supermatrix is of the type \eqref{isotropy_group}
with the following block matrix elements:
\begin{subequations}
\bea
A &=& \ve \hat{\bm x} ~, \qquad \check{\bm b}= - \hat{\bm x}^{-1}~,
\qquad \check \eta = \hat \q ({\bm x}^{\rm T})^{-1} ~, \\
R &=& {\mathbbm 1}_\cN - 2\ri \hat \q \hat{\bm x}^{-1} \hat \q^{\rm T}~.
\label{4.20b}
\eea
\end{subequations}
The $\cN \times \cN$ matrix $R$ is orthogonal, $R^{\rm T} R = {\mathbbm 1}_\cN$,
and unimodular, $\det R = 1$.

Let us denote the two-point analog of the matrix (\ref{4.20b}) as $u_{12}$.
It is defined by
\bea
u_{12} ={\mathbbm 1}_\cN +2\ri \hat \q_{12} \hat{\bm x}_{12}^{-1} \hat \q_{12}^{\rm T}
\label{two-point-u}
\eea
and has the properties
\bea
u_{12}^{\rm T} u_{12} = {\mathbbm 1}_\cN~, \qquad \det u_{12} =1~.
\eea
The sign difference in the right-hand sides of
\eqref{4.20b} and \eqref{two-point-u}  follows from the fact that $\hat{\bm x}_{12} $
has the symmetry property
\bea
\hat{\bm x}_{12} - \hat{\bm x}_{12}^{\rm T} = - 2\ri \hat{\q}_{12}^{\rm T} \hat \q_{12} ~,
\eea
which differs by sign from \eqref{2.24}.

Here the inverse matrix $\hat{\bm x}_{12}^{-1}$ is expressed in
terms of $\check{\bm x}_{12}$ as
\be
\hat{\bm x}_{12}^{-1} = -\frac{\check {\bm x}^{\rm T}_{12}}{{\bm x}_{12}{}^2}~.
\ee
With the use of (\ref{2pt-transf}) one can check that this matrix
transforms as
\begin{subequations}\label{deltaU}
\bea
\widetilde\delta u_{12}^{IJ} = \Lambda^{IK}(z_1)u_{12}^{KJ} - u_{12}^{IK}\Lambda^{KJ}(z_2)
\eea
or, equivalently,
\bea
-(\x_{z_1} + \x_{z_2}) u_{12}^{IJ} +\Lambda^{IK}(z_1)u_{12}^{KJ} - u_{12}^{IK}\Lambda^{KJ}(z_2) =0~.
\eea
\end{subequations}
This shows that $u_{12}^{IJ}$ is an invariant tensor two-point function
of the superconformal group (compare with the transformation law \eqref{primaryTL}
describing a primary superfield).
Therefore this object will naturally appear in correlation
functions of primary superfields with $\sO(\cal N)$ indices.


\subsection{Three-point functions}

Given three superspace points $z_1$, $z_2$ and $z_3$, we construct
the following three-point functions
\begin{subequations}\label{three-points}
\bea
\check{\bm X}_{1}&=&-\hat{\bm x}^{-1}_{21}
 \hat{\bm x}_{23}
 \hat{\bm x}^{-1}_{13}~,\qquad
\check\Theta_{1}=\hat{\bm x}^{-1}_{21}\hat\theta_{12}
-\hat{\bm x}^{-1}_{31}\hat\theta_{13}~,  \label{4.29a} \\
\check{\bm X}_{2}&=&-\hat{\bm x}^{-1}_{32}
 \hat{\bm x}_{31}
 \hat{\bm x}^{-1}_{21}~,\qquad
\check\Theta_{2}=\hat{\bm x}^{-1}_{32}\hat\theta_{23}
-\hat{\bm x}^{-1}_{12}\hat\theta_{21}~,\label{4.29b} \\
\check{\bm X}_{3}&=&-\hat{\bm x}^{-1}_{13}
 \hat{\bm x}_{12}
 \hat{\bm x}^{-1}_{32}~,\qquad
\check\Theta_{3}=\hat{\bm x}^{-1}_{13}\hat\theta_{31}
-\hat{\bm x}^{-1}_{23}\hat\theta_{32}~. \label{4.29c}
\eea
\end{subequations}
The structures in \eqref{4.29b} and \eqref{4.29c} follow from
\eqref{4.29a} by applying cyclic permutations of superspace points.
This is why it suffices to study the properties of \eqref{4.29a}.

With the use of (\ref{2pt-transf}) one can check that
$\check{\bm X}_{1}$ and $\check\Theta_{1}$
transform as tensors at the superspace point $z_1$
and scalars at $z_2$ and $z_3$,
\begin{subequations}
\bea
\widetilde\delta
{\bm X}_{1\alpha\beta}&=&\lambda_\alpha{}^\gamma(z_1) {\bm X}_{1\gamma\beta}
+{\bm X}_{1\alpha\gamma}\lambda^\gamma{}_\beta(z_1)-
\sigma(z_1) {\bm X}_{1\alpha\beta} ~,\\
\widetilde\delta \Theta_{1\alpha I}
&=&\left(\lambda_\alpha{}^\beta(z_1)-\frac12\delta_\alpha{}^\beta\sigma(z_1)
\right)\Theta_{1\beta I}
+\Lambda_{IJ}(z_1)\Theta_{1J\alpha}~.
\eea
\end{subequations}
Thus, they turn out to be essential building blocks for
correlation functions of primary superfields.

Let us consider the squares of the structures in  \eqref{4.29a}
\be
{\bm X}_{1}{}^2 := -\frac12\tr(\hat{\bm X}_{1}\check{\bm X}^{\rm T}_{1})
=\frac{{\bm x}_{23}{}^2}{{\bm x}_{12}{}^2 {\bm
x}_{13}{}^2}~,\qquad
\Theta_{1}^2 := \Theta^\alpha_{1I}\Theta_{1I\alpha}~.
\ee
The variations of these objects involve only the parameter of local
scale transformation
\be
\widetilde{\delta} {\bm X}_{1}{}^2  = -2 \sigma(z_1) {\bm X}_{1}{}^2 ~, \qquad
\widetilde{\delta} \Theta^2_{1} = -\sigma(z_1) \Theta^2_{1}~.
\ee
As a consequence, the combination \cite{Park3}
\be
\frac{\Theta^2_{1}}{\sqrt{{\bm X}_{1}{}^2}}
\label{invariant}
\ee
is a superconformal invariant and the superconformal
symmetry can fix the form of correlation functions only up to
this combination.

The two-point function (\ref{super-interv-X}) has the following
distributive property
\be
{\bm x}_{23}^{\alpha\beta} ={\bm x}_{21}^{\alpha\beta} +{\bm x}_{13}^{\alpha\beta}
-2\ri\theta_{21}^{I\alpha} \theta_{13}^{I\beta}~.
\label{4.34}
\ee
As a consequence, the three-point functions
 \eqref{4.29a}
obey
\be
\varepsilon^{\alpha\beta}{\bm X}_{1\alpha\beta} = \ri\Theta_{1}^2~.
\ee
Hence, similar to the two-point function (\ref{x12-sym}), the decomposition
of the three-point function
${\bm X}_{1}^{\alpha\beta}$ into symmetric and antisymmetric
parts reads
\be
{\bm X}_{1\alpha\beta} = X_{1\alpha\beta}
-\frac{\ri}2\varepsilon_{\alpha\beta}\Theta_{1}^2~,\qquad
X_{1\alpha\beta}=X_{1\b\a}~.
\label{XX}
\ee

Given the symmetric object $X_{1\alpha\beta}$ we construct a
vector by the standard rule,
$X_{1m}=-\frac12\gamma_m^{\alpha\beta}X_{1\alpha\beta}$, and
introduce analogs of the covariant spinor derivative (\ref{D}) and the
corresponding supercharge operator
\be
{\cal D}^I_{(1)\alpha} = \frac\partial{\partial \Theta_{1I}^\alpha }
 +\ri\gamma^m_{\alpha\beta}\Theta_{1}^{I\beta} \frac\partial{\partial X^m_{1}}~,\qquad
{\cal Q}^I_{(1)\alpha} =\ri\frac\partial{\partial \Theta_{1I}^\alpha }
 +\gamma^m_{\alpha\beta} \Theta_{1}^{I\beta} \frac\partial{\partial X^m_{1}}
 ~.
\label{generalized-DQ}
\ee
They obey standard anticommutation relations
\be
\{ {\cal D}^I_\alpha, {\cal D}^J_{\beta} \} =
\{ {\cal Q}^I_\alpha, {\cal Q}^J_{\beta} \}=2
\ri \delta^{IJ}\gamma^m_{\alpha\beta}\frac\partial{\partial
X^m}~,
\ee
where we omit the subscript $(1)$ labeling the superspace point.

There are various
identities that involve the three-point functions ${\bm
X}_{i\,\alpha\beta}$ and $\Theta_{i\,\alpha}$ at different points,
e.g.,
\be
{\bm x}_{13}^{\alpha\alpha'}{\bm X}_{3\alpha'\beta'}{\bm x}_{31}^{\beta'\beta}
=-({\bm X}_{1}^{-1})^{\beta\alpha}=\frac{{\bm X}_{1}^{\alpha\beta}}{{\bm X}_{1}{}^2}~,\qquad
\Theta^I_{1\gamma} {\bm x}_{13}^{\gamma\delta}{\bm X}_{3\delta\beta}
=u^{IJ}_{13} \Theta^J_{3\beta}~.
\label{4.34.1}
\ee
These relations allow us to prove the following properties of three-point
functions (\ref{4.29c})
\bea
&&D^I_{(1)\alpha} \Theta^J_{3\beta} = -{\bm x}^{-1}_{13\beta\alpha}
u^{IJ}_{13}~,\qquad
D^I_{(1)\gamma}{\bm X}_{3\alpha\beta}
=2\ri{\bm x}_{13\alpha\gamma}^{-1} u^{IJ}_{13}\Theta^J_{3\beta}~,
\non\\&&
D^I_{(2)\alpha} \Theta^J_{3\beta} ={\bm x}^{-1}_{23\beta\alpha}
u^{IJ}_{23}~,\qquad
D^I_{(2)\gamma}{\bm X}_{3\alpha\beta}
=2\ri{\bm x}_{23\beta\gamma}^{-1} u^{IJ}_{23}\Theta^J_{3\alpha}~.
\eea
Here $D^I_{(i)\alpha}$ is the conventional covariant spinor
derivative (\ref{D}) which acts on the superspace coordinates at the point
$z_i$.

Given a function $f({\bm X}_{3},\Theta_{3})$  depending on
the objects (\ref{4.29c})
one can prove the following differential identities
\begin{subequations}\label{useful-prop}
\bea
D_{(1)\gamma}^I f({\bm X}_{3},\Theta_{3}) &=&{\bm x}^{-1}_{13\alpha\gamma}u^{IJ}_{13}
 {\cal D}^{J\alpha}_{(3)} f({\bm X}_{3},\Theta_{3})~,
 \label{useful-prop-a}\\
D_{(2)\gamma}^I f({\bm X}_{3},\Theta_{3}) &=& \ri{\bm x}^{-1}_{23\alpha\gamma}u^{IJ}_{23}
 {\cal Q}^{J\alpha}_{(3)} f({\bm X}_{3},\Theta_{3})~.
\label{useful-prop-b}
\eea
\end{subequations}
Note that on the left of these identities there are standard
covariant spinor derivatives (\ref{D}) while on the right there
are generalized derivative and supercharge given in
(\ref{generalized-DQ}). The above properties (\ref{useful-prop})
will be important
in the next sections.

Using the relations (\ref{4.34.1}) one can also check that the
object (\ref{invariant}) is invariant under permutations
of superspace points,
\bea
\frac{\Theta^2_{1}}{\sqrt{{\bm
X}_{1}{}^2}}
=\frac{\Theta^2_{2}}{\sqrt{{\bm
X}_{2}{}^2}}
=\frac{\Theta^2_{3}}{\sqrt{{\bm
X}_{3}{}^2}}~.
\eea

Finally, we introduce the following three-point functions
\bea\label{U}
U_{1}^{IJ}=u_{12}^{IK}u_{23}^{KL}u_{31}^{LJ}~, \qquad
U_{2}^{IJ}=u_{23}^{IK}u_{31}^{KL}u_{12}^{LJ}~, \qquad
U_{3}^{IJ}=u_{31}^{IK}u_{12}^{KL}u_{23}^{LJ}~,
\eea
which have simple transformation properties. One may see  that $U_{1}^{IJ}$
transforms as an $\sO(\cN)$ tensor at superspace point $z_1$
\be
\widetilde \delta U^{IJ}_{1} = \Lambda^{IK}(z_1)U_{1}^{KJ} - U_{1}^{IK}\Lambda^{KJ}(z_1)~.
\ee
By construction, the matrix $U_{1}$ is orthogonal,
$U_{1}^{\rm T} U_{1}= {\mathbbm 1}_{\cN}$,
and unimodular, $\det U_{1}=1$.
It can be expressed in terms of the three-point functions
(\ref{4.29c})
\be
U_{1}^{IJ}=\delta^{IJ}
+2\ri \Theta^I_{1\alpha}({\bm X}_{1}^{-1})^{\alpha\beta}\Theta^J_{1\beta}
=\delta^{IJ}-2\ri\frac{\Theta^I_{1\alpha}{\bm X}_{1}^{\beta\alpha}\Theta^J_{1\beta}}{{\bm X}_{1}{}^2}
~.
\label{U-explicit}
\ee
Analogous results hold for $U_2$ and $U_3$.

The matrices $U_2$ and $U_3$  are related
to $U_{1}$ as
\be
U_{2}^{IJ} = u_{21}^{IK}U_{1}^{KL}u_{12}^{LJ}~,\qquad
U_{3}^{IJ} = u_{31}^{IK}U_{1}^{KL}u_{13}^{LJ}~.
\label{U-relations}
\ee
These properties will be useful in checking the invariance
under permutations of superspace points of correlation functions
of superfields with $\sO(\cN)$ indices.


\section{Correlation functions of primary superfields}
\label{sect-5.1}

Consider a superfield $\Phi_{\cal A}^{\cal I}(z)$ that transforms
in a
representation $T$ of the Lorentz group with respect to its index $\cA$
and in a representation $D$ of the $R$-symmetry group $\sO(\cal N)$
with respect to the index  $\cI$.
Such a superfield is called primary of dimension $q$ if its
superconformal transformation law is
\bea
\delta\Phi_{\cal A}^{\cal I} =
-\xi\Phi_{\cal A}^{\cal I}-q\sigma(z)\Phi_{\cal A}^{\cal I}
+\lambda^{\alpha\beta}(z) (M_{\alpha\beta})_{\cal A}{}^{\cal B}
\Phi_{\cal B}^{\cal I}
+\Lambda_{IJ}(z)(R^{IJ})^{\cal I}{}_{\cal J}\Phi_{\cal A}^{\cal J}~.~~
\label{primaryTL}
\eea
Here $\xi$ is the conformal Killing supervector (\ref{xi}),
and the  $z$-dependent parameters
$\sigma(z)$, $\l^{\alpha\beta}(z)$ and $\Lambda_{IJ}(z)$
associated with $\x$
 are
given in (\ref{Local-param}).
The matrices
$M_{\alpha\beta}$ and $R^{IJ}$ are the Lorentz and $\sO(\cN)$ generators,
respectively.

In the non-supersymmetric case,
the formalism to construct the correlation
functions of primary fields in
conformal field theories in diverse dimensions was developed in \cite{OP} (see also \cite{EO}). In four dimensions,
this approach was generalised to $\cN=1$ superconformal field theories
formulated in superspace
in \cite{Osborn} (see also \cite{Park:1997bq}) as well as to higher
$\cN$ \cite{Park4}. The correlation functions of primary
superfields in three and six dimensions were studied
in \cite{Park3} and \cite{Park6}, respectively.
Here we briefly review
the 3D formalism of \cite{Park3}
as it will be employed further for
constructing correlation functions of conserved current multiplets in
3D superconformal field theories.

The two-point correlation function of the primary superfield
$\Phi_{\cal A}^{\cal I}$ and its conjugate $\bar \F^\cA_\cI $
is fixed by the superconformal symmetry up to a single coefficient $c$ and has the form
\bea
\langle
\Phi^{\cal I}_{\cal A}(z_1)
\bar{\Phi}_{\cal J}^{\cal B}(z_2)
\rangle=c\frac{
T_{\cal A}{}^\cB(\ve \underline{\hat{\bm x}}_{12})
{D}^{\cal I}{}_\cJ(u_{12})
}{({\bm x}_{12}{}^2)^q}~
\label{OO}
\eea
provided the representations $T$ and $D$ are irreducible.
The two-point functions  ${\bm x}_{12}{}^2$, $\underline{\hat{\bm x}}_{12}$
and $u_{12}$ are defined in eqs. (\ref{4.13}) and (\ref{two-point-u}), respectively,
and $\ve = (\ve_{\a\b})$.
The denominator in (\ref{OO}) is
fixed by the dimension of $\F$.

Let
$\Phi$, $\J$ and $\P$
be primary superfields (with indices suppressed)
of dimensions $q_1$, $q_2$
and $q_3$, respectively.
The three-point correlation function for
these superfields can be found with the use of the ansatz
\bea
\langle
\Phi^{{\cal I}_1}_{{\cal A}_1}(z_1)
\J_{{\cal A}_2}^{{\cal I}_2}(z_2)
\P_{{\cal A}_3}^{{\cal I}_3}(z_3) \rangle
&=&\frac{T^{(1)}{}_{{\cal A}_1}{}^{{\cal B}_1}( \ve \underline{ \hat{\bm x}}_{13})
 T^{(2)}{}_{{\cal A}_2}{}^{{\cal B}_2}( \ve\underline{\hat{\bm x}}_{23})
D^{(1)}{}^{{\cal I}_1}{}_{{\cal J}_1}(u_{13})
D^{(2)}{}^{{\cal I}_2}{}_{{\cal J}_2}(u_{23})}{({\bm x}_{13}{}^2)^{q_1} ({\bm x}_{23}{}^2)^{q_2}}
~~~~ \non\\&&
\times H^{{\cal J}_1 {\cal J}_2 {\cal I}_3}_{{\cal B}_1 {\cal B}_2 {\cal A}_3}
 ({\bm X}_{3},\Theta_{3},U_{3})~,
\label{OOO}
\eea
where $H^{{\cal J}_1 {\cal J}_2 {\cal I}_3}_{{\cal B}_1 {\cal B}_2 {\cal
A}_3}$ is a tensor constructed in terms of the three-point
functions (\ref{three-points}) and (\ref{U}).
The functional form of this
tensor is highly constrained by the following conditions:
\begin{itemize}

\item[(i)] It should obey the  scaling property
\bea
H^{{\cal J}_1 {\cal J}_2 {\cal I}_3}_{{\cal B}_1 {\cal B}_2 {\cal A}_3}
 (\lambda^2{\bm X},\lambda\Theta,U)
 =(\lambda^2)^{q_3-q_2-q_1}
 H^{{\cal J}_1 {\cal J}_2 {\cal I}_3}_{{\cal B}_1 {\cal B}_2 {\cal A}_3}
 ({\bm X},\Theta,U)~, \qquad \forall \l \in \mathbb R \setminus \{0\}~~~~
\eea
in order for the correlation function to have the correct transformation law
under the superconformal group.

\item[(ii)] When some of
the superfields $\F$, $\J$ and $\P$
obey differential equations such as the conservation
conditions of conserved current multiplets, the tensor
$H^{{\cal J}_1 {\cal J}_2 {\cal I}_3}_{{\cal B}_1 {\cal B}_2 {\cal A}_3}$
is constrained by certain differential equations as well.
In deriving such equations the identities
\eqref{useful-prop}
may be useful.

\item[(iii)] When two  of
the superfields $\F$, $\J$ and $\P$  (or all of them) coincide,
the tensor $H$
should obey certain constraints originating from the symmetry
under permutations of superspace points, e.g.
\be
\langle \Phi_{{\cal I}}^{{\cal A}}(z_1) \Phi_{{\cal J}}^{{\cal B}}(z_2)
\P_{{\cal K}}^{{\cal C}}(z_3) \rangle =
(-1)^{\epsilon(\Phi)}
\langle \Phi_{{\cal J}}^{{\cal B}}(z_2) \Phi_{{\cal I}}^{{\cal A}}(z_1)
\P_{{\cal K}}^{{\cal C}}(z_3) \rangle~,
\ee
where $\epsilon(\Phi)$ is the Grassmann parity of $\Phi_{{\cal I}}^{{\cal A}}$.
\end{itemize}
These constraints fix the functional form of the tensor $H$
(and, hence, the three-point correlation function) up to
a few arbitrary constants.

The procedure described
reduces the problem of computing thee-point correlation functions
to deriving the single function $H$ subject to the above
mentioned constraints.
 In the next
sections we will apply this procedure to compute the
two- and three-point correlation functions of the supercurrents and the flavour
current multilpets in
superconformal field
theories with $1\leq \cN \leq 3$.


\section{Correlators
in $\cN=1$ superconformal field theory}

To start with, we give an example of a classically $\cN=1$ superconformal field theory.
It is described by $n$ primary real scalar superfields $\vec{\vf} $ of dimension 1/2
with action
\bea
S =  \int \rd^3x \rd^2 \q \, \Big\{
\frac{1}{2} D^\a  \vec{\varphi} \cdot D_\a \vec{\varphi}
+ \frac{\ri}{4} \l  (\vec \varphi \cdot \vec \vf )^2 \Big\} ~,
\label{66.1}
\eea
with $\l$ a coupling constant. This action is invariant under the superconformal transformation
\be
\delta \vec{\vf} = -\xi\vec\vf - \frac 12 \sigma(z)\vec \vf~.
\ee
The supercurrent of this model \cite{Kuzenko:2012ew} is
\be
J_{\a\b\g} = \ri \Big(\vec \vf \cdot D_{(\a} \pa_{\b \g)} \vec \vf
- 3 D_{(\a}\vec \vf \cdot \pa_{\b \g )}\vec \vf \Big)~.
\label{phi-supercurrent}
\ee
The flavour current multiplet reads
\be
J_\alpha^{\bar a} = \ri \Big(\vec \varphi \cdot \Sigma^{\bar a} D_\alpha
\vec \varphi -  D_\alpha\vec \varphi \cdot \Sigma^{\bar a} \vec \varphi \Big)~,
\label{phi-flavor}
\ee
where $\Sigma^{\bar a}$ denotes the generator of the flavour $\sO(n)$ group.
One may check that the currents
(\ref{phi-supercurrent}) and (\ref{phi-flavor})
transform as primary superfields under the superconformal group
and obey the corresponding
conservation laws given in (\ref{1}) and (\ref{2}) on the equations of motion
for $\vec \vf$.

A natural  generalisation of  \eqref{66.1}
is the most general off-shell 3D $\cN=1$  superconformal sigma model
given in \cite{KPT-MvU}.\footnote{On-shell superconformal sigma models
in three dimensions were proposed in \cite{NST,ST,BCSS}.}

\subsection{$\cN=1$ flavour current multiplets}
\label{N1flavour-sect}

In $\cN=1$ supersymmetric field theory, the flavour current multiplet
is described by  a primary real spinor
superfield $L_\alpha$ of dimension 3/2 (with its flavour index suppressed)
which transforms under superconformal group as
\be
\delta L_\alpha = -\xi L_\alpha - \frac 32 \sigma(z)
L_\alpha
+\lambda_\alpha{}^\beta(z) L_\beta
\ee
and obeys the conservation equation
\be
D^\alpha L_\alpha = 0~.
\label{128}
\ee

Let us assume that the superconformal field theory under study
has several flavour current multiplets $L^{\bar a}_\a $,
with $\bar a$  the flavour index.
According to the general formula (\ref{OO}), the two-point function
of such operators
is fixed up to one real coefficient
$a_{{\cal N}=1}$,
\be
\langle L^{\bar a}_\alpha(z_1) L^{\bar b}_\beta(z_2) \rangle =\ri a_{{\cal N}=1}
 \frac{\delta^{\bar a\bar b}{\bm x}_{12\alpha\beta}}{({\bm
 x}_{12}{}^2)^2}~,
\label{2pt-flavour}
\ee
assuming that the flavour group is simple.
With the relation (\ref{antisymmetry}) it is easy to see that
(\ref{2pt-flavour}) obeys the right symmetry property under the
permutation of superspace points, $\langle L^{\bar a}_\alpha(z_1) L^{\bar b}_\beta(z_2) \rangle
=-\langle L^{\bar b}_\beta(z_2) L^{\bar a}_\alpha(z_1)  \rangle$.
Next, using the explicit expression for ${\bm x}_{12\alpha\beta}$ given in
(\ref{super-interv-X}),
one may check that
(\ref{2pt-flavour}) respects
the conservation condition (\ref{128})
\be
D^\alpha_{(1)}\langle L^{\bar a}_\alpha(z_1) L^{\bar b}_\beta(z_2) \rangle
=0~, \qquad
z_1\ne z_2~.
\ee

Consider now the  three-point correlation function
$\langle
L^{\bar a}_\alpha(z_1) L^{\bar b}_\beta(z_2) L^{\bar
c}_\gamma(z_3) \rangle$.
Since the superspace coordinates do not
carry any flavour group indices, the dependence of the correlation
function on $\bar a$, $\bar b$ and $\bar c$ should
factorise in the form of an invariant tensor of the flavour group, which
is  completely antisymmetric,
$f^{\bar a \bar b \bar c}=f^{[\bar a \bar b \bar c]}$,
 or completely symmetric,
 $d^{\bar a \bar b \bar c}=d^{(\bar a \bar b \bar c)}$.
These tensors are defined in terms of the
generators $\Sigma^{\bar a}$ of the flavour group  as follows
\be
[\Sigma^{\bar a},\Sigma^{\bar b}]
=
\ri f^{\bar a\bar b \bar c} \Sigma^{\bar c}~,
\qquad
d^{\bar a\bar b\bar c}= \frac12\tr (\{ \Sigma^{\bar a},\Sigma^{\bar b} \} \Sigma^{\bar
c})~.
\ee
 In principle, the correlator may be a sum of two terms, one of which
 is proportional to $f^{\bar a \bar b \bar c}$
 and the other to $d^{\bar a \bar b \bar c}$.
In four dimensions, contributions with $d^{\bar a\bar b \bar c}$ arise  as a
consequence of anomalies.
In three dimensions, gauge theories are anomaly-free.
Therefore, it
is natural to expect that the part of
$\langle L^{\bar a}_\alpha(z_1) L^{\bar b}_\beta(z_2) L^{\bar
c}_\gamma(z_3) \rangle$
with $d^{\bar a\bar b \bar c}$ should vanish as it was observed in
the non-supersymmetric case studied in \cite{OP}.
Nevertheless, here we
start by considering the most general expression for the correlation
function including
contributions of both types, with $f^{\bar a\bar b \bar
c}$ and $d^{\bar a\bar b\bar c}$, and then show that the latter vanishes upon
imposing all the relevant constrains.

According to the general formula (\ref{OOO}), we have to look for
the three-point correlator in the form
\be
\langle L^{\bar a}_\alpha(z_1) L^{\bar b}_\beta(z_2) L^{\bar c}_\gamma(z_3) \rangle
=\frac{{\bm x}_{13\alpha\alpha'}{\bm x}_{23\beta\beta'}}{
 ({\bm x}_{13}{}^2)^2({\bm x}_{23}{}^2)^2}
\left(f^{\bar a\bar b\bar c}
H_{(f)}^{\alpha'\beta'}{}_\gamma({\bm X}_{3},\Theta_{3})
+d^{\bar a\bar b\bar c}
H_{(d)}^{\alpha'\beta'}{}_\gamma({\bm X}_{3},\Theta_{3})
\right)
~,
\label{3pt-flavour}
\ee
where the tensors $H_{(f,d)}^{\alpha\beta\gamma}$
should obey the following scaling property:
\be
H_{(f,d)}^{\alpha\beta\gamma}(\lambda^2 {\bm X},\lambda\Theta)
=\lambda^{-3} H_{(f,d)}^{\alpha\beta\gamma}({\bm X},\Theta)~.
\label{H-scaling1}
\ee

Recall that the superfield $L^{\bar a}_\alpha$ is Grassmann odd. Hence,
the correlator (\ref{3pt-flavour}) changes its sign when we interchange any pair of
superfields in it, e.g.
\be
\langle
L^{\bar b}_\beta(z_2)L^{\bar a}_\alpha(z_1) L^{\bar c}_\gamma(z_3) \rangle
=-\langle
L^{\bar a}_\alpha(z_1)L^{\bar b}_\beta(z_2) L^{\bar c}_\gamma(z_3) \rangle~.
\label{5.12}
\ee
This equation imposes the following constraint on the tensors
$H_{(f,d)}^{\alpha\beta\gamma}$:
\be
H_{(f)}^{\beta\alpha\gamma}(-{\bm X}^{\rm T},-\Theta)
=H_{(f)}^{\alpha\beta\gamma}({\bm X},\Theta)~,\qquad
H_{(d)}^{\beta\alpha\gamma}(-{\bm X}^{\rm T},-\Theta)
=-H_{(d)}^{\alpha\beta\gamma}({\bm X},\Theta)~.
\label{H-symmetry1}
\ee
The most general expressions for  these tensors
subject to the constraints (\ref{H-scaling1}) and
(\ref{H-symmetry1}) read
\be
H_{(f)}^{\alpha\beta\gamma}=\ri\sum_n c_n H^{\alpha\beta\gamma}_{(f)n}~,
\qquad
H_{(d)}^{\alpha\beta\gamma}=\ri\sum_n d_n H^{\alpha\beta\gamma}_{(d)n}~,
\label{5.14}
\ee
where $c_n$ and $d_n$ are some real coefficients and
\begin{subequations}
\bea
&&
H_{(f)1}^{\alpha\beta\gamma}=\frac{\varepsilon^{\alpha\beta}\Theta^\gamma}{{\bm X}^2}~,
\quad
H_{(f)2}^{\alpha\beta\gamma}=\frac{{\bm X}^{\alpha\beta}\Theta^\gamma}{{\bm
X}^3}~,\quad
H_{(f)3}^{\alpha\beta\gamma}=\frac{\varepsilon^{\beta\gamma}{\bm X}^{\alpha\mu}\Theta_\mu + \varepsilon^{\alpha\gamma}{\bm X}^{\beta\mu}\Theta_\mu}{{\bm
X}^3}~;~~~~~~
\label{5.15} \\
&&
H_{(d)1}^{\alpha\beta\gamma} = \frac{{\bm X}^{\alpha\beta}{\bm X}^{\gamma\delta}
 \Theta_\delta}{{\bm X}^4}~,\quad
H_{(d)2}^{\alpha\beta\gamma} = \frac{\varepsilon^{\alpha\gamma} \Theta^\beta
 + \varepsilon^{\beta\gamma} \Theta^\alpha}{{\bm X}^2} ~,\quad
H_{(d)3}^{\alpha\beta\gamma} = \frac{\varepsilon^{\alpha\beta}{\bm X}^{\gamma\delta}\Theta_\delta}{{\bm X}^3}~.
\eea
\end{subequations}
Recall that we use the notation in which ${\bm X}^2 = -\frac12{\bm X}^{\alpha\beta}{\bm
X}_{\alpha\beta}$ and ${\bm X}^k \equiv ({\bm X}^2)^{k/2}$.

Note that
there is no need to add one more addmisible structure
$\frac1{{\bm
X}^3}({\bm X}^{\alpha\gamma}\Theta^\beta+{\bm X}^{\beta\gamma}\Theta^\alpha )$
to  the list (\ref{5.15}), since  it is linearly dependent of the others,
\be
{\bm X}^{\alpha\gamma}\Theta^\beta + {\bm X}^{\beta\gamma} \Theta^\alpha
 = 2{\bm X}^{\alpha\beta}\Theta^\gamma
 +\varepsilon^{\beta\gamma} {\bm X}^{\alpha\mu}\Theta_\mu
 +\varepsilon^{\alpha\gamma} {\bm X}^{\beta\mu}\Theta_\mu~.
 \label{id-6.15}
\ee

To fix the values of the coefficients $c_n$ and $d_n$ in (\ref{5.14})
we have to take into account the
conservation
condition (\ref{128}),
\be
D^\alpha_{(1)}\langle L^{\bar a}_\alpha(z_1) L^{\bar b}_\beta(z_2) L^{\bar c}_\gamma(z_3)
\rangle =0~.
\ee
Making
use of (\ref{useful-prop-a}), this equation imposes the
following constraint on the tensors $H_{(f,d)}^{\alpha\beta\gamma}$:
\be
{\cal D}_\alpha H_{(f,d)}^{\alpha\beta\gamma}=0~.
\label{5.17}
\ee
Here ${\cal D}_\alpha$ is the generalized covariant spinor
derivative defined in (\ref{generalized-DQ}). The equation
(\ref{5.17})
leads to
the following constraints on the
coefficients $c_n$ and $d_n$:
\bea
c_1=0~, \quad
c_2 + c_3 =0~;\qquad
d_1 = 3d_2~, \quad d_3 =0~.\label{5.18}
\eea

To find further constrains on the coefficients, we recall that the
correlation function changes its sign if we swap any two
superfields in it, e.g.,
\be
\langle
L^{\bar a}_\alpha(z_1) L^{\bar b}_\beta(z_2) L^{\bar c}_\gamma(z_3) \rangle
=-\langle
L^{\bar c}_\gamma(z_3) L^{\bar b}_\beta(z_2)  L^{\bar a}_\alpha(z_1) \rangle~.
\label{5.19}
\ee
Using the identities (\ref{4.34.1}), we find the following
corollaries of (\ref{5.19}):
\be
H_{(f,d)}^{\alpha\beta\gamma}({\bm X}_{3},\Theta_{3})=\pm\frac{
{\bm X}_{3}^{\rho\beta}{\bm x}_{32\rho\rho'}{\bm
x}_{31}{}^\gamma{}_{\gamma'}
 ({\bm x}^{-1}_{13})^{\alpha\alpha'}H_{(f,d)}^{\gamma'\rho'}{}_{\alpha'}
 (-{\bm X}_{1}^{\rm T},-\Theta_{1})}{{\bm X}_{3}{}^4 {\bm
 x}_{13}{}^4}~,
\label{6.20}
\ee
where the right-hand side should be taken with the plus sign
for $H_{(f)}$ and with minus for $H_{(d)}$.
The constraints (\ref{6.20}) are satisfied  under the conditions
\be
c_1=0~, \quad
c_2 + c_3 =0~;\qquad
d_1 =-d_2~, \quad d_3 =0~.
\label{6.21}
\ee
Thus, from (\ref{5.18}) and (\ref{6.21}) we see that all $d$-coefficients
vanish, $d_n=0$, and therefore $H_{(d)}^{\alpha\beta\gamma}=0$.
Furthermore, only one independent coefficient remains among $c_n$,
which we denote by $b_{{\cal N}=1} = c_2 = -c_3$.
Our final expression for the correlator \eqref{3pt-flavour} is
\begin{subequations}\label{H-N1-flavour}
\bea
\langle L^{\bar a}_\alpha(z_1) L^{\bar b}_\beta(z_2) L^{\bar c}_\gamma(z_3) \rangle
&=&f^{\bar a\bar b\bar c}
\frac{{\bm x}_{13\alpha\alpha'}{\bm x}_{23\beta\beta'}}{
 ({\bm x}_{13}{}^2)^2({\bm x}_{23}{}^2)^2}
H_{(f)}^{\alpha'\beta'}{}_\gamma({\bm X}_{3},\Theta_{3}) ~,\\
H_{(f)}^{\alpha\beta\gamma}({\bm X},\Theta) &=&\frac{\ri b_{{\cal N}=1}}{{\bm X}^3}( {\bm X}^{\alpha\beta}\Theta^\gamma
-\varepsilon^{\beta\gamma}{\bm X}^{\alpha\delta}\Theta_\delta
-\varepsilon^{\alpha\gamma}{\bm X}^{\beta\delta}\Theta_\delta
)~.
\eea
\end{subequations}

The superfield operator $ L^{\bar a}_\alpha(z)$ contains
an ordinary conserved flavour current   $L_{m}^{\bar a}(x)$
as its linear in $\q$ component,
 \be
L_{\a \b}^{\bar a} =  D_{\a} L_{\b}^{\bar a}|~, \qquad
L_{\a \b}^{\bar a}= \gamma^m_{\a \b} L_{m}^{\bar a}~, \qquad
\partial^m L_{m}^{\bar a}=0~,
\label{Urr1}
\ee
where $|$ indicates that we have to set $\theta=0$. From (\ref{H-N1-flavour})
we can extract the three-point function
$\langle L_{\a \a'}^{\bar a} (x_1)  L_{\b \b'}^{\bar b} (x_2)
L_{\g \g'}^{\bar c} (x_3) \rangle$
by the rule:
\be
\langle L_{\a \a'}^{\bar a} (x_1)  L_{\b \b'}^{\bar b} (x_2)  L_{\g \g'}^{\bar c} (x_3) \rangle
= - D_{(1) \a}  D_{(2) \b}  D_{(3) \g}
\langle L_{\a' }^{\bar a} (z_1)  L_{\b'}^{\bar b} (z_2)  L_{\g' }^{\bar c} (z_3) \rangle|~.
\label{Urr2}
\ee
It is instructive to compare  the flavour current correlation
function (\ref{H-N1-flavour}) with the corresponding
non-supersymmetric expression found in \cite{OP}.
After a straightforward but lengthy calculation (see Appendix \ref{AppC3} for
the technical details) we find
\be
\langle L_{m}^{\bar a} (x_1) L_{n}^{\bar b}  (x_2) L_{k}^{\bar c} (x_3) \rangle
= \frac{f^{{\bar a} {\bar b} \bar c}}{x_{12}{}^2 x_{23}{}^2  x_{13}{}^2}
I_{m m'} (x_{13})  I_{n n'} (x_{23}) t^{m' n'}{}_k (X_{3})~.
\label{Urr3}
\ee
Here we have defined
\begin{subequations}\label{Urr4}
\bea
I_{mn} (x) &=& \eta_{mn} -\frac{2 x_{m} x_{n}}{x^2}~, \\
 X^m_3&=& \frac{x^m_{13}}{x_{13}{}^2}  -  \frac{x^m_{23}}{x_{23}{}^2}  ~,
 \\
t_{mn k}(X) &=& b_{1} \frac{X_m X_n X_k}{X^3} + b_{2} \frac{X_m \eta_{nk} +  X_n \eta_{m k} -  X_k \eta_{mn}}{X}~.
\eea
\end{subequations}
According to (\ref{C61}), the coefficients $b_{1}$ and $b_{2} $
are given in terms of $b_{\cN=1}$ as follows
\be
b_{1} = b_{2} = 3 b_{\cN=1} ~.
\label{Urr5}
\ee
This is the same result as the one obtained in~\cite{OP}
except for the fact that the two coefficients $b_{1}$ and $b_{2} $,
which were completely independent in the non-supersymmetric case,
 are now equal to each other due to supersymmetry.

As pointed out in~\cite{Giombi:2011rz}, in 3D conformal  field theories an additional
{\it parity violating}\footnote{The parity transformation in question is $x^m \to -x^m$.
The correlator of three flavour currents acquires a minus sign under this transformation.}
structure can arise in the three-point correlator of flavour currents,
\be
\tilde{t}_{mn k}(X) = b_{3}   \frac{-\varepsilon_{nk p} X_m X^p +  \varepsilon_{mk p} X_n X^p + \varepsilon_{m n p} X_k X^p}{X^2}~.
\label{Urr6}
\ee
However, this structure does not appear upon the $\cN=1 \to \cN=0$ reduction
of our result (\ref{H-N1-flavour}) and, hence, it is not consistent with supersymmetry.
The same conclusion holds in all cases considered below in this paper.
 Specifically, the correlators of both flavour current multiplets and supercurrents
contain only parity even contributions.\footnote{One way to check whether a given contribution is even or odd under parity is to
reduce it to the $\cN=0$ case to see whether or not it contains an $\varepsilon_{mnp}$ tensor. This is easy
to see from the general structure of the supersymmetric result without performing the reduction in detail. We will not discuss details of
the reduction of our results to  $\cN=0$ in other sections of this paper.}


\subsection{$\cN=1$ supercurrent}
\label{sect-N1-supercurrent}


The ${\cal N}=1$ supercurrent
is described by a primary symmetric  third-rank spinor
$J_{\a \b \g} =J_{(\a \b \g)} $ of dimension 5/2,
which obeys the conservation law
\be
D^{\a} J_{\a \b \g}=0~.
\label{n1.1.1}
\ee
This conservation  equation is invariant under the superconformal transformation
of $J_{\a \b \g}$, which is
\be
\delta  J_{\a \b \g} =-\xi  J_{\a \b \g} -\frac{5}{2} \sigma(z)  J_{\a \b \g}
+3 \lambda^{\delta}_{\ (\a}(z) J_{\b \g) \d}~.
\label{n1.1.2}
\ee

According to the general discussion in section \ref{sect-5.1},
the two-point function of the supercurrent is given by
\be
\langle  J_{\a \b \g} (z_1)  J^{\a' \b' \g'} (z_2)\rangle =\ri c_{{\cal N}=1}
\frac{ {\bm x}_{12 \a}{}^{( \a'} {\bm x}_{12 \b}{}^{ \b'}
 {\bm x}_{12 \g}{}^{ \g')} }{ ({\bm x}_{12}{}^2)^4}~.
\label{n1.1.3}
\ee
It is easy to show  that the two-point function~\eqref{n1.1.3}
has the right symmetry property under the change of superspace
points,
$\langle  J_{\a \b \g} (z_1)  J_{\a' \b' \g'} (z_2)\rangle
=\langle  J_{\a' \b' \g'} (z_2) J_{\a \b \g} (z_1)  \rangle$, and
satisfies
\be
D^{\a}_{(1)} \langle  J_{\a \b \g} (z_1)  J_{\a' \b' \g'} (z_2)\rangle
=0
~, \qquad
z_1 \neq z_2~.
\label{n1.1.4}
\ee

Similarly, we can write the most general form for
the three-point function that is consistent with the superconformal symmetry.
Let us denote by ${\cal A} = (\a, \a', \a'')$
a symmetric combination of the three spinor indices,
$J_{{\cal A}}\equiv J_{\a \a' \a''}$. Then
\be
\langle J_{{\cal A}} (z_1) J_{{\cal B}} (z_2) J_{{\cal C}} (z_3) \rangle=
\frac{ T_{{\cal A}}^{\ {\cal R}} ({\bm x}_{13}) T_{{\cal B}}^{\ {\cal S}} ({\bm x}_{23}) }{
({\bm x}_{13}{}^2)^4  ({\bm x}_{23}{}^2)^4}
H_{{\cal R} {\cal S} {\cal C} } ({\bm X}_{3}, \Theta_{3})~.
\label{n1.1.5}
\ee
Explicitly,
\be
T_{{\cal A}}{}^{{\cal R}} ({\bm x}) = {\bm x}_{(\a}{}^{ \r} {\bm x}_{\a'}{}^{ \r'} {\bm x}_{\a'')}{}^{ \r''}
\label{n1.1.6}
\ee
and the function\footnote{Here and below we sometimes use  a comma to separate various groups of indices.}
$H_{{\cal A} {\cal B} {\cal C} } ({\bm X}, \Theta)\equiv H_{ (\a \a' \a''),  (\b \b' \b''), (\g \g' \g'')} ({\bm X}, \Theta)$
should satisfy the scaling property
\be
H_{{\cal A} {\cal B} {\cal C} }  (\l^2 {\bm X}, \l \Theta)=
 \l^{-5} H_{{\cal A} {\cal B} {\cal C} } ({\bm X}, \Theta)~.
\label{n1.1.7}
\ee
If we exchange the first and the second superspace points $z_1 \leftrightarrow z_2$ it follows that
${\bm X}_{3} \to - {\bm X}_{3}^{\rm T}$, $\Theta_{3} \to - \Theta_{3}$.
Since the supercurrent $J_{\a \b \g}$ is Grassmann odd the correlation function~\eqref{n1.1.5}
has to change the sign under  $z_1 \leftrightarrow z_2, \ {\cal A} \leftrightarrow {\cal B}$. It implies that
the function $H_{{\cal A} {\cal B} {\cal C} }  ({\bm X}, \Theta)$
has to satisfy
\be
H_{{\cal A} {\cal B} {\cal C} }  (-{\bm X}^{\rm T}, -\Theta)
=- H_{{\cal B} {\cal A} {\cal C} }  ({\bm X}, \Theta)~.
\label{n1.1.8}
\ee
The three-point function~\eqref{n1.1.5} now has the right symmetry property
under $z_1 \leftrightarrow z_2$,
but it does not necessarily has the right symmetry under
$z_1 \leftrightarrow z_3$ and $z_2 \leftrightarrow z_3$.
Additionally, the function $H_{{\cal A} {\cal B} {\cal C} }  ({\bm X}, \Theta)$ is constrained by the conservation law
(\ref{n1.1.1}). Upon the use of the identity (\ref{useful-prop-a}) the latter is translated to
\be
{\cal D}_{\a} H^{ \a \b \g,  \a' \b' \g', \a'' \b'' \g''} ({\bm X}, \Theta)=0~.
\label{n1.1.9}
\ee

Now our aim is to find the most general solution for $H$. The
standard approach used in
4D superconformal field theories with $\cN=1$ \cite{Osborn}
and $\cN=2$ \cite{KT} is based on writing the most general ansatz
in terms of ${\bm X}$ and $\Theta$
consistent with the symmetries and the scaling property~\eqref{n1.1.7} and constrain it by the conservation law~\eqref{n1.1.9}.
However, because of a large number of tensorial indices it appears to be inefficient as such an ansatz would require to analyse quite
a considerable number of possible terms. Hence, we will take a slightly indirect route.

First, let us trade a pair of spinor indices of $H$ for a vector index in each triple. That is, we write
\be
H^{ \a \a' \a'',  \b \b' \b'', \g \g' \g''} =(\g_m)^{\a' \a''} (\g_n)^{\b' \b''} (\g_k)^{\g' \g''} H^{\a m, \b n, \g k}~.
\label{n1.1.10}
\ee
Note that eq.~\eqref{n1.1.10} is not quite correct as it stands because
the left-hand side is fully symmetric
in each triple while the right hand side is symmetric only in $(\a', \a''), \ (\b', \b'')$ and $(\g', \g'')$.
For eq. \eqref{n1.1.10} to make sense, we have to make sure that the antisymmetric part in
$(\a, \a'), \ (\b, \b')$ and $(\g, \g')$ vanishes on the right hand side. That is, we have to impose the following conditions
on $H^{\a m, \b n, \g k}$
\be
(\g_m)_{\a \d} H^{\a m, \b n, \g k} =0~, \quad
(\g_n)_{\b \d} H^{\a m, \b n, \g k} =0~, \quad
(\g_k)_{\g \d} H^{\a m, \b n, \g k} =0~.
\label{n1.1.11}
\ee
From~\eqref{n1.1.9} we still have the conservation law
\be
{\cal D}_{\a} H^{\a m, \b n, \g k} =0~.
\label{n1.1.12}
\ee
Since $H$ is Grassmann odd and since
\be
\Theta^{\a} \Theta^{\b}\Theta^{\g}=0~,
\label{n1.1.13}
\ee
it follows that $H$ must contain only linear $\Theta$-terms.
Then eq.~\eqref{n1.1.12} is equivalent to two independent equations
\begin{subequations}
\bea
&&
\partial_{\a} H^{\a m, \b n, \g k} =0 ~,
\label{n1.1.14.1}
\\
&&
 \Theta^{\d} (\g^t)_{\a \d} \partial_t H^{\a m, \b n, \g k} =0~.
\label{n1.1.14.2}
\eea
\end{subequations}
Let us decompose $H^{\a m, \b n, \g k}$ into symmetric and antisymmetric parts in the first and second pair
of indices
\be
H^{\a m, \b n, \g k} = H^{(\a m, \b n), \g k}+H^{[\a m, \b n], \g k}~.
\label{n1.1.15}
\ee
In our subsequent analysis,
it is more convenient to view $H$ as a function of $X^m$ rather than of ${\bm X}^{\a \b}$.
Then it is easy to see from eqs.~\eqref{n1.1.8}  that
$H^{(\a m, \b n), \g k}$ has to be an even function of $X^m$ while $H^{[\a m, \b n], \g k}$
has to be an odd function. Since even and odd functions cannot mix in the conservation law~\eqref{n1.1.14.1}, \eqref{n1.1.14.2}
$H^{(\a m, \b n), \g k}$ and $H^{[\a m, \b n], \g k}$ must satisfy~\eqref{n1.1.14.1}, \eqref{n1.1.14.2}  separately.
This means that we can consider  $H^{(\a m, \b n), \g k}$ and $H^{[\a m, \b n], \g k}$
independently.

First, we will consider the case of $H^{(\a m, \b n), \g k}$. Due to its symmetry properties, it is the sum of four possible terms:
\begin{enumerate}
\item
$H_1^{(\a m, \b n), \g k} =\varepsilon^{\a \b} \Theta^{\g} A^{[mn], k}$~,
\item
$H_2^{(\a m, \b n), \g k} =\varepsilon^{\a \b} (\gamma_r)^{\g}_{\ \d} \Theta^{\d}  B^{[mn], k, r}$~,
\item
$H_3^{(\a m, \b n), \g k} =(\g_p)^{\a \b} \Theta^{\g}   C^{(mn), k, p}$~,
\item
$H_4^{(\a m, \b n), \g k} =(\g_p)^{\a \b} (\gamma_r)^{\g}_{\ \d} \Theta^{\d}   D^{(mn), k, p,r}$~.
\end{enumerate}
Here we use the fact that every symmetric in $(\a, \b)$ matrix is proportional to a gamma-matrix.
We also  indicated that the matrices $A$ and $B$ are antisymmetric in $(m, n)$ and
$C$ and $D$ are symmetric. The tensors $A, \ B, \ C, \ D$ depend on $X^m$ and are symmetric
under $X^m \to -X^m$.  Now we will impose the conditions~\eqref{n1.1.11} as well as the conservation law~\eqref{n1.1.14.1},  \eqref{n1.1.14.2}.
To begin with, we will impose $\partial_{\a} H^{(\a m, \b n), \g k} =0$.
Then it is easy to see that
\bea
&&
 A^{[mn], k} =0~, \quad B^{[mn], k, r} =0 ~,
 \nonumber \\
&&
\eta_{p r}D^{(mn), k, p, r}=0~, \quad \ve_{rpq} D^{(mn), k, p, r} +\eta_{q q'} C^{(mn), k, q'}=0~.
\label{n1.1.16}
\eea
Hence, $H_1^{(\a m, \b n), \g k} =H_2^{(\a m, \b n), \g k}=0$.

Upon imposing \eqref{n1.1.11} we obtain
\bea
&&
C^{(mn), k, p}= C^{(mnp), k}~, \quad
D^{(mn), k, p, r}= D^{(mnp), (kr)}+ \frac{1}{2} \ve^{k r q} \eta_{q q'} C^{(mnp), q'}~,
\nonumber \\
&&
\eta_{mn}  C^{(mnp), k}=0~, \quad \eta_{mn}  D^{(mnp), (kr)}=0~, \quad
\eta_{kr}  D^{(mnp), (kr)}=0~.
\label{n1.1.17}
\eea
That is we find two symmetric traceless tensors $C^{(mnp), k}$ and $D^{(mnp), (kr)}$.
Substituting now~\eqref{n1.1.17} into~\eqref{n1.1.16} we find that $C^{(mnp), k}$ and $D^{(mnp), (kr)}$
are related to each other as follows
\begin{subequations}\label{n1.1.18}
\bea
&&
\eta_{p r}D^{(mnp), (kr)}+ \frac{1}{2} \ve^{k p q} \eta_{p p'} \eta_{q q'} C^{(mnp'), q'}=0~,\\
&&
\ve^{r p q} \eta_{r r'} \eta_{p p'} D^{(mnp'), (kr')} + C^{(mnq), k} +\frac{1}{2} C^{(mnk), q} -
\frac{1}{2}\eta^{q k} \eta_{p t} C^{(mnp), t}=0~.
\eea
\end{subequations}

Quite remarkably, eqs.~\eqref{n1.1.18} allow us to fully solve for $D^{(mnp), (kr)}$ in terms of $C^{(mnp), k}$.
In order to do this we will decompose $C^{(mnp), k}$ and $D^{(mnp), (kr)}$ into irreducible components.
To understand which irreducible components are relevant it is convenient to trade each
vector index for a pair of spinor ones. Since $C^{(mnp), k}$ and $D^{(mnp), (kr)}$ are symmetric and traceless
they become equivalent to symmetric tensors $C^{(\a_1\dots \a_6), (\b_1 \b_2)}$ and
$D^{(\a_1\dots \a_6), (\b_1 \dots \b_4)}$. Hence, $C$ contains irreducible components (that is, {\it totally} symmetric tensors) of
rank 8, 6, 4 and 2, whereas $D$ contains irreducible components of rank 10, 8, 6, 4 and 2 (note that neither $C$ nor
$D$ contains the rank 0 representation since the number of $\a$ and $\b$ indices is different). Now let us recall that
all irreducible components must be even functions of $X^{\a \b}$. This means that irreducible tensors of rank 10, 6 and 2
must vanish since they contain an odd number of $X^{\a \b}$. Therefore, in both $C$ and $D$ only irreducible components
of rank 8 and rank 4 can contribute. Going back to the vector indices, let us denote the irreducible components of $C$ as
$C_1^{(mnpk)}$ and $C_2^{(mn)}$ and the irreducible components of $D$ as $D_1^{(mnpk)}$ and $D_2^{(mn)}$.
By construction, all these tensors are symmetric and traceless.
It is not hard to construct explicit decompositions of $C^{(mnp), k}$ and $D^{(mnp), (kr)}$ into the irreducible components.
The decomposition of $C^{(mnp), k}$ reads
\bea
C^{(mnp), k} &= &C_1^{(mnp k)} +\eta^{pk} C_2^{(mn)} + \eta^{nk}C_2^{(mp)} + \eta^{mk} C_2^{(np)}
\nonumber \\
&+& \eta^{mn} C_3^{(pk)} + \eta^{mp}C_3^{(nk)} + \eta^{np} C_3^{(mk)}~.
\label{n1.1.19}
\eea
Here we have taken into account that $C^{(mnp), k} $ is symmetric in $(m, n, p)$. Recalling now that it is also traceless (see eq.~\eqref{n1.1.17}),
$\eta_{mn} C^{(mnp), k}=0,$ gives
\be
C_3^{(mn)}= -\frac{2}{5} C_2^{(mn)}~.
\label{n1.1.20}
\ee
Similarly, we have the following decomposition of $D^{(mnp), (kr)}$:
\bea
D^{(mnp), (kr)} &= & \ve^{mks} \eta_{s s'} T^{(np), r, s'} + \ve^{nks} \eta_{s s'} T^{(mp), r, s'} +
\ve^{pks} \eta_{s s'} T^{(mn), r, s'}
\nonumber \\
&+&
 \ve^{mrs} \eta_{s s'} T^{(np), k, s'} + \ve^{nrs} \eta_{s s'} T^{(mp), k, s'} +
\ve^{prs} \eta_{s s'} T^{(mn), k, s'}~,
\label{n1.1.21}
\eea
where $T^{(np), r, s}$ is given in terms of $D_1^{(nprs)}$ and $D_2^{(np)}$ by
\be
T^{(np), r, s} = D_1^{(nprs)} + \eta^{nr} D_2^{(ps)} + \eta^{pr} D_2^{(ns)} +  \eta^{np} D_3^{(rs)}~.
\label{n1.1.22}
\ee
Recalling that $D^{(mnp), (kr)}$ is traceless in each group of indices relates
\be
D_3^{(mn)}= -\frac{2}{5} D_2^{(mn)}~.
\label{n1.1.23}
\ee
Let us point out that symmetry allows us to add in~\eqref{n1.1.22} terms of the form
$\eta^{ns} D_4^{(pr)} + \eta^{ps} D_4^{(nr)} + \eta^{rs} D_5^{(np)}$ with some symmetric traceless tensors
$D_4^{(np)}$ and $D_5^{(np)}$.
However, it is straightforward
to show that such terms will cancel when we substitute them in~\eqref{n1.1.21} and, hence, they can be ignored.
Substituting the irreducible decompositions~\eqref{n1.1.19}, \eqref{n1.1.20}, \eqref{n1.1.21}, \eqref{n1.1.22}, \eqref{n1.1.23} into~\eqref{n1.1.8}
yields the solution
\be
D_1^{(nprs)} =-\frac{3}{10} C_1^{(nprs)}~, \quad D_2^{(np)} =-\frac{1}{8} C_2^{(np)}~.
\label{n1.1.24}
\ee
Thus, the tensor $D$ is fully determined in terms of $C$.

Finally, let us consider the equation~\eqref{n1.1.14.2} which involves the derivative with respect to $X$.
It is possible to show using~\eqref{n1.1.18} that~\eqref{n1.1.14.2} is equivalent to a pair of
simple equations
\be
\partial_m C^{(mnp), k}=0~, \quad \partial_m D^{(mnp), (kr)}=0~.
\label{n1.1.25}
\ee

Now we are ready to construct an explicit solution. It is enough to consider $C^{(mnp), k}$
since $D^{(mnp), (kr)}$ is fully expressed in terms of it.
Using the symmetry in $(m, n, p)$, the scaling property~\eqref{n1.1.7}
and the fact that it is an even function of $X^m$ we have the following most general ansatz
\bea
C^{(mnp), k} &=& \frac{a}{X^3} \Big[\eta^{mn} \eta^{pk} +\eta^{mk} \eta^{np}+ \eta^{mp} \eta^{nk}\Big]
+\frac{b}{X^5} \Big[\eta^{mn} X^p X^k +\eta^{mp} X^n X^k+ \eta^{np} X^m X^k\Big]
\nonumber \\
&+ &
\frac{c}{X^5} \Big[\eta^{pk} X^m X^n +\eta^{nk} X^m X^p+ \eta^{mk} X^n X^p\Big]
+\frac{d}{X^7} X^m X^n X^p X^k~.
\label{n1.1.26}
\eea
Here we adopt the vector notation $X^2= X_m X^m$ and $a, \ b, \ c ,\ d$ are some coefficients.
Imposing $\eta_{mn} C^{(mnp), k}=0$ gives
$c=-5a, \  d=10 a -5 b$. Imposing $\partial_m C^{(mnp), k}=0$ gives $b= 3 a, \ d= -10 b -5 c$.
Thus, we obtain that there is only one independent coefficient which we choose to be $a$ and the remaining
three coefficients are given by
\be
b= 3 a\,, \quad c=d=-5a~.
\label{n1.1.27}
\ee
At this step it is convenient to give particular values to
$a,b,c,d$, say
\be
a=1~,\quad b=3~, \quad c=d=-5~.
\label{partvalues}
\ee
Then, the free parameter, which we denote as $d_{\cN=1}$,
will show up as an overall coefficient
in the final answer for
the correlation function presented below.

It is now straightforward to compute $T^{(np), r, s}$ (and, hence, $D^{(mnp), (kr)}$). Using
eqs.~\eqref{n1.1.22}, ~\eqref{n1.1.23}, ~\eqref{n1.1.24} and the explicit form of $C^{(mnp), k}$
in~\eqref{n1.1.26}, ~\eqref{n1.1.27} we obtain\footnote{
An explicit calculation of $T^{(np), r, s}$ also gives additional
terms containing $ \eta^{n s}$, $ \eta^{p s}$ or $ \eta^{r s}$.
However, all such terms will cancel
when we substitute them into the expression for $D^{(mnp), (kr)}$ in~\eqref{n1.1.21} and, hence, they can be ignored.
It is analogous to the cancellation discussed below~\eqref{n1.1.23}.}
\be
T^{(np), r, s}=\frac{1}{2} \Big[\frac{\eta^{nr} X^p X^s  +\eta^{pr} X^n X^s -\eta^{np} X^r X^s  }{X^5} +
\frac{3 X^n X^p X^r X^s}{X^7}\Big]~.
\label{n1.1.28}
\ee
As the last step, one can check that with $T^{(np), r, s}$ given by~\eqref{n1.1.28} the differential
constraint $\partial_m D^{(mnp), (kr)}=0$ from eq.~\eqref{n1.1.25} is indeed satisfied.
Thus, we have shown that $H^{(\a m, \b n), \g k}$ is fixed by symmetries and by the conservation law up to an overall coefficient $d_{\cN =1}$.

In a similar manner we can consider the antisymmetric part $H^{[\a m, \b n], \g k}$. Fortunately, the consideration is much simpler.
It is not hard to show following the same logic as above that already imposing $\pa_{\a} H^{[\a m, \b n], \g k}=0$ and
eq.~\eqref{n1.1.11} sets $H^{[\a m, \b n], \g k}=0$.

To summarise, we have shown that the three-point function of supercurrents in ${\cal N}=1$ superconformal
theories is fixed up to one overall coefficient $d_{\cN=1}$. The explicit form of the function
$H^{\a m, \b n, \g k}$ is given by
\bea
H^{\a m, \b n, \g k} (X, \Theta)&=&\ri d_{\cN=1} \Big[ (\g_p)^{\a \b} \Theta^{\g}   C^{(mnp), k} +\frac{1}{2} (\g_p)^{\a \b} (\gamma_r)^{\g}_{\ \d} \Theta^{\d}
\ve^{k r q} \eta_{q q'} C^{(mnp), q'}\nonumber \\
&+&(\g_p)^{\a \b} (\gamma_r)^{\g}_{\ \d} \Theta^{\d}   D^{(mnp), (kr)}\Big]~.
\label{n1.1.29}
\eea
The tensors $C^{(mnp), k}$ and $D^{(mnp), (kr)}$ are given by~\eqref{n1.1.26}, \eqref{partvalues} and \eqref{n1.1.21}, \eqref{n1.1.28},
respectively.

Obviously, the correlation function (\ref{n1.1.5}) changes its
sign under permutation of the superspace points $z_1$ and $z_3$ with the simultaneous swap
of indices $\cal A$ and $\cal C$
\be
\langle J_{\cal A}(z_1) J_{\cal B}(z_2) J_{\cal C}(z_3)\rangle
=-\langle J_{\cal C}(z_3) J_{\cal B}(z_2) J_{\cal
A}(z_1)\rangle~.
\ee
As a consequence, the tensor $H$ should obey the following
equation
\bea
H_{\alpha_1\alpha_2\alpha_3,\beta_1\beta_2\beta_3,\gamma_1\gamma_2\gamma_3}({\bm X}_3,\Theta_3)&=&
\frac1{{\bm X}_3{}^8{\bm x}_{13}{}^{8}} ({\bm x}_{13}^{-1})_{\alpha_1}{}^{\alpha_1'}
({\bm x}_{13}^{-1})_{\alpha_2}{}^{\alpha_2'}({\bm x}_{13}^{-1})_{\alpha_3}{}^{\alpha_3'}
{\bm x}_{13}{}^{\gamma_1'}{}_{\gamma_1} {\bm x}_{13}{}^{\gamma_2'}{}_{\gamma_2}
{\bm x}_{13}{}^{\gamma_3'}{}_{\gamma_3}
\non\\&&\times
{\bm x}_{13}^{\beta_1'\delta_1}{\bm X}_{3\delta_1\beta_1}
{\bm x}_{13}^{\beta_2'\delta_2}{\bm X}_{3\delta_2\beta_2}
{\bm x}_{13}^{\beta_3'\delta_3}{\bm X}_{3\delta_3\beta_3}
\non\\&&\times
H_{\gamma_1'\gamma_2'\gamma_3',\beta_1'\beta_2'\beta_3',\alpha_1'\alpha_2'\alpha_3'}(-{\bm X}_1^{\rm T},
 -\Theta_1)~.
 \label{6.53}
\eea
However, it appears to be very difficult to check that the tensor
(\ref{n1.1.29}) obeys this equation because of its complicated
structure. Alternatively, in Appendix C
 we demonstrate that the expression (\ref{n1.1.29}) can be derived as
a result of reduction of the $\cN=2$ supercurrent correlation
function which will be computed in subsection\ \ref{N2supercurrent}.
This will prove that (\ref{n1.1.29}) obeys the required
property  (\ref{6.53}).


\section{Correlators in  $\cN=2$ superconformal field theory}

We start with an example of a classically $\cN=2$ superconformal
field theory.
It is described by  $n$ primary chiral scalars $\F$ (viewed as a column vector)
of dimension 1/2,
$\bar D_\a \F =0$,
and their conjugate antichiral superfields $\F^\dagger$
with action\footnote{For the action
\eqref{77.1} and the associated conserved current multiplets
\eqref{77.2} and \eqref{77.3},
we have employed the complex basis for the superspace Grassmann coordinates
introduced in Appendix \ref{AppendixB}. In the remainder of this section,
 the real basis for the superspace Grassmann coordinates will be used.}
\bea
S =  \int \rd^3x \rd^2 \q  \rd^2 \bar \q \, \F^\dagger \F
+ \left\{  \l \int \rd^3x \rd^2 \q \, (\F^{\rm T} \F)^2 +{\rm c.c.}
\right\}~.
\label{77.1}
\eea
Here $\l$ is a dimensionless coupling constant.
The supercurrent of this model is \cite{DS,KT-M11}
 \bea
J_{\a\b}&=& 2 \bar D_{(\a}{\F}^\dagger D_{\b)}\F
+\hf[D_{(\a},\DB_{\b)}] (\bar \F^\dagger \F)
~. \label{77.2}
\eea
The action is obviously $\sO(n)$ invariant.
The corresponding flavour current multiplet is
\bea
L^{\bar a} = \F^\dagger \S^{\bar a} \F~, \label{77.3}
\eea
with $\S^{\bar a}$ being the generator of the flavour $\sO(n)$ group.
It is not difficult to check that  on-shell
 the currents
(\ref{77.2}) and (\ref{77.3}) obey the $\cN=2$
conservation equations in (\ref{1}) and (\ref{2}), respectively.
In the free case, $\l =0$, the action is $\sU (n)$ invariant; the corresponding
flavour current multiplet is given by  (\ref{77.3}), in which
$\S^{\bar a}$ now stands for the generator of the  $\sU(n)$ group.
The free model is trivially superconformal at the quantum level.

A natural  generalisation of  \eqref{77.1}
is the most general off-shell 3D $\cN=2$  superconformal sigma model
given in \cite{KPT-MvU}.\footnote{For target spaces with $\sU(1)$
isometries, 3D supersymmetric sigma models may be formulated in terms of Abelian vector multiplets described in terms of gauge invariant field strengths.
In the $\cN=2$ case, the field strength of a vector multiplet is a real linear superfield.
The $\cN=2$ superconformal sigma models formulated using real linear superfields
were studied in \cite{BPS,KT-M11}.}

\subsection{${\cN}=2$ flavour current multiplets}

The $\cN=2$ flavour current is described by a primary scalar
$L$ of dimension 1, which means that its superconformal transformation is
\bea
\delta L = -\xi L -\sigma(z) L~.
\eea
This transformation law is uniquely fixed by requiring the conservation equation
\be
(D^{\alpha(I} D_\alpha^{J)}- \frac12 \delta^{IJ}
D^{\alpha K} D_\alpha^{K} )L =0
\label{178}
\ee
to be superconformal.

As in the $\cN=1$ case, we assume that the $\cN=2$ superconformal
field theory under study has a set of flavour current multiplets $L^{\bar a}$
associated with a simple flavour group.
Since the superfields $L^{\bar a}$ carry neither spinor nor
$R$-symmetry indices, their two-point correlation function is
simply
\be
\langle L^{\bar a}(z_1) L^{\bar b}(z_2) \rangle = a_{\cN=2}
\frac{\delta^{\bar a\bar b}}{{\bm x}_{12}{}^2}~,
\ee
where $a_{\cN=2}$ is a free coefficient.
It is straightforward to check that this correlator is symmetric,
$\langle L^{\bar a}(z_1) L^{\bar b}(z_2) \rangle
= \langle  L^{\bar b}(z_2)  L^{\bar a}(z_1) \rangle$,
and  respects the conservation equation (\ref{178}),
\be
(D_{(1)}^{\alpha(I} D_{(1)\alpha}^{J)}- \frac12 \delta^{IJ}
D_{(1)}^{\alpha K} D_{(1)\alpha}^{K} )\langle L^{\bar a}(z_1) L^{\bar b}(z_2) \rangle =0
~,\qquad
z_1\ne z_2~.
\ee

Our next goal is to work out the most general expression for
the three-point function $\langle
L^{\bar a}(z_1) L^{\bar b}(z_2) L^{\bar c}(z_3) \rangle$
compatible with all the physical requirements.
According to (\ref{OOO}), we
have to make the ansatz
\be
\langle
L^{\bar a}(z_1) L^{\bar b}(z_2) L^{\bar c}(z_3) \rangle
=\frac1{{\bm x}_{13}{}^2 {\bm x}_{23}{}^2}
\left[f^{\bar a\bar b\bar c}H_{(f)}({\bm X}_{3},\Theta_{3})
+d^{\bar a\bar b\bar c}H_{(d)}({\bm X}_{3},\Theta_{3})\right]~,
\label{5.57}
\ee
where $f^{\bar a \bar b\bar c}$ and $d^{\bar a\bar b\bar c}$
are antisymmetric and symmetric invariant tensors, respectively.
Both functions $H_{(f)}$ and $H_{(d)}$ should have the same scaling
property
\be
H_{(f,d)}(\lambda^2 {\bm X},\lambda\Theta) = \lambda^{-2}H_{(f,d)}({\bm X},\Theta)~
\label{181}
\ee
and obey the conservation condition
\be
({\cal D}^{\alpha(I}{\cal D}_\alpha^{J)}- \frac12 \delta^{IJ}
{\cal D}^{\alpha K}{\cal D}_\alpha^{K} ) H_{(f,d)}=0~.
\label{182}
\ee
The latter constraint is obtained from (\ref{178}) with the use of
(\ref{useful-prop-a}).

The correlation function (\ref{5.57}) is invariant under
exchange of the superspace points $z_1$ and $z_2$
and the flavour indices $\bar a$ and $\bar b$. As a consequence, the
functions $H_{(f)}$ and $H_{(d)}$ are constrained by
\be
H_{(f)}(-{\bm X}^{\rm T},-\Theta) = -H_{(f)}({\bm X},\Theta)~,\qquad
H_{(d)}(-{\bm X}^{\rm T},-\Theta) = H_{(d)}({\bm X},\Theta)~.
\label{183}
\ee
The general solutions of the equations (\ref{181}), (\ref{182}) and
(\ref{183}) prove to be
\be
H_{(f)}({\bm X},\Theta) =b_{\cN=2}\frac{\ri\varepsilon_{IJ}\Theta^{I}_\alpha X^{\alpha\beta} \Theta^{J}_\beta  }{X^3}~,
\qquad
H_{(d)}({\bm X},\Theta) = \tilde b_{\cN=2}\frac1X~.
\label{5.60}
\ee
Here $b_{\cN=2}$ and $\tilde b_{\cN=2}$ are two real coefficients.
One can also check  that
the functions $H_{(f)}$ and $H_{(d)}$ obey the equations
\begin{subequations}
\bea
H_{(f)}(-{\bm X}_{1}^{\rm T},-\Theta_{1}) &=& - {\bm x}_{13}{}^2 {\bm X}_{3}{}^2
H_{(f)}({\bm X}_{3},\Theta_{3})~,\\
H_{(d)}(-{\bm X}_{1}^{\rm T},-\Theta_{1}) &=& {\bm x}_{13}{}^2 {\bm X}_{3}{}^2
H_{(d)}({\bm X}_{3},\Theta_{3})~,
\eea
\end{subequations}
which are corollaries of the following symmetry property
\be
\langle L^{\bar a}(z_1) L^{\bar b}(z_2) L^{\bar c}(z_3) \rangle =
\langle L^{\bar c}(z_3) L^{\bar b}(z_2) L^{\bar a}(z_1) \rangle~.
\ee

Finally we point out that the functions (\ref{5.60}) can be
rewritten in terms of the covariant object ${\bm X}_{\alpha\beta}$
with the use of (\ref{XX})
\begin{subequations}\label{7.12}
\bea
H_{(f)}({\bm X},\Theta) &=&b_{\cN=2}\frac{\ri\varepsilon_{IJ}\Theta^{\alpha
I}\Theta^{J\beta} {\bm X}_{\alpha\beta} }{{\bm X}^3}~,
\label{7.12a}
\\
H_{(d)}({\bm X},\Theta) &=& \tilde b_{\cN=2}
\frac1{\bm X} \left(  1+ \frac18 \frac{\Theta^4}{{\bm X^2}}
 \right)~. \label{7.12b}
\eea
\end{subequations}
In verifying eq. (\ref{7.12a}),  the $\cN=2$ identity
$\varepsilon^{IJ}\Theta^I_\alpha \Theta^J_\beta \Theta^2 =0$ may
be useful. We point out that the expression in parentheses in \eqref{7.12b}
involves the square of the superconformal invariant \eqref{invariant}.

It should be stressed that the appearance of the $d$-term
in the
flavour current correlation function (\ref{5.57}) is a novel
feature which distinguishes the $\cN=2$ superconformal field theories  from
the $\cN=1$ ones considered in section  \ref{N1flavour-sect} and
from non-supersymmetric ones studied in \cite{OP}. In contrast to
the four-dimensional theories, in three dimensions this part of the
correlation function cannot be considered as an anomaly induced
contribution. To understand the role of this part of the
correlation function it would be interesting to consider some
examples of $\cN=2$ theories in which this contribution is
non-trivial.\footnote{It may be shown that both  $f$- and $d$-terms are generated
in the free model \eqref{77.1} with $\l=0$. }
We leave this issue for further studies.


\subsection{$\cN=2$ supercurrent}
\label{N2supercurrent}


The ${\cal N}=2$ supercurrent is described by a primary symmetric
second-rank spinor $J_{\a \b} = J_{(\a \b)}$ of dimension 2,
hence its superconformal transformation is
\be
\d J_{\a \b}= -\xi  J_{\a \b} -2 \s(z) J_{\a \b}
+2 \lambda_{(\a}{}^\g(z) J_{\b) \g}~.
\label{n2.2}
\ee
This transformation law is uniquely fixed by the condition that
 the supercurrent conservation equation
\be
D^{\a}_I J_{\a \b}=0~
\label{n2.1}
\ee
is superconformal.

According to the general prescription (\ref{OO}), the two-point function
for the supercurrent is given
by
\be
\langle J_{\a \b} (z_1) J^{\a' \b'} (z_2) \rangle = c_{\cN=2}
\frac{{\bm x}_{12 \a}{}^{ (\a'} {\bm x}_{12 \b}{}^{ \b')}}{ ({\bm x}_{12}{}^2)^3}~,
\label{n2.3}
\ee
where $c_{\cN=2} $ is a real coefficient.
It is not difficult to see that this correlator is symmetric,
$ \langle J_{\a \b} (z_1) J_{\a' \b'} (z_2)\rangle
=  \langle J_{\a' \b'} (z_2) J_{\a \b} (z_1)  \rangle$,
and respects the conservation equation \eqref{n2.1},
\be
D^{I\a}_{(1)} \langle J_{\a \b} (z_1) J_{\a' \b'} (z_2)\rangle
=0 ~,\qquad
z_1 \neq z_2~.
\label{n2.4}
\ee

The most general expression for  the three-point function for the supercurrent  is
\be
\langle J_{\a \a'} (z_1)   J_{\b \b'} (z_2) J_{\g \g'} (z_3) \rangle
=
\frac{{\bm x}_{13 \a \r} {\bm x}_{13 \a' \r'}   {\bm x}_{23  \b \s}
 {\bm x}_{23 \b' \s'}  }{ ({\bm x}_{13}{}^2)^3 ({\bm x}_{23}{}^2)^3}
H^{\r \r', \s \s'}{}_{\g \g'} ({\bm X}_{3}, \Theta_{3})~,
\label{n2.5}
\ee
where, by construction, the function $H^{\r \r', \s \s'}{}_{\g \g'} ({\bm X}, \Theta)$
obeys the symmetry property
\be
H^{\r \r', \s \s'}{}_{\g \g'} ({\bm X}, \Theta) = H^{(\r \r'), (\s \s')}{}_{(\g \g')} ({\bm X}, \Theta)~.
\label{n2.6}
\ee
Since both the supercurrent $J_{\a \b}$ and $H^{\a \a', \b \b', \g \g'} $
are Grassmann even,
the three-point function~\eqref{n2.6} has to be symmetric under the exchange $z_1 \leftrightarrow z_2,$
$\a, \a'   \leftrightarrow \b, \b'$.
Hence, $H^{\a \a', \b \b', \g \g'} ({\bm X}, \Theta)$ satisfies the following symmetry property
\be
H^{\a \a', \b \b', \g \g'} (-{\bm X}^{\rm T}, -\Theta) = H^{\b \b', \a \a', \g \g'} ({\bm X}, \Theta)~.
\label{n2.7}
\ee
In addition, $H^{\a \a', \b \b', \g \g'} $ is characterised by  the scaling property
\be
H^{\a \a', \b \b', \g \g'} (\l^2 {\bm X}, \l \Theta)
= \l^{-4} H^{\a \a', \b \b',  \g \g'} ({\bm X}, \Theta)~.
\label{n2.8}
\ee
With the use of eq.\ (\ref{useful-prop-a}),
the supercurrent conservation condition (\ref{n2.1}) is translated
to the following equation for $H^{\a \a', \b \b', \g \g'}$:
\be
{\cal D}^I_{\a} H^{\a \a', \b \b', \g \g'} ({\bm X}, \Theta)=0~.
\label{n2.9}
\ee
Just like in the problem of
the three-point correlator for the $\cN=1$ supercurrent considered in
section \ref{sect-N1-supercurrent},
it is convenient to trade each pair of
spinor indices for vector ones,
\be
H^{\a \a', \b \b', \g \g'}= (\g_m)^{\a \a'} (\g_n)^{\b \b'} (\g_p)^{\g \g'} H^{mnp}~,
\label{n2.10}
\ee
where $H^{mnp} ({\bm X}, \Theta)$ satisfies the same scaling property as~\eqref{n2.8} as well as
\be
H^{mnp} (-{\bm X}^{\rm T}, -\Theta) = H^{nmp} ({\bm X}, \Theta)
\label{n2.11}
\ee
and
\be
(\g^m)^{\a \b}{\cal D}^I_{\a } H_{mnp} ({\bm X}, \Theta)=0~.
\label{n2.12}
\ee

Unlike the ${\cal N}=1$ case,
now
it is not hard to list all possible structures consistent with the symmetry~\eqref{n2.11}
and the scaling property~\eqref{n2.8}. This makes the analysis considerably simpler than in the previous section.
Just like in the ${\cal N}=1$ case,
it is more convenient to view $H$ as function of $X^m$
rather than  ${\bm X}^{\a \b}$.
Then the building blocks which can appear in $H$ are
\bea
&&
\eta_{mn}\,, \quad \varepsilon_{mnp}\,, \quad  X^m =-\frac{1}{2} (\g^m)_{\a \b} X^{\a \b}\,, \quad X= \sqrt{X_m X^m}\,,
 \nonumber \\
&&
(\Theta \Theta)_m = -\frac{\ri}{2} (\g_m)^{\a \b} \Theta_{\a I}   \Theta_{\b J} \varepsilon^{IJ}\,, \quad
\Theta^2 = \Theta^{\a}_I    \Theta_{\a I}~.
\label{n2.13}
\eea
 Note that there is the following ${\cal N}=2$ identity
\be
(\Theta \Theta)_m \Theta^2=0~.
\label{n2.14}
\ee
Taking into account the symmetry property~\eqref{n2.11} and~\eqref{n2.8} we get the following general expression for $H$:
\be
H_{mnp}=
\sum_i A_i H_{i,mnp} + \sum_i B_i {\cal H}_{i,mnp} +\sum_i C_i {\bf H}_{i,mnp}~,
\label{n2.15}
\ee
where $A_i, \ B_i, \ C_i$ are some coefficients and the tensors $H_{i,mnp}$,
${\cal H}_{i,mnp}$, ${\bf H}_{i,mnp}$
are explicitly given by
\bea
&&
H_{1,mnp} = \frac{\eta_{mn} (\Theta \Theta)_p}{X^3}~, \quad
H_{2,mnp} = \frac{X_m X_n (\Theta \Theta)_p}{X^5}~, \nonumber \\
&&
H_{3,mnp} = \frac{X_m X_p  (\Theta \Theta)_n+X_n X_p  (\Theta \Theta)_m }{X^5}~,
\nonumber \\
&&
H_{4,mnp} = \frac{\eta_{mp} (\Theta \Theta)_n+\eta_{np}  (\Theta \Theta)_m }{X^3}~,
\nonumber \\
&&
H_{5,mnp} = \frac{X_m X_n X_p X^q (\Theta \Theta)_q}{X^7}~, \quad
H_{6,mnp} = \frac{\eta_{mn} X_p X^q (\Theta \Theta)_q}{X^5}~,
\nonumber \\
&&
H_{7,mnp} = \frac{\eta_{mp} X_n X^q (\Theta \Theta)_q +\eta_{np} X_m X^q (\Theta \Theta)_q  }{X^5}~;
\label{n2.16}
\eea
\bea
&&
{\cal H}_{1,mnp} = \frac{\ve_{mnr}  X^r (\Theta \Theta)_p}{X^4}~, \quad
{\cal H}_{2,mnp} = \frac{\ve_{mnr}  X_p (\Theta \Theta)^r}{X^4}~, \nonumber \\
&&
{\cal H}_{3,mnp} = \frac{\ve_{mnp}  X^r (\Theta \Theta)_r}{X^4}~, \quad
{\cal H}_{4,mnp} = \frac{\ve_{mnr}  X_p X^r X^q (\Theta \Theta)_q}{X^6}~;
\label{n2.17}
\eea
\bea
&&
{\bf H}_{1,mnp} = \frac{\eta_{np} X_m-\eta_{mp} X_n}{X^3} ~, \quad
{\bf H}_{2,mnp} = \frac{\eta_{np} X_m-\eta_{mp} X_n}{X^4} \Theta^2 ~,
\nonumber \\
&&
{\bf H}_{3,mnp} = \frac{\eta_{np} X_m-\eta_{mp} X_n}{X^5} \Theta^4~.
\label{n2.18}
\eea
Note that $H_{i,mnp}= H_{i,(mn) p}$, ${\cal H}_{i,mnp}= {\cal H}_{i,[mn] p}$,
${\bf H}_{i,mnp}= {\bf H}_{i,[mn] p}$.

It is easy to realize that $H_{i,mnp}$,  ${\cal H}_{i,mnp}$ and  ${\bf H}_{i,mnp}$ do not mix in the equation~\eqref{n2.12}
and, hence, they must satisfy the conservation law independently. Let us now substitute~\eqref{n2.16}, \eqref{n2.17}
and~\eqref{n2.18} in~\eqref{n2.12}. This equation will lead to two types of terms: terms linear in $\Theta$ and
terms proportional to $ \Theta^3$. Clearly, these terms must vanish separately. Let us first consider the terms linear in $\Theta$.
Using the identities
\bea
(\gamma^m)^\alpha_\beta{\cal D}_\alpha^I X^n &=&
 -\ri\eta^{mn}\Theta^I_\beta -\ri\Theta^I_\alpha
 \varepsilon^{mnp}(\gamma_p)^\alpha_\beta~,\non\\
(\gamma^m)^\alpha_\beta{\cal D}_\alpha^I \Theta^2 &=&
2(\gamma^m)^\alpha_\beta \Theta^I_\alpha~,\non\\
(\gamma_m)^\alpha_\beta {\cal D}^I_\alpha (\Theta\Theta)_n
&=&\ri\varepsilon_{IJ}\eta_{mn}\Theta^J_\beta
 +\ri\varepsilon_{IJ}\varepsilon_{mnp}(\gamma^p)_\beta^\alpha
  \Theta_\alpha^J
\eea
it is straightforward to show that
\be
B_1= B_2= B_3= B_4=0~, \quad C_1=C_2=C_3=0~.
\label{n2.20}
\ee
Thus, ${\cal H}_{i,mnp}$ and $ {\bf H}_{i,mnp}$ can be ignored and we can concentrate only on $H_{i,mnp}$ in eq.~\eqref{n2.16}.
Substituting~\eqref{n2.16} in~\eqref{n2.12} and considering only the terms linear in $\Theta$ gives the following constraints on the coefficients $A_i$:
\bea
&&
A_1 - A_4 =0\,, \quad A_3- A_6=0\,, \quad  A_2- A_7=0\,,
\nonumber \\
&& A_1+ 4 A_4 +A_7=0\,, \quad A_2+ 4 A_3 + A_5 +A_6 +A_7=0~.
\label{n2.21}
\eea
Similarly, concentrating on the terms cubic in $\Theta$, after straightforward but lengthy calculations we obtain the following system:
\bea
&&
3 A_1 + A_2+ A_3 + 6 A_4 +  A_6 + 2 A_7=0\,, \quad
3 A_1 + A_2 -A_3 - 3 A_4 +  A_6 -  A_7=0\,,
\nonumber \\
&&
5 A_3+ 3 A_5=0\,, \quad
3 A_1 + A_2+ 2 A_3 + 3 A_4 +  A_5 +A_6 +  A_7=0~.
\label{n2.22}
\eea
To derive the system of equations~\eqref{n2.22},
it is important to make use of the following ${\cal N}=2$ identity
\be
\Theta_{\a I} (\Theta \Theta)_m = \frac{\ri}{2} (\g_m)_{\a \b}\Theta^{\b J} \Theta^2 \ve_{IJ}~,
\label{n2.23}
\ee
which can easily be obtained by differentiating~\eqref{n2.14}.

The systems~\eqref{n2.21} and~\eqref{n2.22} turn out to be consistent and can be solved in terms of one independent
coefficient which we choose to be $A_1 \equiv A$:
\be
A_2=A_5=A_7 = - 5 A\,, \quad A_3= A_6 =3 A\,, \quad A_4= A~.
\label{n2.24}
\ee
Thus, the three-point function of the supercurrent is fixed up to a single coefficient $A$.

Since the three-point function has only one overall coefficient,
our result should
possess the right symmetry properties under the exchange $z_1 \leftrightarrow z_3$,
$z_2 \leftrightarrow z_3$. However,  since the final result is rather simple and contains only a few
terms listed in (\ref{n2.16}), the symmetry under, say, the $z_1 \leftrightarrow z_3$ exchange is not hard to verify. The invariance
of the three-point function
\be
\langle J_{\a \a'} (z_1)   J_{\b \b'} (z_2) J_{\g \g'} (z_3) \rangle=
\langle J_{\g \g'} (z_3)   J_{\b \b'} (z_2) J_{\a \a'} (z_1) \rangle
\label{n2.25}
\ee
implies the following equation\footnote{Due to the identity \eqref{n2.14}
it is trivial to rewrite \eqref{n2.16} in terms of $\bm X$ rather than $X$.}
on the tensor $H$
\bea
H^{\r \r', \s, \s'}{}_{\t \t'}  (-{\bm X}_{1}^{\rm T}, -\Theta_{1}) &=&
{\bm X}_{3}{}^4 {\bm X}_{1}^{\s \l}{\bm X}_{1}^{\s' \l'} {\bm x}_{13 \l \a}
 {\bm x}_{13 \l' \a'}
{\bm x}_{13}^{ \r \g}  {\bm x}_{13}^{ \r' \g'}
 {\bm x}_{13 \t \b}  {\bm x}_{13 \t' \b'}
 \non\\&&\times
H^{\b \b, \a, \a'}{}_{\g \g'} ({\bm X}_{3}, \Theta_{3})~.
\label{n2.26}
\eea
Using the formulae
\eqref{4.34.1}
we can relate ${\bm X}_{3}$
with ${\bm X}_{1}$ and $\Theta_{3}$ with $\Theta_{1}$
\be
{\bm X}_{1 \a \b} =-\frac{{\bm x}_{13 \a \a'}
{\bm x}_{13 \b \b'} {\bm X}_{3}^{\a' \b'}}{{\bm x}_{13}{}^4 {\bm X}_{3}{}^2}~, \quad
\Theta_{1 \a}^{ I} = \frac{ u_{13}^{IJ}  {\bm x}_{13 \a \b} {\bm X}_{3}^{\b \g}
\Theta_{3 \g J}}{{\bm x}_{13}{}^2 {\bm X}_{3}{}^2}~.
\label{n2.27}
\ee
It is now straightforward to substitute eq.~\eqref{n2.16} into~\eqref{n2.26} and verify that it is indeed fulfilled
if the coefficients $A_i$ satisfy~\eqref{n2.24}. More precisely, eq.~\eqref{n2.26} constrains the coefficients $A_i$ as follows
\be
A_1 -A_4=0\,, \quad 2 A_1 + A_2+ A_3=0\,, \quad 2 A_1 + A_6 +A_7=0~.
\label{n2.28}
\ee
The system of equations~\eqref{n2.28} is weaker than the system~\eqref{n2.21}, \eqref{n2.22} and is contained there.
That is why the conservation law alone fully constrains the coefficients.

To conclude this section, we rewrite explicitly the final result
for the tensor $H$ in terms of the
objects (\ref{three-points}):
\bea
H^{\alpha\alpha',\beta\beta',\gamma\gamma'}&=&
\ri d_{\cN=2} \bigg\{
\frac{2}{{\bm X}^3}\left[
\varepsilon^{\alpha(\beta}\varepsilon^{\beta')\alpha'}\Theta_I^\gamma\Theta_J^{\gamma'}
+\varepsilon^{\alpha(\gamma}\varepsilon^{\gamma')\alpha'}\Theta_I^\beta \Theta_J^{\beta'}
+\varepsilon^{\beta(\gamma}\varepsilon^{\gamma')\beta'}\Theta_I^\alpha\Theta_J^{\alpha'}
\right] \varepsilon^{IJ}\non\\&&
+\frac{1}{{\bm X}^5}\left[
3{\bm X}^{\alpha\alpha'}{\bm X}^{\gamma\gamma'}\Theta_I^\beta \Theta_J^{\beta'}
+3{\bm X}^{\beta\beta'}{\bm X}^{\gamma\gamma'}\Theta_I^\alpha \Theta_J^{\alpha'}
-5{\bm X}^{\alpha\alpha'}{\bm X}^{\beta\beta'}\Theta_I^\gamma\Theta_J^{\gamma'}
\right]\varepsilon^{IJ}\non\\&&
+\frac{1}{{\bm X}^5}\left[
5\varepsilon^{\alpha(\gamma}\varepsilon^{\gamma')\alpha'}{\bm X}^{\beta\beta'}
+5\varepsilon^{\beta(\gamma}\varepsilon^{\gamma')\beta'} {\bm X}^{\alpha\alpha'}
-3\varepsilon^{\alpha(\beta}\varepsilon^{\beta')\alpha'} {\bm X}^{\gamma\gamma'}
\right]{\bm X}^{\delta\delta'}\Theta^I_\delta
\Theta^J_{\delta'}\varepsilon_{IJ}\non\\&&
+\frac52\frac{1}{{\bm X}^7}{\bm X}^{\alpha\alpha'}{\bm X}^{\beta\beta'}{\bm X}^{\gamma\gamma'}
 {\bm X}^{\delta\delta'}\Theta^I_\delta \Theta^J_{\delta'}\varepsilon_{IJ}
 \bigg\} ~.
\label{7.41}
\eea
Here we have denoted the overall coefficient by $d_{\cN=2}$.


\section{Correlators in $\cN=3$ superconformal field theory}

The off-shell $\cN=3$ superconformal sigma model in three dimensions
proposed in \cite{KPT-MvU}
is a nontrivial example of classically $\cN=3$ superconformal theories.
 Its formulation is based on the
projective superspace techniques \cite{KLR,LR-projective}
(see \cite{K-Lectures} for a review). An alternative approach to describe
off-shell $\cN=3$ hypermultiplets in three dimensions  \cite{Zupnik97}
is provided by the harmonic superspace formalism
\cite{GIKOS,GIOS}, see, e.g., \cite{Buchbinder:2008vi} for the formulation
of the ABJM models \cite{ABJM}  in $\cN=3$ harmonic superspace.
In the present paper, we will not discuss the harmonic and the projective
superspace formulations for the off-shell hypermultiplet as it goes  beyond
our goals.
Here we will simply provide examples of $\cN=3$ supercurrent and
flavour current multiplets, and for this it suffices to consider a free on-shell massless
$\cN=3$ hypermultiplet.\footnote{Although the    harmonic and the projective
 formulations for the free hypermultiplet differ off the mass shell,
they lead to the same on-shell superfield.}
It is described by a primary superfield $q^i$,
and its conjugate $\bar q_i$,  subject
to the equation of motion \cite{Zupnik97}
\bea
D^{(ij}_\alpha q^{k)} =0~,
\label{hyper-constraint}
\eea
which is the 3D $\cN=3$ analogue of the famous 4D $\cN=2$
hypermultiplet constraints due to Sohnius \cite{Sohnius-hyper}.
Here $D^{ij}_\a$ is obtained from $D^I_\a$ by replacing its isovector index
with  a pair  of isospinor ones by the general rule \cite{KPT-MvU}
\bea
Z^I ~\to ~ Z_i{}^j :=\frac{\rm i}{\sqrt 2} (\vec{Z} \cdot \vec{\s})_i{}^j
\equiv Z^I (\t^I)_i{}^j~, \qquad Z_i{}^i=0~,
\label{N=3conversion}
\eea
with $\vec \s$ being the Pauli matrices. The hypermultiplet $q^i$ transforms
in the defining representation
of $\sSU(2)$, which is the double cover of the $R$-symmetry group $\sSO(3)$.
The $\sSU(2)$ indices are raised and lowered with the
antisymmetric tensors $\varepsilon^{ij}$ and $\varepsilon_{ij}$,
$\varepsilon^{12}=-\varepsilon_{12}=1$.

Let us consider a system of $n$ free on-shell hypermultiplets.
It is described by
a column $n$-vector  ${\bm q}^i$ constrained by \eqref{hyper-constraint}
and its conjugate ${\bm q}^\dagger_i$.
The supercurrent $J_\alpha$  and a flavour current multiplet $L^{\bar a}_{ij}$ are
given by
\begin{subequations}\label{hyper-currents}
\bea
J_\alpha &=&
\ri \,
 {\bm q}^\dagger_i \stackrel{\longleftrightarrow} {D^{ij}_\alpha }{\bm q}_j ~,\\
L^{\bar a}_{ ij}&=&
\ri \,
 {\bm q}^\dagger_{(i} \S^{\bar a}{\bm q}_{j)}
~,
\eea
\end{subequations}
where $\Sigma^{\bar a}$ is a flavour group generator.
With the use of (\ref{hyper-constraint})
it is possible to check that the operators (\ref{hyper-currents})
obey the $\cN=3$ conservation equations given in (\ref1) and (\ref2). In
the $\sSU(2)$ notation,  these equations read
\begin{subequations}
\bea
D^{ij\,\alpha} J_\alpha &=&0~, \\
D^{(ij}_\alpha L^{kl)} &=&0~.
\eea
\end{subequations}

We now turn to
studying the correlation functions of
the supercurrent and flavour current multiplets in  quantum $\cN=3$
superconformal models.

\subsection{$\cN=3$ flavour current multiplets}
\label{sect-N3-flavour}


The $\cN=3$ flavour current multiplet  is described by a primary real isovector $L^{I}$
of dimension 1, which transforms under the superconformal group as
\be
\delta L^{I} = -\xi L^{I} -\sigma(z) L^{I}
+\Lambda^{IJ}(z)L^{J}
\ee
and obeys the conservation equation
\be
D^{(I}_\alpha L^{J)}-\frac13 \delta^{IJ}D^K_\alpha L^{K} =0~.
\label{N3-flavour-conserv}
\ee

Similar to the $\cN=1$ and $\cN=2$ cases, we assume that the $\cN=3$ superconformal
field theory under study has a set of flavour current multiplets $L^{I\bar a}$
associated with a simple flavour group.
According to (\ref{OO}), the two-point correlator for these multiplets is
\be
\langle L^{I\bar a}(z_1) L^{J\bar b}(z_2) \rangle
= a_{\cN=3} \frac{\delta^{\bar a\bar b} u^{IJ}_{12}}{{\bm x}_{12}{}^2}~,
\ee
with some coefficient $a_{\cN=3}$.
It may be  checked that this two-function is symmetric,
$\langle L^{I\bar a}(z_1) L^{J\bar b}(z_2) \rangle
= \langle   L^{J\bar b}(z_2) L^{I\bar a}(z_1) \rangle$,
and respects the conservation law (\ref{N3-flavour-conserv}),
\be
D^{(K}_{(1)\alpha} \langle L^{I)\bar a}(z_1) L^{J\bar b}(z_2) \rangle
-\frac13 \delta^{KI} D^L_{(1)\alpha}
\langle L^{L\bar a}(z_1) L^{J\bar b}(z_2) \rangle =0~, \qquad
z_1\ne z_2~.
\ee

For the three-point function
$\langle L^{I\bar a}(z_1) L^{J\bar b}(z_2) L^{K\bar c}(z_3) \rangle$,
we follow (\ref{OOO}) and make the ansatz
\be
\langle L^{I\bar a}(z_1) L^{J\bar b}(z_2) L^{K\bar c}(z_3) \rangle
=\frac{u_{13}^{II'}u_{23}^{JJ'}}{{\bm x}_{13}{}^2{\bm x}_{23}{}^2}
\left(f^{\bar a\bar b\bar c} H_{(f)}^{I'J'K}({\bm X}_{3},\Theta_{3})
+d^{\bar a\bar b\bar c} H_{(d)}^{I'J'K}({\bm X}_{3},\Theta_{3})
\right)
~,
\label{N3flavour}
\ee
where $f^{\bar a\bar b\bar c}$ and $d^{\bar a\bar b\bar c}$ are
antisymmetric and symmetric invariant tensors of the flavour group.
The functions $H_{(f,d)}^{IJK}$ should obey the following scaling property
\be
H_{(f,d)}^{IJK}(\lambda^2 {\bm X},\lambda \Theta) =\lambda^{-2}
H_{(f,d)}^{IJK}({\bm X},\Theta)~.
\label{H-scaling3}
\ee

The three-point function under consideration has to possess the symmetry property
\be
\langle L^{I\bar a}(z_1) L^{J\bar b}(z_2) L^{K\bar c}(z_3) \rangle
=\langle L^{J\bar b}(z_2) L^{I\bar a}(z_1) L^{K\bar c}(z_3) \rangle~,
\ee
which
implies the following constraints for $H_{(f,d)}$
\be
H_{(f)}^{IJK}({\bm X},\Theta) = -H_{(f)}^{JIK}(-{\bm X}^{\rm
T},-\Theta)~,\quad
H_{(d)}^{IJK}({\bm X},\Theta) = H_{(d)}^{JIK}(-{\bm X}^{\rm T},-\Theta)~.
\label{H12}
\ee
The most general solution of the equations (\ref{H-scaling3}) and
(\ref{H12}) can be written as
\be
H_{(d)}^{IJK}=\sum_n b_n H_{(d)n}^{IJK}~,\qquad
H_{(f)}^{IJK}=\sum_{n}c_n H^{IJK}_{(f)n}
+\sum_n d_n {\cal H}_{(f)n}^{IJK}
\label{Hflavour-N3}
~,
\ee
where $b_n$, $c_n$ and $d_n$ are some coefficients and the tensors
$H_{(d)n}^{IJK}$, $H^{IJK}_{(f)n}$ and ${\cal H}^{IJK}_{(f)n}$ are
\bea
H_{(d)1}^{IJK} &= & \frac{\varepsilon_{IJL}A_{LK}}{X^3}~,\quad
H_{(d)2}^{IJK}=
\frac{\varepsilon^{IKL}B^{JL}+\varepsilon^{JKL}B^{IL}}{X^2}~,\non\\
H_{(d)3}^{IJK}&=&\frac{\varepsilon^{IJL}A_{LK}\Theta^2}{X^4}~,\quad
H_{(d)4}^{IJK}=\frac{\varepsilon^{IKL}B^{JL}+\varepsilon^{JKL}B^{IL}}{X^3}
\Theta^2~;
\eea
\bea
H_{(f)1}^{IJK}&=&\frac{\varepsilon^{IJK}}{X}~,\qquad
H_{(f)2}^{IJK}=-\frac12\frac{A^{IL}\varepsilon^{LJK}+A^{JL}\varepsilon^{LIK}}{X^3}
~, \non\\
H_{(f)3}^{IJK}&=&-\frac12\frac{\delta^{IJ}\varepsilon^{KMN}A^{MN}}{X^3}~,
\qquad
H_{(f)4}^{IJK}=\frac14
\frac{A^{IJ}\varepsilon^{KMN}A^{MN}}{X^5} ~,\non\\
H_{(f)5}^{IJK}&=&-\frac12\frac{\varepsilon^{IJL}B^{KL}\Theta^2}{X^3}~,
\qquad
H_{(f)6}^{IJK}=-\frac12\frac{\varepsilon^{IJK}\Theta^4}{X^3}~;
\label{H_n-flavour}
\eea
\bea
{\cal H}_{(f)1}^{IJK}&=&\frac{\varepsilon^{IJK}\Theta^2}{X^2}~,
\qquad
{\cal H}_{(f)2}^{IJK}=\frac{\varepsilon^{IJL}B^{LK}}{X^2}~,\non\\
{\cal H}_{(f)3}^{IJK}&=&-\frac12\frac{\varepsilon^{IJK}\Theta^6}{X^4}~,
\qquad
{\cal
H}_{(f)4}^{IJK}=-\frac12\frac{\delta^{IJ}\varepsilon^{KMN}A^{MN}\Theta^2}{X^4}~,\non\\
{\cal H}_{(f)5}^{IJK}&=&-\frac12\frac{\delta^{IK}\varepsilon^{JMN}A^{MN}+\delta^{JK}
 \varepsilon^{IMN}A^{MN}}{X^4}\Theta^2~,\non\\
{\cal H}_{(f)6}^{IJK}&=&-\frac12\frac{B^{IJ}\varepsilon^{KMN}A^{MN}}{X^4}~.
\label{calH}
\eea
Here we have introduced
\bea
A^{IJ}:=\ri\Theta^{I\alpha}X_{\alpha\beta}\Theta^{J\beta} = -A^{JI}~,\qquad
B^{IJ}:=\Theta^{I\alpha}\Theta^J_\alpha = B^{JI}~.
\label{AB}
\eea

In principle, the set (\ref{calH})  could be extended by one more term
\be
{\cal
H}_{(f)7}^{IJK}=-\frac12\frac{B^{IK}\varepsilon^{JMN}A^{MN}+B^{JK}\varepsilon^{IMN}A^{MN}}{X^4}~,
\ee
which obeys both equations (\ref{H-scaling3}) and (\ref{H12}). However
this term is linearly dependent
of the others,
\be
{\cal H}_{(f)7}^{IJK} = {\cal H}_{(f)4}^{IJK} + {\cal H}_{(f)5}^{IJK} -
{\cal H}_{(f)6}^{IJK}~,
\ee
and therefore
the list (\ref{calH}) is complete.

The tensor $H_{(f)}^{IJK}$ in (\ref{Hflavour-N3}) is determined by two
sectors with functions $H^{IJK}_{(f)n}$ and ${\cal H}^{IJK}_{(f)n}$. As will
be seen further, these pieces should independently obey the
constraints imposed by the conservation condition
and symmetry of the correlation function under the permutation of
superspace points.

With the aid of
the identity (\ref{useful-prop-a}), the
supercurrent conservation law (\ref{N3-flavour-conserv}) leads to
the following constraint on $H_{(f,d)}^{IJK}$
\be
{\cal D}_\alpha^{(I}H_{(f,d)}^{J)KL}  -\frac13 \delta^{IJ}{\cal
D}_\alpha^{M}H_{(f,d)}^{MKL} =0~.
\label{flavour-conserv}
\ee
Computing the derivatives of the tensors (\ref{H_n-flavour}) and
(\ref{calH}) and substituting them in the equation (\ref{flavour-conserv}) we find the
following constraints on the coefficients $c_n$ and $d_n$:
\begin{subequations}
\bea
&&
\mbox{all }
 b_i=0~;\label{allbi=0}\\
&&c_2 = 2 c_1~, \quad
c_3 =  c_1~,\quad
c_5=-2c_6~,\quad
c_4=-4c_6~;  \label{5.106b}\\
&&d_1=d_2=d_3=d_4=0~,\quad
2d_5+d_6=0~.
\label{5.106}
\eea
\end{subequations}
In deriving these equations we have used the following $\cN=3$ identities:
\begin{subequations}
\bea
\Theta^{I\alpha}\Theta^{J\beta}\Theta^{K\gamma}
\Theta^{L\delta} \varepsilon_{JKL}
&=&-\frac12\varepsilon^{\alpha\beta}\Theta^2\Theta^{J\gamma}\Theta^{K\delta}
 \varepsilon_{IJK}
-\frac12\varepsilon^{\alpha\gamma}\Theta^2\Theta^{J\beta}\Theta^{K\delta}
 \varepsilon_{IJK}\\&&
-\frac12\varepsilon^{\alpha\delta}\Theta^2\Theta^{J\beta}\Theta^{K\gamma}
 \varepsilon_{IJK}~,\non\\
\Theta^{I\alpha} \Theta^J_\alpha\Theta^{K\beta}\Theta^L_\gamma
\varepsilon_{JKL} &=& 2\Theta^2 \Theta^{J\beta} \Theta^K_\gamma
\varepsilon_{IJK}~,
\label{227}
\eea
\end{subequations}
which
are differential consequences of the more general
$\cN=3$ identity
\be
\Theta^2 \varepsilon_{IJK}\Theta^I_\alpha \Theta^J_\beta
\Theta^K_\gamma =0~.
\label{N3id}
\ee

As is seen from (\ref{allbi=0}), the part of the correlation
function with the symmetric tensor $d^{\bar a\bar b\bar c}$
vanishes since $H_{(d)}^{IJK}=0$. In the rest of this section we
concentrate on the derivation of the part of the flavour current
correlator with the antisymmetric tensor $f^{\bar a\bar b\bar c}$.

The three-point correlation function has to possess the symmetry property
\be
\langle L^{I\bar a}(z_1) L^{J\bar b}(z_2) L^{K\bar c}(z_3) \rangle
=\langle L^{K\bar c}(z_3) L^{J\bar b}(z_2) L^{I\bar a}(z_1) \rangle~.
\ee
It imposes the following constraint on the tensor $H_{(f)}^{IJK}$
\be
H_{(f)}^{IJK}(-{\bm X}_{1}^{\rm T},-\Theta_{1})=
- {\bm x}_{13}{}^2{\bm X}_{3}{}^2 u_{13}^{II'} u_{13}^{JL}U_{3}^{LJ'}u_{13}^{KK'}
H_{(f)}^{K'J'I'}({\bm X}_{3},\Theta_{3})~,
\label{5.108}
\ee
as a consequence of (\ref{U-relations}).
This equation gives additional relations among coefficients $c_i$
and $d_i$, which are:
\bea
&&
c_2= 2 c_1- c_5+\frac12c_4~,\quad
c_3= c_1-\frac{3}{2}c_5+\frac{3}{4} c_4~,\quad
c_6=-\frac5{12}c_1~;\non\\
&&
d_2=d_6=0~,\quad
d_5 = d_4+2 d_1~.
\eea
Comparing these equations with \eqref{5.106b} and
(\ref{5.106}) we see that all
coefficients $d_i$ vanish while all $c_i$ can be expressed in
terms of $c_1 \equiv  b_{\cN=3}$,
\bea
c_2 = c_3  = b_{\cN=3}~, \qquad
c_4  = 3 c_5 = -4 c_6 = \frac 53 b_{\cN=3}~, \qquad
d_i=0~.
\eea
Taking into account these relations, we can
rewrite the resulting expression for the tensor
$H_{(f)}^{IJK}$ (\ref{Hflavour-N3}) in terms of the matrix (\ref{U}).
Our final result for the three-point function is
\begin{subequations} \label{8.28}
\bea
\langle L^{I\bar a}(z_1) L^{J\bar b}(z_2) L^{K\bar c}(z_3) \rangle
=
f^{\bar a\bar b\bar c}
\frac{u_{13}^{II'}u_{23}^{JJ'}}{{\bm x}_{13}{}^2{\bm x}_{23}{}^2}
H_{(f)}^{I'J'K}({\bm X}_{3},\Theta_{3})
\eea
where
\bea
H_{(f)}^{IJK}&=&\frac{b_{\cN=3}}{\bm X}\Big[
\varepsilon^{IJK}
-U^{LJ}\varepsilon^{LIK}
+U^{IL}\varepsilon^{LJK}
\non\\&&
-\frac1{16}(\delta^{IJ}\varepsilon^{KMN}U^{MN}
+\varepsilon^{IMN}U^{MN}U^{KJ}
+\varepsilon^{JMN}U^{MN}U^{IK})
\non\\&&
+\frac5{16}(
U^{IJ}\varepsilon^{KMN}U^{MN}
+\delta^{IK}\varepsilon^{JMN}U^{MN}
+\delta^{JK}\varepsilon^{IMN}U^{MN})
\Big]~.
\eea
\end{subequations}
Here we have used the following relation between the matrix $U^{IJ}$
given by (\ref{U-explicit}) and the composites in (\ref{AB})
\be
U^{IJ}=\delta^{IJ}-2\frac{A^{IJ}}{{\bm X}^2}+\frac{B^{IJ}\Theta^2}{{\bm X}^2}~.
\ee

The three-point function \eqref{8.28} looks more complicated than its 4D $\cN=2$
counterpart \cite{KT}. The reason for that is the isovector notation used for the $R$-symmetry indices. Switching to the isospinor notation, following the prescription \eqref{N=3conversion},
should simplify the structure of the correlation function.

\subsection{$\cN=3$ supercurrent}

The $\cN=3$ supercurrent is described by a primary real spinor  $J_\alpha$
of dimension 3/2, which is characterised by the superconformal
transformation law
\be
\delta J_\alpha  = -\xi J_\a
-\frac32 \sigma(z) J_\alpha
+\lambda_\alpha{}^\beta(z) J_\beta
\ee
and obeys
the conservation equation
\be
D^{I\alpha} J_\alpha =0~.
\label{N3-conserv-law}
\ee

According to (\ref{OO}), the two-point correlation function for the supercurrent
reads
\be
\langle
J_\alpha(z_1) J_\beta(z_2)
 \rangle = \ri c_{\cN=3} \frac{{\bm x}_{12\alpha\beta}}{({\bm
 x}_{12}{}^2)^2}~,
\ee
with $c_{\cN=3}$ a parameter.
It is antisymmetric,
$\langle J_\alpha(z_1) J_\beta(z_2) \rangle
=- \langle J_\beta(z_2) J_\alpha(z_1)  \rangle  $,
and respects
the conservation equation (\ref{N3-conserv-law}),
\be
 D_{(1)}^{I\alpha} \langle
J_\alpha(z_1) J_\beta(z_2)
 \rangle= 0~, \qquad
 z_1\ne z_2~.
\ee

In accordance with (\ref{OOO}),  the three-point correlator
for  the supercurrent has the form
\be
\langle
J_\alpha(z_1)J_\beta(z_2) J_\gamma(z_3) \rangle
=\frac{{\bm x}_{13\alpha\alpha'}{\bm x}_{23\beta\beta'}}{({\bm x}_{13}{}^2)^2({\bm x}_{23}{}^2)^2}
H^{\alpha'\beta'}{}_\gamma({\bm X}_{3},\Theta_{3})~,
\ee
where $H$ should have the following scaling property
\be
H^{\alpha\beta\gamma}(\lambda^2 {\bm X},\lambda\Theta)
=\lambda^{-3} H^{\alpha\beta\gamma}({\bm X},\Theta)~.
\label{N3-scaling}
\ee
Due to
\be
\langle
J_\beta(z_2)J_\alpha(z_1) J_\gamma(z_3) \rangle
=-\langle
J_\alpha(z_1)J_\beta(z_2) J_\gamma(z_3) \rangle~,
\ee
the tensor $H$ should obey the following symmetry property
\be
H^{\beta\alpha\gamma}(-{\bm X}^{\rm T},-\Theta)
=-H^{\alpha\beta\gamma}({\bm X},\Theta)~.
\label{sym}
\ee
The most general form for $H$ compatible with the
relations (\ref{N3-scaling}) and (\ref{sym}) is
\be
H^{\alpha\beta\gamma}({\bm X},\Theta)= \sum c_i
H_{i}^{\alpha\beta\gamma}({\bm X},\Theta)~,
\ee
where $c_{i}$ are some coefficients and
\bea
H_1^{\alpha\beta}{}_\gamma({\bm X},\Theta)&=&
\frac{{\bm X}^{\alpha\alpha'}{\bm X}^{\beta'\beta}\Theta^I_{\alpha'}\Theta^J_{\beta'}\Theta^K_{\gamma}
 \varepsilon_{IJK}}{{\bm X}^5}~,\non\\
H_{2}^{\alpha\beta}{}_\gamma({\bm X},\Theta)&=&
\frac{{\bm X}^{\beta\alpha}{\bm X}^{\mu\nu}\Theta^I_{\mu}\Theta^J_{\nu}
 \Theta^K_{\gamma}\varepsilon_{IJK}}{{\bm X}^5}~,\non\\
H_{3}^{\alpha\beta}{}_\gamma({\bm X},\Theta)&=&
\frac{(\delta^\beta_\gamma {\bm X}^{\alpha\rho}
 +\delta^\alpha_\gamma {\bm X}^{\rho\beta}){\bm X}^{\mu\nu}\Theta^I_\mu \Theta^J_\nu\Theta^K_\rho
  \varepsilon_{IJK}}{{\bm X}^5} ~,\non\\
H_{4}^{\alpha\beta}{}_\gamma({\bm X},\Theta)&=&
\frac{{\bm X}^{\beta'\beta}\Theta^{I\alpha}\Theta^J_{\beta'}\Theta^{K}_{\gamma}}{{\bm X}^4}
 \varepsilon_{IJK}
-\frac{{\bm X}^{\alpha\alpha'}\Theta^I_{\alpha'}\Theta^{J\beta}\Theta^{K}_{\gamma}}{{\bm X}^4}
 \varepsilon_{IJK}~.
 \label{tensors}
\eea
Here we have listed all linearly independent structures.
Note that owing to the identity (\ref{N3id})
there are no terms of order $O(\Theta^5)$.

Now we have to impose the constraint
\be
{\cal D}^I_\alpha H^{\alpha\beta\gamma} = 0~,
\label{156}
\ee
which follows from the conservation law
(\ref{N3-conserv-law}). In deriving (\ref{156}), the identity
(\ref{useful-prop-a}) has been used. At order $O(\Theta^2)$ the
equation (\ref{156}) gives
\be
 -c_1 +2 c_2 =0~,\quad
c_2-c_3=0~, \quad c_4 =0~,
\label{5.128}
\ee
while collecting the terms of order $O(\Theta^4)$ we find
\be
-c_1+2c_2=0~.
\label{5.129}
\ee
In the derivation of these equations we have used the identities
(\ref{227}).
The general solution of \eqref{5.128}, \eqref{5.129} reads
\be
c_4 =0~,\quad
c_2 = c_3 = d_{\cN=3}~,\quad
c_1 = 2d_{\cN=3}~,
\ee
where $d_{\cN=3}$ is a single free coefficient.

As a result, the tensor $H^{\alpha\beta\gamma}$ has the following
explicit form
\bea
H^{\alpha\beta}{}_\gamma({\bm X},\Theta)&=&\frac{ d_{\cN=3}}{{\bm X}^5}
\Big[
(\delta^\beta_\gamma {\bm X}^{\alpha\rho}
 +\delta^\alpha_\gamma {\bm X}^{\rho\beta}){\bm X}^{\mu\nu}\Theta^I_\mu \Theta^J_\nu\Theta^K_\rho
  \varepsilon_{IJK}\non\\&&
 +{\bm X}^{\beta\alpha}{\bm X}^{\mu\nu}\Theta^I_{\mu}\Theta^J_{\nu}
 \Theta^K_{\gamma}\varepsilon_{IJK}
+2{\bm X}^{\alpha\mu}{\bm X}^{\nu\beta}\Theta^I_{\mu}\Theta^J_{\nu}\Theta^K_{\gamma}
 \varepsilon_{IJK}\Big]~.
\eea
One can also check that this expression obeys the equation
\be
H^{\mu\nu}{}_\alpha(-{\bm X}_{1}^{\rm T},-\Theta_{1})
={\bm X}_{3}{}^2{\bm x}_{13}^{\mu\mu'}{\bm x}_{13\alpha\alpha'}
{\bm x}_{13}^{\nu\nu'}{\bm X}_{3\nu'\rho}
H^{\alpha'\rho}{}_{\mu'}({\bm X}_{3},\Theta_{3})~,
\ee
which is a consequence of the symmetry property
\be
\langle
J_\gamma(z_3)J_\beta(z_2) J_\alpha(z_1) \rangle
=-\langle
J_\alpha(z_1)J_\beta(z_2) J_\gamma(z_3) \rangle~.
\ee


\section{Concluding remarks}

In this paper, we have demonstrated  that
for three-dimensional  $\cN$-extended superconformal field theories  with
$1\leq \cN \leq 3$ each of the two-point
and three-point functions for the  supercurrent
is fixed by the symmetries and by the conservation law
up to a single overall coefficient.
In particular, our results  imply that each of the two- and three-point functions for
the stress-energy tensor in 3D superconformal theories are fixed up to one coefficient.
It is natural to expect that the coefficients
in the two- and three-point functions for the supercurrent
are related to each other through a Ward identity
just like in 4D (super)conformal theories~\cite{OP, Osborn}.
Although the required Ward identities may be derived
using the known prepotential formulations for 3D $\cN=1$ supergravity
 \cite{GGRS} and 3D $\cN=2$ supergravity \cite{ZP,Kuzenko12},
we postpone the study of such a relation for future work.
The fact that such correlation functions are constrained up to an overall coefficient makes 3D  superconformal theories similar to the well-studied case of 2D conformal field theory.

We have also proved that the  three-point function of the
flavour current multiplets is determined
by a single functional form in the $\cN=1$ and $\cN=3$ cases.
The specific feature of the $\cN=2$ case is that two independent structures are allowed
for the three-point function of the flavour current multiplets, but only one of them
contributes to the three-point function of the conserved  currents contained
in these multiplets. This is explicitly demonstrated in Appendix C.

As was shown in~\cite{Giombi:2011rz, Giombi:2011kc}, 3D non-supersymmetric conformal theories can have certain {\it odd parity}
contributions to three-point functions of the stress-energy tensors and flavour currents. They do not exist for an arbitrary number of space-time
dimensions but are a specific 3D (and, perhaps, in general, an odd-dimensional)  feature.
Such terms do not arise in free conformal field theories
but can appear in interacting Chern-Simons theories coupled to parity violating matter. Some general constructions of the parity
violating terms in ${\cal N}=1$ superconformal theories were later discussed in~\cite{ Nizami:2013tpa}. There it was shown
that in some examples correlators of conserved currents can also contain parity odd contributions.
In our approach we did not distinguish whether
various allowed structures are even or odd under parity.
For ${\cal N}=1, 2, 3$ we always assumed the most general ansatz.
Hence,  our analysis demonstrates that odd parity contributions do not appear
in the supersymmetric cases for both the flavour current and supercurrent correlators.\footnote{It is not difficult to show that our results are
in complete agreement with~\cite{ Nizami:2013tpa}.
For example, the correlator of three ${\cal N}=1$ supercurrents corresponds
to $\langle J_{3/2}  J_{3/2}  J_{3/2} \rangle$ in~\cite{ Nizami:2013tpa}.
This correlator admits only one parity odd structure respecting
the proper symmetry under the permutation of the three points but
this structure is inconsistent with the conservation law.}
This is explicitly proved in Appendix D for the $\cN=1$ flavour current multiplets.

It is an interesting problem to
generalise the present results to the superconformal theories with ${\cal N} \geq 4$ supersymmetry. This is also postponed for future work.

We hope that the techniques developed in our paper will be useful
in the context of generalised higher spin superconformal theories formulated
on hyper-superspaces, see \cite{Florakis:2014aaa} and references therein.

Recently, the so-called superembedding formalism in four dimensions
\cite{Goldberger:2011yp,Maio,Goldberger:2012xb},
which was originally introduced by Siegel \cite{Siegel9395}
and fully elaborated in \cite{K-compactified12} under the name bi-supertwistor formalism,
has been applied to compute
correlation functions of multiplets containing conserved currents in 4D $\cN=1$
superconformal theories \cite{Goldberger:2012xb,Poland1,Poland2}.\footnote{So far,
the thee-point function for the $\cN=1$ supercurrent
originally computed by Osborn \cite{Osborn}
has not been re-derived within the superembedding approach.}
The 3D $\cN$-extended bi-supertwistor formalism was presented in \cite{K-compactified12}.
It would be of interest to apply this approach for an alternative computation
of the correlation functions of the supercurrent and flavour current multiplets
derived in our paper. The results given in section 3 might be useful for that.
\\


\noindent
{\bf Acknowledgements:}\\
We are grateful to Joseph Novak for comments on the manuscript.
This work is supported in part by the ARC DP project
 DP140103925. The work of E.I.B. was also supported by the ARC Future Fellowship FT120100466.


\appendix

\section{3D notation and conventions}
We mostly follow the notation and conventions adopted in
\cite{KPT-MvU}. In particular, the Minkowski metric is
$\eta_{mn}=\mbox{diag}(-1,1,1)$.

The spinor indices are  raised and lowered using
the $\sSL(2,{\mathbb R})$ invariant tensors
\bea
\ve_{\a\b}=\left(\begin{array}{cc}0~&-1\\1~&0\end{array}\right)~,\qquad
\ve^{\a\b}=\left(\begin{array}{cc}0~&1\\-1~&0\end{array}\right)~,\qquad
\ve^{\a\g}\ve_{\g\b}=\d^\a_\b
\eea
by the standard rule:
\bea
\psi^{\a}=\ve^{\a\b}\psi_\b~, \qquad \psi_{\a}=\ve_{\a\b}\psi^\b~.
\label{A2}
\eea

We employ real gamma-matrices,  $\g_m := \big( (\g_m)_\a{}^\b \big)$, which are
expressed in terms of the Pauli matrices as
$\gamma_0=-\ri\sigma_2$, $\gamma_1=\sigma_3$,
$\gamma_2=-\sigma_1$. They obey the algebra
\be
\gamma_m \gamma_n=\eta_{mn}{\mathbbm 1} + \varepsilon_{mnp}
\gamma^p~,
\label{A3}
\ee
where the Levi-Civita tensor is normalised as
$\varepsilon^{012}=-\varepsilon_{012}=1$. The completeness
relation for the gamma-matrices reads
\be
(\gamma^m)_{\alpha\beta}(\gamma_m)^{\rho\sigma}=-(\delta_\alpha^\rho\delta_\beta^\sigma
+\delta_\alpha^\sigma\delta_\beta^\rho)~.
\label{A4}
\ee
Here $(\gamma_m)^{\alpha\beta}$ and $(\gamma_m)_{\alpha\beta}$
are obtained from $\g_m=(\g_m)_\a{}^{\b}$ by the rules (\ref{A2}).

Given a three-vector $x_m$,
it  can be equivalently described by a symmetric second-rank spinor $x_{\a\b}$
defined as
\bea
x_{\a\b}:=(\g^m)_{\a\b}x_m=x_{\b\a}~,\qquad
x_m=-\hf(\g_m)^{\a\b}x_{\a\b}~.
\eea
The same convention is adopted for the spacetime derivatives,
\be
\partial_{\alpha\beta}=(\gamma^m)_{\alpha\beta}\partial_m~,
\qquad
\partial_m=-\frac12 (\gamma_m)^{\alpha\beta}\partial_{\alpha\beta}
\ee
such that
\be
\partial_m x^n=\delta_m^n~,\qquad
\partial_{\alpha\beta}x^{\gamma\delta}
=-(\delta_\alpha^\gamma\delta_\beta^\delta+\delta_\alpha^\delta\delta_\beta^\gamma)~.
\ee
Note also that the square of a vector in terms of spinor indices
reads
\be
x^2 = x^m x_m = -\frac12x^{\alpha\beta}x_{\alpha\beta}~.
\ee


\section{$\cN=2$ correlation functions in chiral basis}
\label{AppendixB}

The case $\cN=2$ is special since
the $R$-symmetry
group $\sSO(2)$ is isomorphic to $\sU(1)$, and one can define a chiral subspace of
the full superspace on which the superconformal group $\sOSp(2|4;\mathbb{R})$
acts by holomorphic transformations.
This appendix is devoted to a brief discussion of
the correlation functions involving (anti)chiral
superfields.

\subsection{(Anti)chiral two-point functions}

Instead of the real Grassmann coordinates
$\theta^{I\alpha}=(\theta^{1\alpha},\theta^{2\alpha})$, we
introduce new complex variables,
\be
\theta^\alpha=\frac1{\sqrt2}(\theta^{1\alpha}+\ri\theta^{2\alpha})~,\qquad
\bar\theta^\alpha =
\frac1{\sqrt2}(\theta^{1\alpha}-\ri\theta^{2\alpha})~,
\label{4.24}
\ee
which have definite $\sU(1)$ charges with respect to the $R$-symmetry group.
The corresponding spinor covariant derivatives
\bea
D_\alpha = \frac1{\sqrt2}(D^1_\alpha - \ri D^2_\alpha)~,\qquad
\bar D_\alpha  =-\frac1{\sqrt2}(D^1_\alpha + \ri
D^2_\alpha)
\eea
obey the anti-commutation relations
\bea
 \{D_\alpha ,  D_\beta \} = 0 ~, \quad
  \{\bar D_\alpha , \bar D_\beta \} =  0 ~, \quad
  \{D_\alpha , \bar D_\beta \} =
-2\ri\partial_{\alpha\beta}~,
\eea
which guarantee the existence of a chiral subspace of the full superspace.
The crucial features of the chiral subspace are that (i) it is invariant under the
$\cN=2$ super-Poincar\'e group; and (ii) its bosonic  $y^a$ and fermionic  $\q^\a$
coordinates are annihilated by the operators $\bar D_\g$.
Its bosonic coordinate is
\bea
y^{\alpha\beta} = x^{\alpha\beta} + 2\ri \bar\theta^{(\alpha}
\theta^{\beta)}~.
\eea

The superconformal transformation law of the real superspace coordinates,
eq. (\ref{3.10}), implies that the superconformal group acts by holomorphic
transformations on the chiral subspace.
The  superconformal variations
 of the chiral coordinates
$y^{\alpha\beta}$ and $\theta^\alpha$ are
\begin{subequations}
\bea
\delta y^{\alpha\beta}&=&
a^{\alpha\beta} +4\ri \bar\epsilon^{(\alpha}\theta^{\beta)}
 -\lambda^\alpha{}_\gamma y^{\gamma\beta}
 -y^{\alpha\gamma}\lambda_\gamma{}^\beta
 +\sigma y^{\alpha\beta}
 + y^{\alpha\gamma}y^{\beta\delta}b_{\gamma\delta}
 \non\\&&
 +\ri y^{\alpha\gamma} \theta_\gamma \bar\eta^\beta
 +\ri y^{\alpha\gamma}\theta^\beta\bar\eta_\gamma
 +2\ri y^{\beta\gamma}\theta^\alpha\bar\eta_\gamma
 +\ri y^{\alpha\beta}\theta^\gamma \bar\eta_\gamma~,\\
\delta\theta^\alpha&=&\epsilon^\alpha
- \theta^\beta \lambda_\beta{}^\alpha
+\frac12\sigma \theta^\alpha
 -\ri \Lambda\theta^\alpha
 +y^{\alpha\gamma}b_{\gamma\beta}\theta^\beta
 -  y^{\alpha\beta}\eta_\beta
 +\ri\theta^2 \bar\eta^\alpha ~.
\eea
\end{subequations}
The parameter $\Lambda$ of
$\sU(1)$ $R$-symmetry is related to the $\sSO(2)$ parameters
$\Lambda_{IJ}$ as $\Lambda_{IJ}=\varepsilon_{IJ} \Lambda$, where
$\varepsilon_{IJ}$ is the antisymmetric tensor.

Let us consider the following $Q$-supersymmetric two-point function
\be
y_{1 2}^{\alpha\beta} = (x_1-x_2)^{\alpha\beta}
-2\ri \theta_1^{(\alpha}\bar\theta^{\beta)}_{12}
+2\ri \theta_{12}^{(\alpha} \bar\theta_2^{\beta)}~,
\ee
which is chiral in its first superspace argument and antichiral in
the other,
\be
\bar D_{(1)\gamma} y_{12}^{\alpha\beta} = D_{(2)\gamma} y_{12}^{\alpha\beta} =0~.
\ee
It is related to the two-point function (\ref{super-interv-X})
as
\be
y_{12}^{\alpha\beta} ={\bm x}_{12}^{\alpha\beta}+2\ri\bar\theta^\alpha_{12}\theta^\beta_{12}~.
\ee
For its square we have
\be
y_{12}{}^2 = {\bm x}_{12}{}^2+\ri\theta_{12}^{I\alpha}({\bm x}_{12})_{\alpha\beta}\theta_{12}^{I\beta}
  +\varepsilon_{IJ}\theta_{12}^{I\alpha}({\bm x}_{12})_{\alpha\beta}
   \theta_{12}^{J\beta}~.
\label{yy}
\ee
Using this formula, we get the conjugation
rule for $y_{12}{}^2$:
\be
\overline{y_{12}{}^2} \equiv \bar y_{12}{}^2= y_{21}{}^2~.
\ee
One also finds
the following useful identity for the product of $y_{12}{}^2$ and $\bar y_{12}{}^2$
\be
y_{12}{}^2 \bar y_{12}{}^2 = ({\bm x}_{12}{}^2)^2~.
\ee

Now, consider the orthogonal matrix $u^{IJ}_{12}$ given by
(\ref{two-point-u}) and transforming by the rule (\ref{deltaU}).
In this transformation, the matrix $\Lambda_{IJ}(z)$ is antisymmetric,
and thus has one independent component which we denote by
$\Lambda(z)$,
\be
\Lambda_{IJ}(z)=\varepsilon_{IJ} \Lambda(z)~.
\label{Lambda}
\ee
Given the matrix $u_{12}^{IJ}$, we construct a complex scalar
two-point function $v_{12}$
\be
v_{12}=\frac12( u_{12}^{II}+\ri \varepsilon_{IJ}u_{12}^{IJ})~,
\label{v}
\ee
which transforms as
\be
\widetilde\delta v_{12} = \ri (\Lambda(z_1)-\Lambda(z_2)) v_{12}~.
\label{delta-v}
\ee

Using the explicit expressions for the two-point functions
(\ref{super-interv-X}) and (\ref{two-point-u}) it can be shown
that (\ref{yy}) and (\ref{v}) are related to each other as
\be
y_{12}{}^2 = {\bm x}_{12}{}^2 v_{12}~.
\label{y2}
\ee
Aapplying (\ref{delta-x2}) and (\ref{delta-v}), we find the
transformation of (\ref{yy}) to be
\be
\widetilde\delta y_{12}{}^2 = ({\bm \sigma}(z_1)+{\bm \sigma}(z_2)) y_{12}{}^2~,
\label{delta-y2}
\ee
where the chiral superfield $\bm \sigma$ includes parameters of
local scale and $\sU(1)$ transformations \cite{KPT-MvU}
\be
{\bm \sigma}(z) = \sigma(z)+\ri\Lambda(z)~,\qquad
\bar D_\alpha {\bm \sigma}(z)=0~.
\label{sigma-bold}
\ee
Thus  (\ref{y2}) is a natural (anti)chiral
generalisation of ${\bm x}_{12}{}^2$ which can serve as a
building block for correlation functions involving (anti)chiral
superfields.

\subsection{$\cN=2$ correlation functions with (anti)chiral superfields}

In this section, we consider some simple correlation functions
which involve chiral and antichiral primary superfields. First, we
will consider a two-point correlator with chiral and antichiral
superfields,  and then we will derive the general expression for
the three-point correlation
function with chiral, antichiral and linear superfields.

Let $\Phi$ be a chiral superfield of dimension $q$ with no spinor
indices. Its superconformal transformation reads
\be
\delta \Phi(z) = -\xi \Phi(z) -q{\bm \sigma}(z)\Phi(z)~,
\ee
where $\bm \sigma$ is given by (\ref{sigma-bold}). This transformation
preserves chirality since $\bm \sigma$ is chiral. Using
the two-point function (\ref{yy}) it is straightforward to construct
the two-point correlator of the chiral superfield $\Phi$ and
its conjugate $\bar\Phi$,
\be
\langle \Phi(z_1)\bar\Phi(z_2)\rangle = \frac{
c}{(y_{12}{}^2)^q}~,
\ee
where $c$ is an arbitrary coefficient.
Owing to (\ref{4.35}) this expression automatically possesses
correct chirality properties with respect to both arguments and has the right
transformation rule because of (\ref{delta-y2}).

As an example of a three-point function, we consider the
correlator of a linear superfield $G^{\bar a}$, a chiral superfield
$\Phi^{\bar a}$ and an antichiral one $\bar \Phi^{\bar a}$. Here the
index $\bar a$ can be considered as a flavour group index.
In this case these superfields can be identified with $\cN=2$
superfield components of the $\cN=3$ flavour current
studied in sect.~\ref{sect-N3-flavour}. Assuming
that all these three superfields have dimension one, we look for
the correlation function with the use of the standard ansatz
\be
\langle G^{\bar a}(z_1) \Phi^{\bar b}(z_2) \bar\Phi^{\bar c}(z_3)\rangle=
\frac1{{\bm x}_{13}{}^2 y_{23}{}^2}
[ f^{\bar a\bar b\bar c}H_{(f)}({\bm X}_3,\Theta_3)
+d^{\bar a\bar b\bar c}H_{(d)}({\bm X}_3,\Theta_3)]~,
\label{corr-chiral}
\ee
where the functions $H_{(f,d)}$ should have the following homogeneity property
\be
H_{(f,d)}(\lambda^2 {\bm X},\lambda\Theta) = \lambda^{-1} H_{(f,d)}({\bm X},\Theta)~.
\ee

Using the identity (\ref{useful-prop-a}), the linearity
of the superfield $G^{\bar a}$, $D^2 G^{\bar a} = \bar D^2 G^{\bar
a}=0$, turns into the following equations for the functions $H_{(f,d)}$
\be
{\cal D}^2 H_{(f,d)} = \bar {\cal D}^2 H_{(f,d)}=0~,
\label{linearityN2}
\ee
where
\be
{\cal D}_\alpha =
\frac\partial{\partial\Theta^\alpha}+\ri\bar\Theta^\beta\frac\partial{\partial X^{\alpha\beta}}
~,\qquad
\bar {\cal D}_\alpha =
-\frac\partial{\partial\bar\Theta^\alpha}-\ri\Theta^\beta\frac\partial{\partial
X^{\alpha\beta}}~.
\label{5.98}
\ee
The objects $\Theta_\alpha$ and $\bar \Theta_\alpha$ are
expressed in terms of $\Theta^I_\alpha$ by the rule (\ref{4.24}).
One can check that the equations (\ref{5.98}) being rewritten in terms of the derivatives
${\cal D}^I_\alpha$ are equivalent to (\ref{182}). Therefore
the solution of (\ref{linearityN2}) is
\be
H_{(f,d)}({\bm X},\Theta) =\frac{\ri c_{(f,d)1}}{X}
+ c_{(f,d)2}\frac{\ri\varepsilon_{IJ}\Theta^{I}_\alpha X^{\alpha\beta} \Theta^{J}_\beta  }{X^3}~,
\ee
where $c_{(f,d)1}$ and $c_{(f,d)2}$ are some complex coefficients.

Obviously, the correlation function (\ref{corr-chiral}) obeys the
reality condition
\be
\langle G^{\bar a}(z_1) \Phi^{\bar b}(z_2) \bar\Phi^{\bar c}(z_3)\rangle^*
=\langle G^{\bar a}(z_1) \Phi^{\bar c}(z_3) \bar \Phi^{\bar b}(z_2)\rangle~.
\ee
The latter leads to the constraints for the functions $H_{(f,d)}$:
\be
H_{(f)}(-{\bm X}_2^{\rm T},-\Theta_2)=-\frac{{\bm x}_{12}{}^2}{{\bm x}_{13}{}^2}\bar
H_{(f)}({\bm X}_3,\Theta_3)~,\quad
H_{(d)}(-{\bm X}_2^{\rm T},-\Theta_2)=\frac{{\bm x}_{12}{}^2}{{\bm x}_{13}{}^2}\bar
H_{(d)}({\bm X}_3,\Theta_3)~.
\ee
This equation implies the following reality properties of the constants
$c_{(f,d)1}$ and $c_{(f,d)2}$ in (\ref{5.60})
\be
\bar c_{(f)1} = c_{(f)1}~,\quad \bar c_{(f)2} = c_{(f)2}~,\quad
\bar c_{(d)1} =- c_{(d)1}~,\quad \bar c_{(d)2} = -c_{(d)2}~.
\ee

Finally, we have to take into account the chirality of the
correlation function with respect to the second argument
\be
\bar D_{(2)\alpha} H_{(f,d)}({\bm X}_3,\Theta_3) = 0~.
\ee
With the use of the identity (\ref{useful-prop-b}) the latter
equation gives the following constraint to the functions $H_{(f,d)}$
\be
\bar{\cal Q}_\alpha H_{(f,d)}({\bm X},\Theta)=0~,
\qquad \bar {\cal Q}_\alpha = -\ri\frac\partial{\partial\Theta^\alpha}-\gamma^m_{\alpha\beta}\Theta^\beta\frac\partial{\partial
X^m}~.
\ee
This equation is satisfied if the coefficients $c_{(f,d)1}$ and $c_{(f,d)2}$ in
(\ref{5.60}) are related to each other as
\be
c_{(f,d)1}= 2c_{(f,d)2}\equiv 2 c_{(f,d)}~.
\ee
As a result, each of the functions $H_{(f)}$ and $H_{(d)}$
has one free coefficient
\be
H_{(f,d)}=\ri c_{(f,d)}\left(\frac2X+\frac{\varepsilon_{IJ}\Theta^{I}_\alpha
X^{\alpha\beta}\Theta^J_\beta}{X^3} \right)
=\ri c_{(f,d)} \left(
\frac2{\bm X}+\frac{\Theta^4}{4{\bm X}^3}
+\frac{\varepsilon_{IJ}\Theta^{I}_\alpha
{\bm X}^{\alpha\beta}\Theta^J_\beta}{{\bm X}^3}
\right)~.
\label{B29}
\ee
Here we used the relation (\ref{XX}) to represent the function $H$
in terms of covariant objects (\ref{three-points}). Note that
in (\ref{B29}) the coefficient $c_{(f)}$ is real while $c_{(d)}$
is imaginary.

\section{$\cN=2\to \cN=1$ superspace reduction}

This appendix is devoted to the $\cN=2\to\cN=1$ superspace reduction of
the three-point functions for the $\cN=2$ supercurrent and flavour current multiplets.

\subsection{The supercurrent correlation function}
\label{appC}

As discussed in section 1, every $\cN=2$ superconformal field theory is
a special $\cN=1$ superconformal field theory.
The $\cN=1$ supercurrent $J_{\alpha\beta\gamma}$ for this theory is related to its
$\cN=2$ supercurrent $J_{\alpha\beta}$ by the first equation in
(\ref{1.7b}). As a consequence, the $\cN=1$ supercurrent
correlation function (\ref{n1.1.5}) appears as a result of the
$\cN=2\to\cN=1$ reduction from (\ref{n2.5})
\be
\langle J_{\alpha\alpha'\alpha''}(z_1)
J_{\beta\beta'\beta''}(z_2) J_{\gamma\gamma'\gamma''}(z_3) \rangle
=-\ri D^{\bf 2}_{(1)\alpha} D^{\bf 2}_{(2)\beta} D^{\bf 2}_{(3)\gamma }
\langle J_{\alpha'\alpha''}(z_1)
J_{\beta'\beta''}(z_2) J_{\gamma'\gamma''}(z_3) \rangle |~.
\label{C1}
\ee
Recall that here the symbol $|$ means that we have to set
$\theta^{\bf 2}_\alpha$ to zero after computing the derivatives.
In this section we denote the values of the $\sSO(2)$ indices $I={\bf1},{\bf2}$
with boldface font to distinguish them from indices of other types.

According to (\ref{n1.1.5}), the $\cN=1$ supercurrent correlation function
is expressed in terms of the tensor
$H^{\alpha\alpha'\alpha'',\beta\beta'\beta'',\gamma\gamma'\gamma''}$
(or $H^{\alpha m,\beta n, \gamma k}$ if we trade the pairs of the spinor indices
into the vector ones by the rule (\ref{n1.1.10})) which was found in the
form (\ref{n1.1.29}) with the tensors $C$ and $D$ given by
(\ref{n1.1.26}), \eqref{partvalues} and (\ref{n1.1.21}), (\ref{n1.1.28}), respectively.
In its turn, the $\cN=2$ supercurrent correlation function is
represented by the tensor $H^{\alpha\alpha',\beta\beta',\gamma\gamma'}$
given explicitly by (\ref{7.41}). In this section we will show
that the former can be derived from the latter by means of the
equation (\ref{C1}).

To start with, we point out that the expression (\ref{7.41}) for
the tensor $H^{\alpha\alpha',\beta\beta',\gamma\gamma'}$ with
spinor indices converted into Lorentz ones can be rewritten as
\be
H^{mnk} = -  \ri d_{\cN=2}\, \gamma_p^{\mu\nu} \Theta^{\bf 1}_\mu \Theta^{\bf 2}_\nu\,
C^{mnp,k}~,
\label{C2}
\ee
where the tensor $C^{mnp,k}$ has
the form
\bea
C^{mnp,k}&=&\frac1{X^3} (\eta^{mn}\eta^{kp}+\eta^{mk}\eta^{np}+\eta^{nk}\eta^{mp})
+\frac3{X^5} (X^mX^k\eta^{np} + X^n X^k \eta^{mp}+ X^p X^k \eta^{mn})
\non\\&&
-\frac5{X^5} (X^m X^n \eta^{pk}+X^n X^p \eta^{mk} + X^m X^p \eta^{nk})
-\frac5{X^7} X^m X^n X^p X^k~.
\label{C3}
\eea
In the formula (\ref{C2}) the dependence on the Grassmann
variables is only through the factor $\gamma_p^{\mu\nu} \Theta^{\bf 1}_\mu
\Theta^{\bf 2}_\nu$ while the rest is described by the tensor (\ref{C3})
which is a function of $X$.
It is interesting to note that~\eqref{C3} coincides with the similar tensor in eqs.~\eqref{n1.1.26}, \eqref{partvalues}
which was encountered in sect.\  \ref{sect-N1-supercurrent}. As was already shown there,
this tensor is
symmetric and traceless over the first three indices
\be
C^{mnp,k} = C^{(mnp),k}~, \qquad
\eta_{mn}C^{mnp,k}  = 0~,
\label{C4}
\ee
and obeys the differential equation
\be
\partial_m C^{mnp,k} =0~,
\label{C5}
\ee
where $\partial_m = \frac\partial{\partial X^m}$. We also showed in sect.\  \ref{sect-N1-supercurrent}
that the equations (\ref{C4}) and (\ref{C5}) define the form of
the tensor (\ref{C3}) uniquely, up to an overall coefficient.

Now we substitute the expression (\ref{n2.5}) for the $\cN=2$
supercurrent correlation functions into (\ref{C1}) and represent
it in the following form
\bea
&&\langle J_{\alpha\alpha'\alpha''}(z_1) J_{\beta\beta'\beta''}(z_2)
 J_{\gamma\gamma'\gamma''}(z_3) \rangle \non\\
 &=&
 -\ri D^{\bf 2}_{(1)\alpha} D^{\bf2}_{(2)\beta} D^{\bf2}_{(3)\gamma}
 \frac{{\bm x}_{13\alpha'\rho'}{\bm x}_{13\alpha''\rho''} {\bm x}_{23\beta'\sigma'}{\bm x}_{23\beta''\sigma''}}{
  {\bm x}_{13}{}^6 {\bm x}_{23}{}^6}
  H^{\rho'\rho'',\sigma'\sigma''}{}_{\gamma'\gamma''}(X_3,\Theta_3)|
  \non\\
 &=& A+B\,,
 \label{C6}
\eea
where in these two terms $A$ and $B$ the derivatives are
distributed as follows
\bea
A&=&\ri \frac{{\bm x}_{13\alpha'\rho'}{\bm x}_{13\alpha''\rho''}}{{\bm x}_{13}{}^6}
  \left( D^{\bf2}_{(3)\gamma} D^{\bf2}_{(2)\beta}\frac{{\bm x}_{23\beta'\sigma'}{\bm x}_{23\beta''\sigma''}
  }{{\bm x}_{23}{}^6 }\right)
  D^{\bf 2}_{(1)\alpha}
  H^{\rho'\rho'',\sigma'\sigma''}{}_{\gamma'\gamma''}|
   \non\\&&
  -\ri \frac{{\bm x}_{23\beta'\sigma'}{\bm x}_{23\beta''\sigma''} }{{\bm x}_{23}{}^6}
  \left( D^{\bf 2}_{(3)\gamma} D^{\bf 2}_{(1)\alpha}
  \frac{{\bm x}_{13\alpha'\rho'}{\bm x}_{13\alpha''\rho''}}{{\bm x}_{13}{}^6} \right)
  D^{\bf 2}_{(2)\beta}H^{\rho'\rho'',\sigma'\sigma''}{}_{\gamma'\gamma''}
  |~,
  \label{C7}\\
B&=&  \ri
 \frac{{\bm x}_{13\alpha'\rho'}{\bm x}_{13\alpha''\rho''}{\bm x}_{23\beta'\sigma'}{\bm x}_{23\beta''\sigma''}}{
 {\bm x}_{13}{}^6  {\bm x}_{23}{}^6}
  D^{\bf 2}_{(3)\gamma} D^{\bf2}_{(2)\beta} D^{\bf2}_{(1)\alpha}
  H^{\rho'\rho'',\sigma'\sigma''}{}_{\gamma'\gamma''}|~.
  \label{C8}
\eea
It is easy to see that the terms with the covariant spinor
derivatives distributed in other ways vanish since the expressions
like $D^{\bf2}_{(1)\alpha} {\bm x}_{13\beta\beta'} = - 2\ri\theta^{\bf2}_{13\beta'}
\varepsilon_{\beta\alpha}$ die in the $|$-projection.
We will analyse the $A$ and $B$ sectors separately.

We begin by considering the $A$ term.
Using the explicit expression for ${\bm x}$, eq. (\ref{super-interv-X}), we find
\bea
D^{\bf2}_{(3)\gamma}D^{\bf2}_{(1)\alpha}
\frac{{\bm x}_{13\alpha'\rho'}{\bm x}_{13\alpha''\rho''}}{{\bm x}_{13}{}^6}|&=&
2\ri\frac{\varepsilon_{\rho'\gamma}\varepsilon_{\alpha'\alpha}{\bm x}_{13\alpha''\rho''}
 +\varepsilon_{\rho''\gamma} \varepsilon_{\alpha''\alpha}{\bm x}_{13\alpha'\rho'}}{{\bm x}_{13}{}^6}
+6\ri\frac{{\bm x}_{13\alpha\gamma}{\bm x}_{13\alpha'\rho'}{\bm x}_{13\alpha''\rho''}}{{\bm x}_{13}{}^8}
~,\non\\
D^{\bf2}_{(3)\gamma}D^{\bf2}_{(2)\beta}
\frac{{\bm x}_{23\beta'\sigma'}{\bm x}_{23\beta''\sigma''}}{{\bm x}_{23}^6}|&=&
2\ri\frac{\varepsilon_{\sigma'\gamma}\varepsilon_{\beta'\beta}{\bm x}_{13\beta''\sigma''}
 +\varepsilon_{\sigma''\gamma} \varepsilon_{\beta''\beta}{\bm x}_{23\beta'\sigma'}}{{\bm x}_{23}{}^6}
+6\ri\frac{{\bm x}_{23\beta\gamma}{\bm x}_{23\beta'\sigma'}{\bm x}_{13\beta''\sigma''}}{{\bm x}_{23}{}^8}~.
\non\\
\label{C9}
\eea
Next, with the use of (\ref{useful-prop}) we get
\bea
D^{\bf2}_{(1)\alpha} H^{mnk} |&=& ({\bm x}_{13}^{-1})_{\rho\alpha} {\cal D}^{{\bf2}\rho}
H^{mnk}|
 =-\frac{{\bm x}_{13\alpha\rho}}{{\bm x}_{13}{}^2}{\cal D}^{{\bf2}\rho} H^{mnk} |~,\non\\
D^{\bf2}_{(2)\beta} H^{mnk} | &=& \ri({\bm x}^{-1}_{23})_{\rho\beta}{\cal Q}^{{\bf2}\rho}H^{mnk} |
=\frac{{\bm x}_{23\beta\rho}}{{\bm x}_{23}{}^2} {\cal D}^{{\bf2}\rho} H^{mnk} |~.
\label{C10}
\eea
Now we substitute (\ref{C9}) and (\ref{C10}) into (\ref{C7}) and
apply simple identities like
\be
\varepsilon_{\sigma'\gamma}\varepsilon_{\beta'\beta}=\frac1{{\bm x}_{23}{}^2}
 ({\bm x}_{23\beta\sigma'}{\bm x}_{23\beta'\gamma}-{\bm x}_{23\beta\gamma}{\bm x}_{23\beta'\sigma'})
\ee
to represent the $A$ sector of the supercurrent correlation function
in the form
\be
A=\frac{{\bm x}_{13\alpha'\rho'}{\bm x}_{13\alpha''\rho''}{\bm x}_{13\alpha\rho}}{{\bm x}_{13}{}^8}
\frac{{\bm x}_{23\beta\sigma}{\bm x}_{23\beta'\sigma'}{\bm x}_{23\beta''\sigma''}}{{\bm x}_{23}{}^8}
H_{(A)}^{\rho\rho'\rho'',\sigma\sigma'\sigma''}{}_{\gamma\gamma'\gamma''}(X_3,\Theta_3)~,
\ee
where
\be
H_{(A)}^{\alpha\alpha'\alpha'',\beta\beta'\beta''}{}_{\gamma\gamma'\gamma''}
=6\delta_\gamma^{(\underline{\beta}} {\cal D}^{{\bf2}\alpha}
H^{\alpha'\alpha'',\underline{\beta}'\underline{\beta}'')}{}_{\gamma'\gamma''}
+6\delta_\gamma^{(\underline{\alpha}} {\cal D}^{{\bf2}\beta}
H^{\underline{\alpha}'\underline{\alpha}''),\beta'\beta''}{}_{\gamma'\gamma''} |~.
\label{212+}
\ee
Here the symmetrisation involves only the underlined indices.

We stress that the covariant spinor derivatives ${\cal D}^{\bf2}_\alpha$ in
the expression (\ref{212+}) act only on the Grassmann variable $\Theta^{\bf2}_\nu$ and
do not hit the $X$-dependent tensor $C^{mnp,k}$
since the action of the covariant spinor derivative on any
combination of $X^m$ is proportional to $\Theta^{\bf2}$ which dies in
the $|$-projection. Hence, after converting pairs of spinor
indices into vector ones and using identities with
three-dimensional gamma-matrices (\ref{A3},\ref{A4}),
the expression (\ref{212+}) can be rewritten in the form
\bea
H_{(A)}^{\alpha m,\beta n,\gamma k }&=&\ri d_{\cN=2}\big[-6 (\gamma_p)^{\alpha\beta}
\Theta^\gamma C^{mnp,k}
-(\gamma^p)^{\alpha\beta}(\gamma^r)^{\gamma\delta}\Theta_\delta
 \eta_{pp'}\eta_{qq'}\eta_{rr'}
 (4\varepsilon^{p'q'r'}C^{mnq,k}
  \non\\&&
 + \varepsilon^{np'q'} C^{mqr',k}
  +\varepsilon^{nq'r'}C^{mqp',k}
  +\varepsilon^{mq'p'}C^{qnr',k}
  +\varepsilon^{mq'r'}C^{qnp',k})\big]~,
 \label{C14}
\eea
where $\Theta_\delta \equiv \Theta^{\bf1}_\delta$.
The first term here coincides (up to the factor $-6$) with the
corresponding term in (\ref{n1.1.29}). To match the other terms we
need to consider also contributions to $H^{\alpha m,\beta n,\gamma k
}$ from the $B$ part given by (\ref{C8}).

In the $B$ sector of the correlation function
we need to compute three covariant spinor derivatives of the
tensor (\ref{C2}),
\be
D^{\bf2}_{(3)\gamma} D^{\bf2}_{(2)\beta} D^{\bf2}_{(1)\alpha}
  H^{mnk}|
=\ri({\bm x}_{23}^{-1})^\sigma{}_{\beta} ({\bm x}_{13}^{-1})^{\rho}{}_{\alpha}
D^{\bf2}_{(3)\gamma}[{\cal Q}^{\bf2}_\sigma {\cal D}^{\bf2}_\rho
+u^{\bf21}_{23}{\cal Q}^{\bf1}_\sigma{\cal D}^{\bf2}_\rho
+u^{\bf21}_{13}{\cal Q}^{\bf2}_\sigma{\cal D}^{\bf1}_\rho]  H^{mnk}|~.
\label{211_}
\ee
In this expression, $u^{\bf21}_{23}$ and $u^{\bf21}_{13}$ are components
of the matrix (\ref{two-point-u}) which appear in (\ref{211_})
owing to the identities (\ref{useful-prop}).
The factor $({\bm x}_{23}^{-1})^\sigma{}_{\beta} ({\bm
x}_{13}^{-1})^{\rho}{}_{\alpha}$ in the right-hand side of (\ref{211_}) is
the right one which is required to form the expression
(\ref{n1.1.5}). Now we have to analyse the remaining piece of this
expression.

Using the explicit expressions (\ref{generalized-DQ}) for
${\cal Q}^{\bf2}_\sigma$ and $ {\cal D}^{\bf2}_\rho$ the first term in
the right-hand side of (\ref{211_}) can be rewritten as
\bea
&&
D^{\bf2}_{(3)\gamma}{\cal Q}^{\bf2}_{\sigma} {\cal D}^{\bf2}_{\rho} H^{mnk}| \non\\
&=&
D^{\bf2}_{(3)\gamma}
\big[ \Theta^{{\bf2}\mu}
\frac\partial{\partial X^{\sigma\mu}}\frac\partial{\partial\Theta^{{\bf2}\rho}}
+\Theta^{{\bf2}\mu}
\frac\partial{\partial X^{\rho\mu}}\frac\partial{\partial\Theta^{{\bf2}\sigma}}
-\Theta^{{\bf2}\mu}\frac\partial{ \partial X^{\rho\sigma} } \frac\partial{\partial\Theta^{{\bf2}\mu}
}\big]H^{mnk}|
\non\\
&=& [({\bm x}^{-1}_{13})^\mu{}_\gamma - ({\bm x}^{-1}_{23})^\mu{}_\gamma]
 \left(   \frac\partial{\partial X^{\sigma\mu}} \frac\partial{\partial\Theta^{{\bf2}\rho}}
+\frac\partial{\partial X^{\rho\mu}} \frac\partial{\partial\Theta^{{\bf2}\sigma}}
 - \frac\partial{\partial X^{\rho\sigma}} \frac\partial{\partial \Theta^{{\bf2}\mu} }
\right)
 H^{mnk}|~.~~~
\label{C16}
\eea
To get the last line we used the fact that in the $|$-projection
only those terms survive in which the derivative $D^{\bf2}_{(3)\gamma}$
acts on $\Theta^{{\bf2}\mu}$ and produces the factor $[({\bm x}^{-1}_{13})^\mu{}_\gamma -
({\bm x}^{-1}_{23})^\mu{}_\gamma]$. However, the latter structure
is non-covariant in the sense that it cannot be expressed solely
in terms of ${\bm X}_{\alpha\beta}$ and $\Theta^I_\alpha$. Indeed, using
the identity (\ref{4.34}) and the definition of ${\bm X}_{3\alpha\beta}$
(\ref{three-points}), we represent the factor
$[({\bm x}^{-1}_{13})^\mu{}_\gamma - ({\bm x}^{-1}_{23})^\mu{}_\gamma]$ in
(\ref{C16}) as
\be
({\bm x}_{13}^{-1})_{\alpha\beta} - ({\bm x}_{23}^{-1})_{\alpha\beta}
=-{\bm X}_{3\alpha\beta} +\ri\frac{\varepsilon_{\alpha\beta}}{{\bm x}_{23}{}^2}
\theta_{23}^2
+2\ri ({\bm x}^{-1}_{13})_{\alpha\mu} \theta^\mu_{13}
 \theta^{\nu}_{32} ({\bm x}_{32}^{-1})_{\nu\beta}~.
\label{C17}
\ee
The last two terms here are non-covariant. Therefore, they must
cancel against similar terms coming from the last two terms in
(\ref{211_})
\bea
&&
D^{\bf2}_{(3)\gamma}[u^{\bf21}_{23}{\cal Q}^{\bf1}_\sigma{\cal D}^{\bf2}_\rho
+u^{\bf21}_{13}{\cal Q}^{\bf2}_\sigma{\cal D}^{\bf1}_\rho]  H^{mnk}|
=2\Theta^{\bf1}_{\gamma}\frac\partial{\partial\Theta^{{\bf1}\sigma}}
 \frac\partial{\partial\Theta^{{\bf2}\rho}} H^{mnk}
\non\\&& \qquad
-2\ri[({\bm x}_{23}^{-1})_{\gamma\mu} \theta^{\mu}_{23}\Theta^{{\bf1}\nu}
 \frac\partial{\partial X^{\sigma\nu}} \frac\partial{\partial\Theta^{{\bf2}\rho}}
+({\bm x}_{13}^{-1})_{\gamma\mu} \theta_{13}^{\mu}
  \Theta^{{\bf1}\nu}
  \frac\partial{\partial X^{\rho\nu}}
  \frac\partial{\partial\Theta^{{\bf2}\sigma}}]H^{mnk}|~.
\eea
To prove the cancellation of non-covariant terms we have to
use the fact the tensor $H$ is linear in $\Theta^{\bf1}$ and can be
represented in the form $H=\Theta^{\bf1}_\kappa h^\kappa$, for some
$h^\kappa$.
Here we suppress all indices of the tensor $H$ as they do not play
role in this consideration. Then, using the explicit expression for
$\Theta^{\bf1}_\alpha$ (\ref{three-points}), we observe that the
non-covariant terms have the same structure and cancel against
each other
\bea
\ri({\bm x}^{-1}_{13\mu\gamma}-{\bm x}^{-1}_{23\mu\gamma})\Theta^{\bf1}_{\kappa}h^\kappa
&=&-\ri{\bm X}_{3\mu\gamma} H
+\left(\frac{\theta_{23}^2}{{\bm x}_{23}^2}{\bm x}^{-1}_{13\gamma\kappa}\theta_{13}^\kappa
+\frac{\theta_{13}^2}{{\bm x}_{13}^2}{\bm x}^{-1}_{23\gamma\kappa}\theta_{23}^\kappa
\right)h_\mu~,\non\\
2{\bm x}^{-1}_{23\gamma\mu}\theta^\mu_{23}\Theta^{{\bf1}\nu}\Theta^{\bf1}_{\kappa}h^\kappa
&=&-\left(
\frac{\theta_{13}^2}{{\bm x}_{13}^2}{\bm x}^{-1}_{23\gamma\mu}\theta_{23}^\mu
 + \frac{\theta_{23}^2}{{\bm x}_{23}^2}
{\bm x}^{-1}_{13\gamma\mu}\theta_{13}^\mu
\right)h^\nu~,\non\\
2{\bm x}^{-1}_{13\gamma\mu}\theta^\mu_{13}\Theta^{{\bf1}\nu}\Theta^{\bf1}_{\kappa}h^\kappa
&=&-\left(
\frac{\theta_{13}^2}{{\bm x}_{13}^2}{\bm x}^{-1}_{23\gamma\mu}\theta_{23}^\mu
 + \frac{\theta_{23}^2}{{\bm x}_{23}^2}
{\bm x}^{-1}_{13\gamma\mu}\theta_{13}^\mu
\right)h^\nu~.
\eea

Thus in the expression (\ref{211_}) only covariant terms remain
\bea
D^{\bf2}_{(3)\gamma} D^{\bf2}_{(2)\beta} D^{\bf2}_{(1)\alpha}
  H^{mnk}|
&=&
\ri({\bm x}_{23}^{-1})^\sigma{}_{\beta} ({\bm x}_{13}^{-1})^\rho{}_{\alpha}
\bigg(
2\Theta^{\bf1}_\gamma \frac\partial{\partial\Theta^{{\bf1}\sigma}}
 \frac\partial{\partial\Theta^{{\bf2}\rho}}
+X^\mu_\gamma
\frac\partial{\partial X^{\rho\sigma}}\frac\partial{\partial\Theta^{{\bf2}\mu}}
\non\\&&
-X^\mu_\gamma\frac\partial{\partial X^{\sigma\mu}}\frac\partial{\partial\Theta^{{\bf2}\rho}}
-X^\mu_\gamma\frac\partial{\partial X^{\rho\mu}} \frac\partial{\partial\Theta^{{\bf2}\sigma}}
\bigg) H^{mnk}|~.
\label{C20}
\eea

To summarise, the  $B$ part of the correlation function (\ref{C8})
can be represented in the form
\be
B=\frac{{\bm x}_{13\alpha'\rho'}{\bm x}_{13\alpha''\rho''}{\bm x}_{13\alpha\rho}}{{\bm x}_{13}{}^8}
\frac{{\bm x}_{23\beta\sigma}{\bm x}_{23\beta'\sigma'}{\bm x}_{23\beta''\sigma''}}{{\bm x}_{23}{}^8}
H_{(B)}^{\rho\rho'\rho'',\sigma\sigma'\sigma''}{}_{\gamma\gamma'\gamma''}
(X_3,\Theta_3)
~,
\label{C21}
\ee
where the tensor $H_{(B)}$, after converting the pairs of spinor indices
into the vector ones, is expressed in terms of the derivatives of
(\ref{C2}) as follows
\bea
H_{(B)}^{\alpha m,\beta n}{}_{\gamma k}
 =
\bigg(&-&2 \Theta^{\bf1}_\gamma \frac\partial{\partial\Theta^{\bf1}_\beta}
 \frac\partial{\partial\Theta^{\bf2}_\alpha}
+ X_{\mu\gamma}\frac\partial{\partial X_{\beta\mu}}
 \frac\partial{\partial\Theta^{\bf2}_\alpha}
\non \\
&&\qquad
+X_{\mu\gamma}\frac\partial{\partial X_{\alpha\mu}}
 \frac\partial{\partial\Theta^{\bf2}_\beta}
-X_{\mu\gamma}\frac\partial{\partial X_{\alpha\beta}}
\frac\partial{\partial\Theta^{\bf2}_\mu}
 \bigg)H^{mn}{}_k |~.~~~
\label{222+}
\eea
Now we substitute here the tensor $H^{mnk}$ in the form (\ref{C2})
and compute the derivatives over the Grassmann variables. As a
result, with the use of identities with three-dimensional
gamma-matrices (\ref{A3},\ref{A4}), we find
\bea
\label{C22}
H_{(B)}^{\alpha m,\beta n,\gamma k}
 &=&\ri d_{\cN=2}\big[(\gamma^p)^{\alpha\beta} \Theta^\gamma C^{mnp,k}
 -3(\gamma^p)^{\alpha\beta}(\gamma^r)^{\gamma\rho} \Theta_\rho
  \varepsilon_{qpr}
   C^{mnq,k}\\&&
 -(\gamma^p)^{\alpha\beta}(\gamma^t)^{\gamma\rho} \Theta_\rho
  \varepsilon_{lst} X^l\partial^s C^{mnp,k}
 -(\gamma^p)^{\alpha\beta}(\gamma_s)^{\gamma\rho} \Theta_\rho
  \varepsilon_{qpl} X^l\partial^s C^{mnq,k}
  \non\\&&
 +(\gamma^p)^{\alpha\beta}(\gamma_l)^{\gamma\rho} \Theta_\rho
  \varepsilon_{qps} X^l \partial^s C^{mnq,k}
 +(\gamma^p)^{\alpha\beta}(\gamma^r)^{\gamma\rho} \Theta_\rho
  \varepsilon_{qsr} X^q\partial_p C^{mns,k}\big].
\non
\eea
In deriving this expression we have also used the simple relation
\be
X^l \partial_l C^{mnp,k}=-3C^{mnp,k}~,
\label{simple-relation}
\ee
which reflects the fact that the tensor
$C$ is homogeneous of degree $-3$ with respect to $X_m$.

The final result of computing the correlation function of $\cN=1$
supercurrent is given by the sum of the tensors (\ref{C14}) and
(\ref{C22}). It can be represented in the form similar to
(\ref{n1.1.29}):
\be
H^{\alpha m,\beta n,\gamma k} = -5\ri\, d_{\cN=2}\left( \gamma_p^{\alpha\beta}
\Theta^\gamma C^{mnp,k} + \gamma_p^{\alpha\beta} \gamma_r^{\gamma\delta}
 \Theta_\delta D^{(mn),k, p, r}\right)~,
\label{C23}
\ee
where
\bea
D^{(mn), k, p, r}&=&\frac15\big(
\varepsilon^{pqr}\eta_{qq'}C^{mnq',k}
 +\varepsilon^{nqk}\eta_{qq'}C^{mq'r,t}
 +\varepsilon^{nqr}\eta_{qq'}C^{mq'p,k}
 +\varepsilon^{mqp}\eta_{qq'}C^{q'nr,k}\non\\&&
 +\varepsilon^{mqr}\eta_{qq'}C^{q'np,k}
 +\varepsilon^{lsr}X_l\partial_s C^{mnp,k}
 +\varepsilon^{qpl} \eta_{qq'} X_l\partial^r C^{mnq',k}
 \non\\&&
 -\varepsilon^{qpl} \eta_{qq'} X^r\partial_l C^{mnq',k}
 -\varepsilon^{lqr} \eta_{qq'} X_q\partial^p  C^{mnq',k}
 \big)~.
\label{C24}
\eea

Our final task is to match the tensor (\ref{C24}) with the last
two terms in (\ref{n1.1.29}). A straightforward comparison is rather complicated and
we will give an indirect proof.
For this we will show that~\eqref{C24} satisfies all the same
equations as $D^{(mnp), k, r} $ from sect.\  \ref{sect-N1-supercurrent}.

First, one can show that~\eqref{C24} is symmetric and traceless in $(m, n, p)$
(though this symmetry is not manifest)
\be
D^{(mn), k, p, r} = D^{(mnp), k, r} \,, \quad \eta_{mn} D^{(mnp), k, r} =0\,.
\label{UR1}
\ee
Now we split $D^{(mnp), k, r}$ in~\eqref{C24} into the symmetric and antisymmetric parts in $(k, r)$.
Using the explicit form of the tensor $C^{mnp,k}$ in~\eqref{C3} one can show that
\be
\varepsilon^{krq}\eta_{kk'}\eta_{rr'}D^{(mnp), k', r'} = - C^{mnp,q} \,,  \quad \eta_{kr} D^{(mnp), k, r}=0\,.
\label{UR2}
\ee
This implies that $D^{(mnp), k, r}$ can be written as follows
\be
D^{(mnp),k, r} = D^{(mnp), (kr)} +\frac12 \varepsilon^{krq}\eta_{qq'} C^{mnp,q'}\,, \quad \eta_{kr} D^{(mnp), (k r)}=0\,.
\label{UR3}
\ee
We see that the antisymmetric part in eq.~\eqref{UR3}  precisely agrees with that of the tensor $D^{(mnp),k, r} $
in sect.\ \ref{sect-N1-supercurrent}, see eq.~\eqref{n1.1.17}. We are now left to match the symmetric part $D^{(mnp), (kr)} $.

To continue,  we contract~\eqref{C24} with $\eta_{pr}$ and $\varepsilon_{prq}$ to obtain
\be
\eta_{p r} D^{(mnp), k, r} =0\,, \quad \varepsilon_{rpq}  D^{(mnp), k, r} + \eta_{q q'} C^{mnq', k}=0\,,
\label{UR4}
\ee
which, using~\eqref{UR3}, imply~\eqref{n1.1.18}. Finally, using eqs.~\eqref{C3}, \eqref{C4}, \eqref{C5} we find that
\be
\partial_m D^{(mnp),k, r} =0~, \quad \partial_m D^{(mnp), (k r)} =0\,.
\label{UR5}
\ee

As a result, we found that $D^{(mnp), (k r)}$ satisfies exactly the same equations as in sect.\  \ref{sect-N1-supercurrent}.
On the other hand, we have shown in sect.\  \ref{sect-N1-supercurrent} that these equations allow us to fully
solve for  $D$ in terms of $C$ and such a solution is unique. Since the tensor $C$ in~\eqref{C2} coincides with the
one from sect.\  \ref{sect-N1-supercurrent} we conclude that~\eqref{C24} is the same as $D^{(mnp), k, r}$ found in sect.\  \ref{sect-N1-supercurrent}.
This completes our proof.

As a byproduct of the above computations,
we find that the
coefficients in the three-point functions of $\cN=1$ and $\cN=2$
supercurrents derived in the sections \ref{sect-N1-supercurrent}
and \ref{N2supercurrent} are related to each other as
\be
d_{\cN=1} = -5 d_{\cN=2}~.
\ee

\subsection{The flavour current correlation function}
\label{AppC2}
The $\cN=2\to\cN=1$ superspace reduction of the flavour current
correlation functions given by (\ref{5.57}) and (\ref{3pt-flavour}) goes
the same way as in the previous section. Therefore here we mention only
the essential details of this derivation.

Recall that the $\cN=1$ flavour current multiplet $L_\alpha$ appears as a
component of the $\cN=2$ flavour current superfield $L$ as in eq.\
(\ref{1.10b}). Hence, the corresponding relation for the correlation
functions reads
\be
\langle
L^{\bar a}_\alpha(z_1) L^{\bar b}_\beta(z_2) L^{\bar c}_\gamma(z_3) \rangle =
-\ri D^{\bf 2}_{(1)\alpha} D^{\bf 2}_{(2)\beta} D^{\bf 2}_{(3)\gamma}
\langle
L^{\bar a}(z_1) L^{\bar b}(z_2) L^{\bar c}(z_3) \rangle |~,
\label{C33}
\ee
where $|$ means that we set $\theta^{\bf 2}_\alpha=0$ at each superspace point.
Recall that the $\cN=2$ flavour current correlation function
(\ref{5.57}) consists of two parts which include tensors $f^{\bar a\bar b \bar
c}$ and $d^{\bar a\bar b\bar c}$ and both functions $H_{(f)}$ and
$H_{(d)}$ are non-trivial, see (\ref{7.12}). One could expect that
the corresponding $\cN=1$ correlator appearing in (\ref{C33}) may
include both such parts. However, as we will show further, the
part with the symmetric tensor $d^{\bar a \bar b \bar c}$ vanishes
upon this reduction and does not contribute to the $\cN=1$ flavour
current correlator.

Substituting (\ref{5.57}) into (\ref{C33}) we
represent the latter as a sum of the two pieces
\be
\langle
L_\alpha(z_1) L_\beta(z_2) L_\gamma(z_3) \rangle = f^{\bar a\bar b\bar c}(A_{(f)} + B_{(f)})
+d^{\bar a\bar b\bar c}(A_{(d)} + B_{(d)})~,
\ee
where
\bea
A_{(f,d)}&=& \frac{\ri}{{\bm x}_{13}{}^2} \left(
D^{\bf 2}_{(3)\gamma}D^{\bf 2}_{(2)\beta}\frac1{{\bm x}_{23}{}^2}
\right)D^{\bf 2}_{(1)\alpha} H_{(f,d)}(X_3,\Theta_3)
\non\\&&
-\frac{\ri }{{\bm x}_{23}{}^2} \left(
D^{\bf 2}_{(3)\gamma}D^{\bf 2}_{(1)\alpha}\frac1{{\bm x}_{13}{}^2}
\right)D^{\bf 2}_{(2)\beta} H_{(f,d)}(X_3,\Theta_3) |~,\non \\
B_{(f,d)}&=&  \frac{\ri}{{\bm x}_{13}{}^2 {\bm x}_{23}{}^2}
D^{\bf 2}_{(3)\gamma} D^{\bf 2}_{(2)\beta} D^{\bf 2}_{(1)\alpha}
H_{(f,d)}(X_3,\Theta_3)|~.
\label{C35}
\eea
The functions $H_{(f,d)}$ are given in (\ref{5.60}).

In the $A$ part, we apply the following equations:
\be
D^{\bf 2}_{(3)\gamma}D^{\bf 2}_{(2)\beta}\frac1{{\bm x}_{23}{}^2}|
= 2\ri\frac{{\bm x}_{23\beta\gamma}}{{\bm x}_{23}{}^4}~,\qquad
D^{\bf 2}_{(3)\gamma}D^{\bf 2}_{(1)\alpha}\frac1{{\bm
x}_{13}{}^2}|
= 2\ri\frac{{\bm x}_{23\alpha\gamma}}{{\bm x}_{12}{}^4}~.
\ee
Then using analogs of the equations (\ref{C10}), we represent the $A$ sector in the form
\be
A_{(f,d)}= \frac{{\bm x}_{13\alpha\alpha'} {\bm x}_{23\beta\beta'}}{{\bm x}_{13}{}^4 {\bm x}_{23}{}^4}
H_{(A,f,d)}^{\alpha'\beta'}{}_\gamma(X_3,\Theta_3)~,
\ee
where
\begin{subequations}
\bea
\label{C38}
H_{(A,f)}^{\alpha\beta\gamma} &=& 2\varepsilon^{\gamma\beta}{\cal D^{{\bf
2}\alpha}}H_{(f)} + 2\varepsilon^{\gamma\alpha} {\cal D}^{{\bf 2}\beta}
H_{(f)}| = b_{\cN=2}\frac{4\ri}{X^3}(\varepsilon^{\gamma\beta} X^{\alpha\rho}\Theta_\rho
+\varepsilon^{\gamma\alpha}
X^{\beta\rho}\Theta_\rho )~,~~~~~\\
H_{(A,d)}^{\alpha\beta\gamma} &=& 2\varepsilon^{\gamma\beta}{\cal D^{{\bf
2}\alpha}}H_{(d)} + 2\varepsilon^{\gamma\alpha} {\cal D}^{{\bf 2}\beta}
H_{(d)}| = 0~.
\label{C38a}
\eea
\end{subequations}

Now we consider the part $B$ in (\ref{C35}). Computation of this
piece goes similarly to the analysis given in the previous section. Indeed, the
equations (\ref{211_})--(\ref{C20}) remain exactly the same with
the only modification that we have to discard the indices $m,n,k$
in the tensor $H$. Thus, we can immediately write down the
analog of (\ref{C21}):
\be
B_{(f,d)} = \frac{{\bm x}_{13\alpha\alpha'} {\bm x}_{23\beta\beta'}}{{\bm x}_{13}{}^4 {\bm x}_{23}{}^4}
H_{(B,f,d)}^{\alpha'\beta'}{}_\gamma(X_3,\Theta_3)~,
\ee
where
\bea
H_{(B,f,d)}^{\alpha \beta }{}_{\gamma}
 =
\Big(-2 \Theta^{\bf1}_\gamma \frac\partial{\partial\Theta^{\bf1}_\beta}
 \frac\partial{\partial\Theta^{\bf2}_\alpha}
&+& X_{\mu\gamma}\frac\partial{\partial X_{\beta\mu}}
 \frac\partial{\partial\Theta^{\bf2}_\alpha}
+X_{\mu\gamma}\frac\partial{\partial X_{\alpha\mu}}
 \frac\partial{\partial\Theta^{\bf2}_\beta} \non \\
&-&X_{\mu\gamma}\frac\partial{\partial X_{\alpha\beta}}
\frac\partial{\partial\Theta^{\bf2}_\mu}
 \Big)H_{(f,d)} |~.
 \label{C40_}
\eea
Substituting the function (\ref{5.60}) into (\ref{C40_}) and computing the
derivatives we find
\begin{subequations}
\bea
H_{(B,f)}^{\alpha\beta\gamma} &=& b_{\cN=2} \frac{\ri}{X^3}(
10X^{\alpha\beta} \Theta^\gamma - 4 X^{\alpha\gamma}\Theta^\beta
-4X^{\beta\gamma}\Theta^\alpha\non\\&&
-6\varepsilon^{\gamma\alpha}X^{\beta\rho}\Theta_\rho
-6\varepsilon^{\gamma\beta}X^{\alpha\rho}\Theta_\rho)~,
\label{C41}\\
H^{\alpha\beta\gamma}_{(B,d)} &=&0~.
\label{HBd}
\eea
\end{subequations}

The equations (\ref{C38a}) and (\ref{HBd}) show that the $\cN=1$
flavour current correlation function does not receive
contributions with the symmetric tensor $d^{\bar a\bar b\bar c}$,
\be
H_{(d)}^{\alpha\beta\gamma} =H_{(A,d)}^{\alpha\beta\gamma} +
H_{(B,d)}^{\alpha\beta\gamma} =0~.
\ee
The other part with the antisymmetric tensor $f^{\bar a\bar b \bar c}$
is non-trivial. It is given by the sum of the expressions (\ref{C38}) and
(\ref{C41})
\bea
H_{(f)}^{\alpha\beta\gamma} &=&H_{(A,f)}^{\alpha\beta\gamma} +
H_{(B,f)}^{\alpha\beta\gamma} \non\\&=& b_{\cN=2} \frac{2\ri}{X^3}(
5X^{\alpha\beta} \Theta^\gamma - 2 X^{\alpha\gamma}\Theta^\beta
-2X^{\beta\gamma}\Theta^\alpha
-\varepsilon^{\gamma\alpha}X^{\beta\rho}\Theta_\rho
-\varepsilon^{\gamma\beta}X^{\alpha\rho}\Theta_\rho)~.~~~~
\label{C42}
\eea
Finally, applying the identity (\ref{id-6.15}), the tensor
(\ref{C42}) can be brought to the form
\bea
H_{(f)}^{\alpha\beta\gamma}(X,\Theta) &=& b_{\cN=2} \frac{2\ri}{X^3}(
X^{\alpha\beta} \Theta^\gamma
-\varepsilon^{\alpha\gamma}X^{\beta\rho}\Theta_\rho
-\varepsilon^{\beta\gamma}X^{\alpha\rho}\Theta_\rho)
\non\\
&=& b_{\cN=2} \frac{2\ri}{{\bm X}^3}(
{\bm X}^{\alpha\beta} \Theta^\gamma
-\varepsilon^{\alpha\gamma}{\bm X}^{\beta\rho}\Theta_\rho
-\varepsilon^{\beta\gamma}{\bm X}^{\alpha\rho}\Theta_\rho)~.
\label{C43}
\eea
Comparing the last expression with (\ref{H-N1-flavour}) we
conclude that the coefficients $b_{\cN=1}$ and $b_{\cN=2}$ are
related to each other as
\be
b_{\cN=1} = 2 b_{\cN=2}~.
\ee

\section{Component reduction}
\label{AppC3}

The correlation functions of the ener\-gy-mo\-men\-tum tensor
and  flavour currents originate  as components in the $\q$-expansion of the
  correlation functions for the supercurrent  and flavour current multiplets,
 respectively.
 In this section, we consider a
particular example in which we demonstrate how to derive the
correlation function of the flavour current from the
corresponding superfield correlator obtained in section
\ref{N1flavour-sect}.

We start with the $\cN=1$ flavour current correlator in the form (\ref{3pt-flavour}). Substituting
the latter into (\ref{Urr2}), we represent the correlation
function as a sum of two pieces
\be
\langle L_{\a \a'}^{\bar a} (x_1)  L_{\b \b'}^{\bar b} (x_2)  L_{\g \g'}^{\bar c} (x_3) \rangle
=A+B~,
\ee
where
\bea
A&=&f^{\bar a\bar b \bar c}\frac{{\bm x}_{13\alpha'\alpha''}}{{\bm x}_{13}{}^4}
\left( D_{(3)\gamma} D_{(2)\beta}\frac{{\bm x}_{23\beta'\beta''}}{{\bm x}_{23}{}^4}
\right)
D_{(1)\alpha}H^{\alpha''\beta''}{}_{\gamma'}(X_3,\Theta_3)
\non\\&&
-f^{\bar a\bar b \bar c}\frac{{\bm x}_{23\beta'\beta''}}{{\bm x}_{23}{}^4}
\left( D_{(3)\gamma} D_{(1)\alpha}\frac{{\bm x}_{13\alpha'\alpha''}}{{\bm x}_{13}{}^4}
\right)
D_{(2)\beta}H^{\alpha''\beta''}{}_{\gamma'}(X_3,\Theta_3)|
 ~,\label{C46-A}\\
B&=&f^{\bar a\bar b\bar c}\frac{{\bm x}_{13\alpha'\alpha''}{\bm x}_{23\beta'\beta''}}{
{\bm x}_{13}{}^4 {\bm x}_{23}{}^4}
 D_{(3)\gamma} D_{(2)\beta} D_{(1)\alpha}
 H^{\alpha''\beta''}{}_{\gamma'}(X_3,\Theta_3)|~.\label{C46-B}
\eea
In the  $A$ sector, we compute the derivatives of the objects ${\bm x}_{13}$ and ${\bm x}_{23}$
using their explicit form (\ref{super-interv-X}),
\bea
D_{(3)\gamma}D_{(1)\alpha} \frac{{\bm x}_{13\alpha'\alpha''}}{{\bm
x}_{13}{}^4}|
&=&\frac{2\ri}{{\bm x}_{13}{}^6}({\bm x}_{13\alpha\alpha''}{\bm x}_{13\alpha'\gamma}
 +{\bm x}_{13\alpha\gamma}{\bm x}_{13\alpha'\alpha''})~,\non\\
D_{(3)\gamma}D_{(2)\beta} \frac{{\bm x}_{23\beta'\beta''}}{{\bm
x}_{23}{}^4}|
&=&\frac{2\ri}{{\bm x}_{23}{}^6}({\bm x}_{23\beta\beta''}{\bm x}_{23\beta'\gamma}
 +{\bm x}_{23\beta\gamma}{\bm x}_{23\beta'\beta''})~.
 \label{C48}
\eea
With the use of (\ref{useful-prop}), the derivatives of the tensor
$H$ in (\ref{C46-A}) can be written as
\be
D_{(1)\alpha} H^{\alpha''\beta''}{}_{\gamma'}| = -\frac{{\bm
x}_{13\alpha\rho}}{{\bm x}_{13}{}^2}
{\cal D}^\rho H^{\alpha''\beta''}{}_{\gamma'}|~,\quad
D_{(2)\beta} H^{\alpha''\beta''}{}_{\gamma'}| =
\frac{{\bm x}_{23\beta\rho}}{{\bm x}_{23}{}^2}
{\cal D}^\rho H^{\alpha''\beta''}{}_{\gamma'}|~.
\label{C49}
\ee
Substituting (\ref{C48}) and (\ref{C49}) into (\ref{C46-A}) we
represent the  $A$ part in the form
\be
A=\frac{{\bm x}_{13\alpha\rho}{\bm x}_{13\alpha'\rho'}{\bm x}_{23\beta\sigma}
 {\bm x}_{23\beta'\sigma'}}{{\bm x}_{13}{}^6 {\bm x}_{23}{}^6}
  f^{\bar a\bar b \bar c}
  H_{(A)}^{\rho\rho',\sigma\sigma'}{}_{\gamma\gamma'}(X_3)~,
\ee
where
\be
H_{(A)}^{\rho\rho',\sigma\sigma'}{}_{\gamma\gamma'}=
 -4\ri \delta_\gamma^{(\underline{\sigma}} {\cal D}^\rho
 H^{\rho'\underline{\sigma}')}{}_{\gamma'}
 -4\ri \delta_\gamma^{(\underline{\rho}} {\cal D}^\sigma
 H^{\underline{\rho}')\sigma'}{}_{\gamma'}|~.
\label{C51}
\ee
The symmetrisation here involves only the underlined indices.

Consider the  $B$ part given by (\ref{C46-B}). Similarly as in eqs.\
(\ref{211_}) and (\ref{C16}), we find
\bea
&&D_{(3)\gamma} D_{(2)\beta} D_{(1)\alpha} H^{\alpha'\beta'\gamma'} = \ri ({\bm
x}^{-1}_{23})^\sigma{}_\beta ({\bm x}^{-1}_{13})^\rho{}_\alpha
D_{(3)\gamma} {\cal Q}_\sigma {\cal D}_\rho
H^{\alpha'\beta'\gamma'}\non\\
&=&-\ri ({\bm
x}^{-1}_{23})^\sigma{}_\beta ({\bm x}^{-1}_{13})^\rho{}_\alpha
 X^\mu_\gamma  \left(   \frac\partial{\partial X^{\sigma\mu}} \frac\partial{\partial\Theta^{\rho}}
+\frac\partial{\partial X^{\rho\mu}} \frac\partial{\partial\Theta^{\sigma}}
 - \frac\partial{\partial X^{\rho\sigma}} \frac\partial{\partial \Theta^{\mu} }
\right)H^{\alpha'\beta'\gamma'}|~.
\label{C53}
\eea
Here we used an analog of the identity (\ref{C17}) in which all
Grassmann variables vanish. Substituting (\ref{C53}) into
(\ref{C46-B}) we represent the  $B$ part of the correlation
function in the form
\be
B=\frac{{\bm x}_{13\alpha\rho}{\bm x}_{13\alpha'\rho'}{\bm x}_{23\beta\sigma}
 {\bm x}_{23\beta'\sigma'}}{{\bm x}_{13}{}^6 {\bm x}_{23}{}^6}
  f^{\bar a\bar b \bar c}
  H_{(B)}^{\rho\rho',\sigma\sigma'}{}_{\gamma\gamma'}(X_3)~,
\ee
where
\be
H_{(B)\rho\rho',\sigma\sigma',\gamma\gamma'}
=-\ri X^\mu_\gamma  \left(   \frac\partial{\partial X^{\sigma\mu}} \frac\partial{\partial\Theta^{\rho}}
+\frac\partial{\partial X^{\rho\mu}} \frac\partial{\partial\Theta^{\sigma}}
 - \frac\partial{\partial X^{\rho\sigma}} \frac\partial{\partial \Theta^{\mu} }
\right)H_{\rho'\sigma'\gamma'}|~.
\label{C55}
\ee

Now we substitute the tensor (\ref{H-N1-flavour}) into (\ref{C51})
and (\ref{C55}), and after computing derivatives, we find
\bea
H_{\rho\rho',\sigma\sigma',\gamma\gamma'}&=&H_{(A)\rho\rho',\sigma\sigma',\gamma\gamma'}
+H_{(B)\rho\rho',\sigma\sigma',\gamma\gamma'}\non\\
&=&\frac{3b_{\cN=1}}{X^5}X_{\rho\sigma}X_{\rho'\sigma'}X_{\gamma\gamma'}
-\frac{b_{\cN=1}}{X^3}\big[
X_{\rho'\sigma'}(\varepsilon_{\gamma\sigma}\varepsilon_{\rho\gamma'}
 +\varepsilon_{\gamma\rho}\varepsilon_{\sigma\gamma'})\non\\&&
 -5X_{\rho\sigma}(\varepsilon_{\gamma\sigma'}\varepsilon_{\rho'\gamma'}
  +\varepsilon_{\gamma\rho'}\varepsilon_{\sigma'\gamma'})
 +X_{\rho\rho'}(\varepsilon_{\gamma\sigma}\varepsilon_{\sigma'\gamma'}
  -2\varepsilon_{\gamma\sigma'}\varepsilon_{\sigma\gamma'})\non\\&&
 +X_{\sigma\sigma'}(\varepsilon_{\gamma\rho}\varepsilon_{\rho'\gamma'}
  -2\varepsilon_{\gamma\rho'}\varepsilon_{\rho\gamma'})
 +X_{\rho\sigma'}(\varepsilon_{\gamma\sigma}\varepsilon_{\rho'\gamma'}
  -2\varepsilon_{\gamma\rho'}\varepsilon_{\sigma\gamma'})\non\\&&
 +X_{\sigma\rho'}(\varepsilon_{\gamma\rho}\varepsilon_{\sigma'\gamma'}
  -2\varepsilon_{\gamma\sigma'}\varepsilon_{\rho\gamma'})
 -X_{\gamma\gamma'}(\varepsilon_{\rho\rho'}\varepsilon_{\sigma\sigma'}
  +\varepsilon_{\rho\sigma'}\varepsilon_{\sigma\rho'})\non\\&&
 +2X_{\rho\gamma}(\varepsilon_{\sigma\sigma'}\varepsilon_{\rho'\sigma'}
  +\varepsilon_{\sigma\rho'}\varepsilon_{\sigma'\gamma'})
 +2X_{\sigma\gamma}(\varepsilon_{\rho\rho'}\varepsilon_{\sigma'\gamma'}
  +\varepsilon_{\rho\sigma'}\varepsilon_{\rho'\gamma'})\non\\&&
 +X_{\sigma'\gamma}(\varepsilon_{\rho\rho'}\varepsilon_{\sigma\gamma'}
  +\varepsilon_{\sigma\rho'}\varepsilon_{\rho\gamma'})
 +X_{\gamma\rho'}(\varepsilon_{\sigma\sigma'}\varepsilon_{\rho\gamma'}
  +\varepsilon_{\sigma\gamma'}\varepsilon_{\rho\sigma'})\big]~.
\eea
This tensor defines the flavour current three-point correlation
function,
\be
\langle L_{\a \a'}^{\bar a} (x_1)  L_{\b \b'}^{\bar b} (x_2)  L_{\g \g'}^{\bar c} (x_3) \rangle
=\frac{{\bm x}_{13\alpha\rho}{\bm x}_{13\alpha'\rho'}{\bm x}_{23\beta\sigma}
 {\bm x}_{23\beta'\sigma'}}{{\bm x}_{13}{}^6 {\bm x}_{23}{}^6}
  f^{\bar a\bar b \bar c}
  H^{\rho\rho',\sigma\sigma'}{}_{\gamma\gamma'}(X_3)~.
\label{C56}
\ee

It is instructive to convert the pairs of spinor indices into
vector ones in the correlation function (\ref{C56}) to compare it
with the corresponding expression obtained in \cite{OP}. Using the
identity \footnote{Note that in the case $\cN=0$ the objects
(\ref{super-interv-X}) and (\ref{sym-interval}) coincide and
we do not distinguish ${\bm x}_{13}$ from $x_{13}$ in what follows.}
\be
-\frac12\gamma_m^{\alpha\alpha'}\gamma_n^{\beta\beta'}
 \frac{{\bm x}_{13\alpha\beta} {\bm x}_{13\alpha'\beta'}}{{\bm x}_{13}{}^2}
 = \eta_{mn}-2\frac{x_{13m}x_{13n}}{x_{13}{}^2}\equiv
 I_{mn}(x_{13})~,
\ee
we find
\bea
\langle
L^{\bar a}_m(x_1) L^{\bar b}_n(x_2) L^{\bar c}_k(x_3)\rangle&=&
-\frac18 \gamma_m^{\alpha\alpha'}\gamma_n^{\beta\beta'}
 \gamma_k^{\gamma\gamma'}
 \langle L_{\a \a'}^{\bar a} (x_1)  L_{\b \b'}^{\bar b} (x_2)  L_{\g \g'}^{\bar c} (x_3) \rangle
\non\\&=&\frac{I_{mm'}(x_{13})I_{nn'}(x_{23})}{x_{13}{}^4 x_{23}{}^4}
f^{\bar a\bar b\bar c}
H^{m'n'}{}_k(X_3)~,
\label{C58}
\eea
where
\bea
H_{mnk}(X)&=&-\frac18\gamma_m^{\rho\rho'}\gamma_n^{\sigma\sigma'}\gamma_k^{\gamma\gamma'}
 H_{\rho\rho',\sigma\sigma',\gamma\gamma'}(X)\non\\
&=&3d_{\cN=1} \left(
\frac{X_m X_n X_k}{X^5} + \frac{\eta_{nk}X_m - \eta_{mk}X_n -\eta_{mn}X_k}{X^3}
\right)~.
\eea
Finally, using the identity $X_3{}^2 = \frac{x_{12}{}^2}{x_{23}{}^2
x_{13}{}^2}$ we represent the denominator in (\ref{C58}) in a
symmetric form with respect to the indices labelling spacetime
points
\bea
\langle
L^{\bar a}_m(x_1) L^{\bar b}_n(x_2) L^{\bar c}_k(x_3)\rangle=
\frac{I_{mm'}(x_{13})I_{nn'}(x_{23})}{x_{12}{}^2 x_{13}{}^2 x_{23}{}^2}
f^{\bar a\bar b\bar c}\,
t^{m'n'}{}_k(X_3)~,
\label{C60}
\eea
where
\bea
t_{mnk}(X)= X^2 H_{mnk}(X)=3d_{\cN=1} \left(
\frac{X_m X_n X_k}{X^3} + \frac{\eta_{nk}X_m - \eta_{mk}X_n -\eta_{mn}X_k}{X}
\right)~.~~~
\label{C61}
\eea

\begin{footnotesize}

\end{footnotesize}

\end{document}